\documentclass[fleqn]{report}
\usepackage{epsfig}
\begin{document}           
\pagestyle{headings}
\begin{center}

\vspace{3.cm}
{\huge \bf Hadronic Particle Production 

\vspace{.5cm}
in Nucleus-Nucleus Collisions}

\vspace{1.cm}
{\large P. Senger$^1$ and H. Str\"{o}bele$^2$}  

\vspace{1.cm}
{\large
$^1$ Gesellschaft f\"{u}r Schwerionenforschung, Darmstadt}

\vspace{.5cm}
{\large
$^2$ Fachbereich Physik\\
Johann Wolfgang Goethe-Universit\"{a}t Frankfurt/Main
}

\vspace{4.cm}

{\bf abstract}
\end{center}

Data on hadronic particle production in symmetric nuclear collisions
from SIS/BEVALAC to SPS energies are reviewed. The main emphasis is put
on the production of pions, kaons, and antibaryons. 
Global features  will be discussed in 
terms of rapidity and transverse momentum
distributions  and the total energy stored in  produced particles.  
Pion and kaon production probabilities
are studied as function of beam energy and their distribution in polar
and azimuthal angle.
Special emphasis is put on medium effects expected for kaons in dense
nuclear matter at low energies.
An enhanced strange particle yield is found
at all energies, its explanation at SPS energies is still controversial.
Experimental data on antibaryon and multistrange hyperon  production is
less complete and does not allow for similar systematic studies.

% {\large to appear in publication in J. Phys. G,  Nucl. Part. Phys. }
\newpage
\tableofcontents
\indent
\chapter{Introduction}
Relativistic nucleus-nucleus collisions offer the unique possibiliy
to produce hadronic matter at high temperatures and densities in the 
laboratory.
At center-of-mass energies in the range from  0.2 to 10 GeV per nucleon,
temperatures of 50 to about 
200 MeV and baryonic densities up to 10 times normal 
nuclear matter density are reached. The variation of 
density with energy allows to extract information about the nuclear 
equation of state which is essential for the understanding of 
the dynamics of supernovae and the formation of neutron stars. 
The hadrons inside the dense baryonic medium are expected to
change their free particle properties due to 
the onset of chiral symmetry restoration. 
At energy densities above a few GeV/fm$^3$ the 
hadrons in the reaction zone can no longer exist as discrete entities;
they probably dissolve into a plasma of quarks and gluons. Such a state of 
matter is predicted to prevail in the interior of neutron stars and believed
to have existed in the early universe.
Hence, heavy ion reactions at relativistic energies are an essential tool
for the investigation of cardinal questions in  nuclear and subnuclear physics.

\vspace{.4cm}
Nuclear collisions are complex multibody reactions in which three stages
may be identified:(1) the interpenetration of the nuclei with highly non 
equilibrium hadronic - and at high energies partonic - interactions;
(2) the 'burning' of the fireball with its evolution towards chemical and 
thermal equilibrium; (3) the 'freeze-out' of the final state hadrons. The 
crucial questions concern the second and third stage:
does the system ''live'' long enough to  
behave as bulk matter  and do 
the  produced particles contain information about the  matter they originate 
from?  The experimental data on the 
ratios of produced particles and their distributions 
in phase space indeed indicate that they are emitted from an expanding
source which is close to thermal and chemical equilibrium.  
Once the steady state features of nuclear collisions are 
established, fundamental aspects of strong interactions come into focus:
the modification of hadron properties in the hadronic medium 
and the deconfinement of 
the quarks with its implications for the understanding of the physical 
vacuum.

\vspace{.4cm}

There are various experimental observables which give access to 
the physics of a nuclear collision. The multiplicities and momenta 
of emitted nucleons and  of nuclear fragments prevail at low energies. 
Electromagnetic radiation and lepton pairs are 
especially interesting because they are 
penetrating probes which, however, are strongly suppressed relative 
to hadronic signals. Newly produced hadrons, which are the 
subject of this article, dominate the final state at 
high energies. Here they are the main carriers of energy and entropy.
Near production threshold, i.e. at low energy, they are 
interesting observables because their emission characteristics are 
strongly affected by the interplay between medium and threshold effects.   
In the following we review hadronic particle production 
in nucleus-nucleus collisions at c.m. energies in  
the range from a few hundred MeV to 10 GeV per nucleon.

\vspace{.4cm}
The formation of a new hadron implies the creation of at least one 
$q\overline{q}$ pair and often the rearrangement with the 
primordial quarks in the initial nucleons. This complex process may be 
modified in  A+A collisions by medium effects like 
mean field potentials, Fermi motion, Pauli blocking of certain final
states normally attained in N-N collisions, and more subtle microscopic effects
like reabsorption or the scattering between hadronic resonances.      
Interactions involving resonances can change 
considerably the effective accessible phase space and, thus, particle
production thresholds in nuclear collisions.
A detailed picture of the large hadronic system formed in central 
nucleus-nucleus collision at relativistic energies has to account for the large
differences in densities of the different quark flavours. 
For not too high energies there will always
be a high population of the light $u$ and $d$ quarks which are brought in
by the projectile nucleons. Thus the reabsorption
of the antiquark of a newly created light $q\overline{q}$ pair is a
frequent process. If, however, a heavy quark pair like $s \overline{s}$
or $c \overline{c}$ etc. is produced, its annihilation is strongly supressed
as long as the density of the corresponding quark species is low.
This is equally true for mesons containing a heavy antiquark, for example the
K$^+$ meson. Thus K$^+$ mesons will leave 
the reaction zone unaffected by reabsorption.     
The approach to chemical equilibrium of the light quarks will 
be especially boosted and thus be much faster than for the heavy quarks
not only because of the smaller mass but also because of the favourable
environment rich in $u$ and $d$ quarks. 
The result of the equilibration for the different quark
flavours and also for the hadron abundances can be worked out in a bulk matter
scenario (using the laws of thermodynamics or hydrodynamics).
Microscopic transport models can in addition analyse the approach
to equilibrium. Both lead to predictions of particle 
ratios in the final state, which are observable in experiments.

This list of important and interesting subjects pertinent to nuclear 
collisions demonstrates not only the scientific potential of the field but
also its problems: the superposition, attenuation and amplification of 
many different phenomena will make it difficult to extract parameters of
individual elementary processes.  

\vspace{.4cm}
Particle production in nuclear collisions has been a subject of scientific 
research already for two decades \cite{stock}. 
Relatively light projectiles like
He, C, Ne, up to Ar had been available at the BEVALAC in Berkeley, at 
the Synchrophasotron in Dubna, and at the ISR in Geneva (He and D only). 
The main topics of research were collective effects and the
search for exotic phenomena. The latter was unsuccessful whereas the former  
opened up an avenue to study the bulk properties of nuclear matter far from
stability. In particular predictions  \cite{stoecker}
were confirmed that nuclear matter 
will exhibit hydrodynamic behavior if compressed to several times nuclear 
ground state density and if heated to temperatures in excess of
a few tens of MeV \cite{gustaf,renford,greiner,reisdorf}. 
Particle production experiments had their share in the 
research programme, but the main emphasis was put on the emission 
characteristics of the nucleons and the nuclear fragments \cite{bartke}.
At ultrarelativistic energies, however, particle production was found 
to be the dominant feature of nucleus-nucleus collisions \cite{stachel92}.

\vspace{.4cm}
Recently even the heaviest stable nuclei have been accelerated at the
BEVALAC, at the SIS of GSI in Darmstadt, at the AGS in Brookhaven,
and at the SPS of CERN in Geneva. New high sensitivity experiments at
all laboratories and the high energy beams available at CERN 
lead to new theoretical and experimental efforts 
to understand particle production in nuclear collisions.
Two extreme configurations will be of special importance:
At low energies particle production occurs near to the threshold and 
is thus especially sensitive to collective effects. Low energy in this 
context means that the c.m. kinetic energy per nucleon of the projectile
is of the same order of magnitude as the mass of the produced particle.
At high energy densities, parton effects should come into play.
In principle this is true also for high energy nucleon-nucleon interactions,
however, in nuclear collisions the origin of partonic effects is not a
large momentum transfer, but rather the overlap of the hadronic
particles. Therefore, subnucleonic effects in relativistic heavy 
ion collisions will set in at lower c.m. energies per nucleon than in
the interactions of two nucleons.

\vspace{.4cm}
In this article, beam energies of a few hundred  to a  few hundred thousand  
MeV per nucleon are considered. 
This range translates into c.m. kinetic 
energies (per nucleon) from approximately 200 MeV (BEVALAC/SIS) to 
about 10 GeV (CERN-SPS). 
We concentrate on hadronic particle production 
in high-energy nucleus-nucleus collisions; often we will compare its
characteristic features to the ones of nucleon-nucleon (N-N) interactions.  
The main emphasis will be put on the production of 
pions, kaons and antibaryons.
Other particles like heavy non-strange mesons
as well as nucleons will be mentioned only if necessary to round up 
the picture. Hydrodynamic-like phenomena resulting from pressure, expansion, 
and directed flow prevail at low energies   
and are seen almost exclusively in 
the nucleons, light fragments and their correlations. For a recent review 
of this subject see \cite{reisdorf}. Here we will cover only related 
phenomena observed in the meson spectra at low beam energies. 
Although there is a large body of
data on asymmetric colliding systems like light projectile on heavy target,
we concentrate on symmetric collision systems.

\chapter{Global features}
In this chapter we will consider the features of nucleus-nucleus collisions
in general and distinguish those which are unique to nuclear reactions from
others which are also observable in hadronic (i.e.  nucleon-nucleon)
interactions. 

A simple, although important feature of nuclear collisions is the fact
that the direction and magnitude of the impact parameter can be measured 
for each individual collision event. The experimental observable
for the centrality of the collision is 
usually the number of nucleons which
participate in the collision (A$_{part}$). 
Independent of the size of the collision system, 
A$_{part}$ determines the reaction volume and the total center-of-mass 
energy.  For impact parameter integrated (inclusive)
cross sections  and symmetric collision systems, 
the mean number of interacting nucleons can be estimated to 
$<$A$_{part}> \approx$  A/2
with  A the mass number of the colliding nuclei \cite{cugnon1,huefner}. 
The mean multiplicity of a particle species is given by the ratio of 
the inclusive production cross section to the reaction cross section 
($<M>=\sigma_{incl}/\sigma_R$) or determined directly from an
event-by-event measurement. For comparisons of different size systems
it is convenient to use the normalized mean multiplicity
$<M>/A_{part}$ (see section 3.1). 
The direction of the impact parameter  defines the reaction plane 
and can be determined by the sum of the transverse momentum vectors of the 
projectile spectator fragments \cite{dan_ody}.   

\vspace{.4cm}
The most important features of nuclear collisions are
collective effects. They 
are related to the dynamics and the bulk properties of the reaction zone like
nuclear stopping, the built-up of pressure, the energy density, 
thermal and chemical equilibration, the  decompressional flow.
Some aspects of collective phenomena can be studied by means of
global observables like 
the phase space distributions of the produced particles,
the inelasticity of the collisions,
the ratios of different particle species  and particle correlations. 
The latter are used to obtain information about the
space-time configuration of the reaction zone and are not a subject of
of this review (see e.g.\cite{HBT}). The phase space
distributions carry information about the degree of thermalisation and
collective motion and will be considered  in the following section.

\begin{figure}[ht]
\vspace{0.cm}
\begin{minipage}[t]{10cm}
\hspace{ 0.cm}\mbox{\epsfig{file=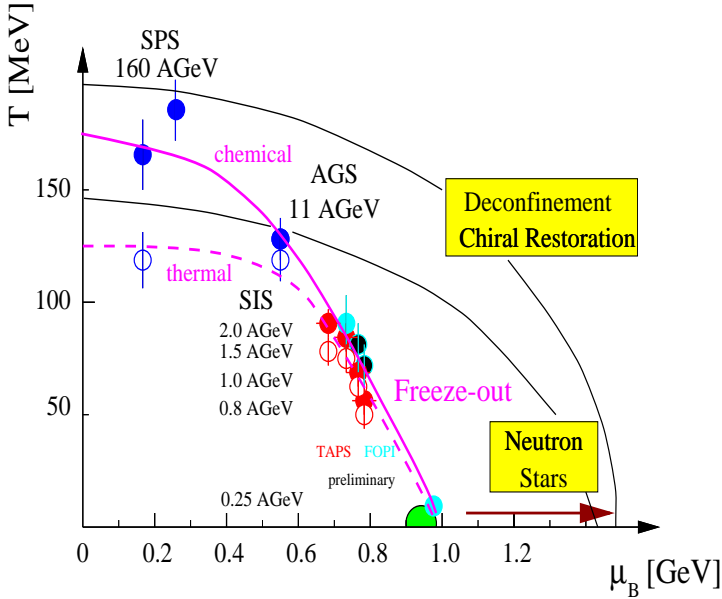,width=9.5cm,height=8cm}}
\end{minipage}
\begin{minipage}[t]{6cm}
\vspace{-7.5cm}
\caption{Phase diagram of hadronic matter: the 
temperature T and  the baryochemical potentials $\mu_B$ are derived from 
yield ratios of particles produced in nucleus-nucleus collisions
at different incident energies. The thick solid line through the data points
represents the curve for chemical freeze-out of hadronic matter. 
Open data points (connected by the dashed line) represent parameter pairs 
for thermal freeze-out. The figure is taken from 
\protect\cite{averbeck2}, the results are from
\protect\cite{averbeck2,pbm1,pbm2,hong,stock2}.
}
\label{t_my}
\end{minipage}
\end{figure}

\vspace{.4cm}
Particle ratios can be employed 
to evaluate the degree of chemical equilibration observed
in the final state of hadronic collisions. 
For nuclear matter in equilibrium, the particle composition 
is characterized by a temperature T and various chemical potentials. 
The particle number densities can be 
calculated considering a grand-canonical ensemble 
(as the particle number is not conserved) of noninteracting 
fermions and bosons:    

\begin{equation}
\rho_i = \frac{g_i}{2\pi^2} \int \limits_0^{\infty}
\frac{p^2 dp}{exp[ (E_i -\mu_BB_i -\mu_SS_i)/T ] \pm 1}
\label{rho_i}
\end{equation}
where $g_i$ is the spin-isospin degeneracy, $E_i$ the total energy in
the local restframe, $B_i$ ($S_i$) the baryon (strangeness) number and 
$\mu_B$ ($\mu_S$) the baryon (strange) chemical potential.  
In general one has to modify expression \ref{rho_i} in order account 
for the finite size of the system, for the volume of the particles
(''surface'' and ''excluded volume'' corrections, see \cite{pbm1}) and 
for resonance decays which occur in a later stage of the collision.
The results of such an analysis for particle
production data from  nuclear collisions at SIS 
\cite{averbeck2}, AGS \cite{pbm1,Spieles}, and SPS \cite{Spieles,pbm2} energies
show consistently that the relative abundances of different particle species
are close to what is expected for a system in chemical equilibrium.
Fig.~\ref{t_my} presents the resulting phase diagram of hadronic matter 
which  suggests  that for SPS and AGS energies the collision 
system at freeze-out is close to the phase transition to the 
quark-gluon-plasma.

Moreover it has recently been shown that 
hadronic final states in high energy e$^+$+e$^-$ and p+p interactions
also fulfil the criteria of chemical equilibrium \cite{becattini} 
if a suppression of strange particles is assumed.
This behavior is expected if particle production is a microscopic 
process involving many particles and interactions. 
The appearance of a new hadron is always preceeded by the 
creation of a quark-antiquark pair. In the next step one of the quarks is
normally exchanged with another one from the surrounding medium
(hadron or string). Finally the quarks get dressed to become constituent
quarks. It is conceivable that this complicated 
multistep and multibody process probes the available
phase space which will lead to a thermodynamic 
behavior of the produced particles. 

\vspace{.4cm}
The finding of thermodynamic behavior in both nucleus-nucleus and 
nucleon-nucleon
collisions can be interpreted in the following way: All hadrons are created
by processes involving the partonic medium which constitutes the interior
of hadrons. In hadronic interactions strong constraints on energy,
momentum, and quantum numbers affect the configuration of the final state
hadrons. In central A+A collisions the constraints are losened and a
grand canonical ensemble emerges. 
Therefore, the resulting hadronic fireball is close to thermal and 
chemical equilibrium because 
the prehadronic partonic
processes randomly filled phase space according to the laws of statistics
rather than because of many elastic and inelastic collisions.

\section{Rapidity distributions}

Particle spectra are often 
treated separately in longitudinal and transverse directions. This is because 
longitudinal velocities are mostly relativistic. The corresponding momenta 
need therefore a relativistic treatment which is handled most easily when 
using the rapidity 
\begin{equation}
y=0.5\cdot ln\{(E+p_L)/(E-p_L)\}
\end{equation}
\noindent
with $E$ and $p_L$ being the energy and longitudinal momentum of the particle.
Such a representation guarantees that the shape of the corresponding 
distribution is independent of the Lorentz frame~\footnote{
On the other hand  it implies that
a possible kinematical correlation of the considered particles 
with the projectiles in their c.m.-frame 
may be hidden. In such a case the Feynman x 
variable x$_F$=p$_L$/p$_{max}$ is more appropriate}. 

Rapidity and transverse momentum distributions allow to address some 
important questions on the 
reaction dynamics and on properties of the particle emitting source:
Are the nucleons slowed down to the extend that their motion
is entirely thermal or is nuclear matter
becoming transparent at high energies, is the source spherical symmetric 
or elongated, is there a transverse and longitudinal expansion, 
is the source thermally equilibrated?  

\begin{figure}[hpt]
\vspace{1.5cm}
\hspace{ 1.cm}\mbox{\epsfig{file=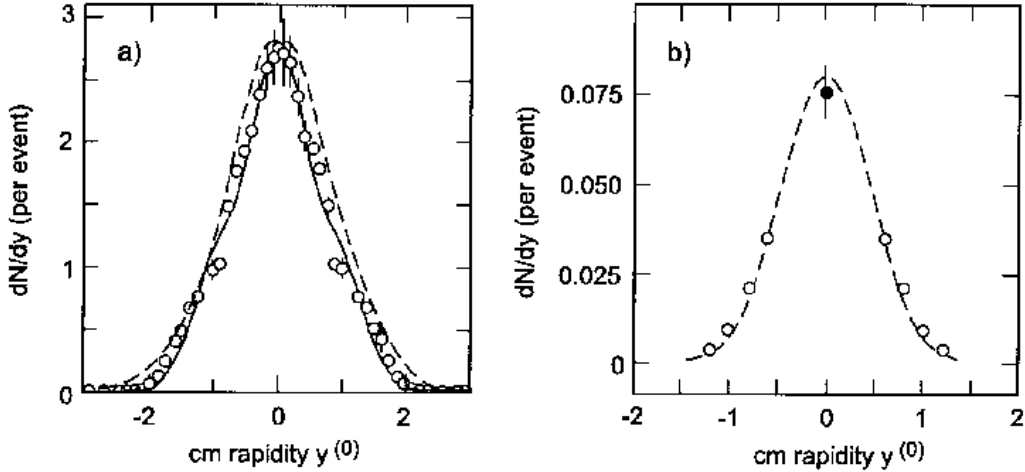,width=13.5cm}}
\caption{Rapidity density distributions of $\pi^-$ mesons from 
central Au+Au collisions at 1.06 AGeV (left) 
\protect\cite{pelte2} and of K$^+$ mesons 
from Ni+Ni collisions at 1.93 AGeV (right)
\protect\cite{best}. The data are measured in the 
backward hemisphere ($\theta^{\pi}_{lab} > 30^0$, $\theta^{K^+}_{lab} > 44^0$) 
and symmetrized with respect to midrapidity. The solid line (left)
is a fit to the data assuming three thermal pion sources
\protect\cite{pelte2}.The dashed lines correspond to a thermal source 
at midrapidity with two temperatures (left) and one temperature (right).      
}
\label{dndy_sis}
\end{figure}

\vspace{.4cm}
Fig.~\ref{dndy_sis} shows rapidity density distributions  
dN/dy$^{(0)}$ of charged pions 
for very central collisions (corresponding to a cross section of 100 mb) of  
Au+Au at 1.06 AGeV \cite{pelte1} 
and for positively charged kaons in Ni+Ni at 1.93 AGeV
\cite{best}. 
The data are measured in the backward 
hemisphere only and symmetrized with respect to midrapidity
(y$^{(0)}$=y/y$_{proj}$ with y$_{proj}$ being the 
center-of-mass rapidity of the projectile and  y$^{(0)}$=0 at midrapidity).

The distributions have a gaussian-like shape.
For the pions the full width at half height is approximately a factor 
of 1.8 larger than the beam rapidity (y$^{beam}$=0.67 in the c.m.s. at 1 AGeV).
This pertains also to central Ni+Ni collisions at 1.93 AGeV (not shown).
The corresponding number for K$^+$ is about 1.2 for Ni+Ni collisions
at 1.93 AGeV. 

For the Ni+Ni system, the width of the pion and kaon 
rapidity distributions is in fairly good  
agreement with the assumption of an isotropically
emitting thermal source \cite{hong,best,pelte2}. 
For  Au+Au collisions at 1.06 AGeV, the width of the pion rapidity distribution
is found to be slightly 
narrower than expected for an isotropic thermal source located at midrapidity 
(dashed lines: fit with two temperatures). 
For a detailled discussion of the pion data see section 3.2.

\begin{figure}[ht]
\vspace{-0.5cm}
\begin{minipage}[t]{11cm}
\mbox{\epsfig{file=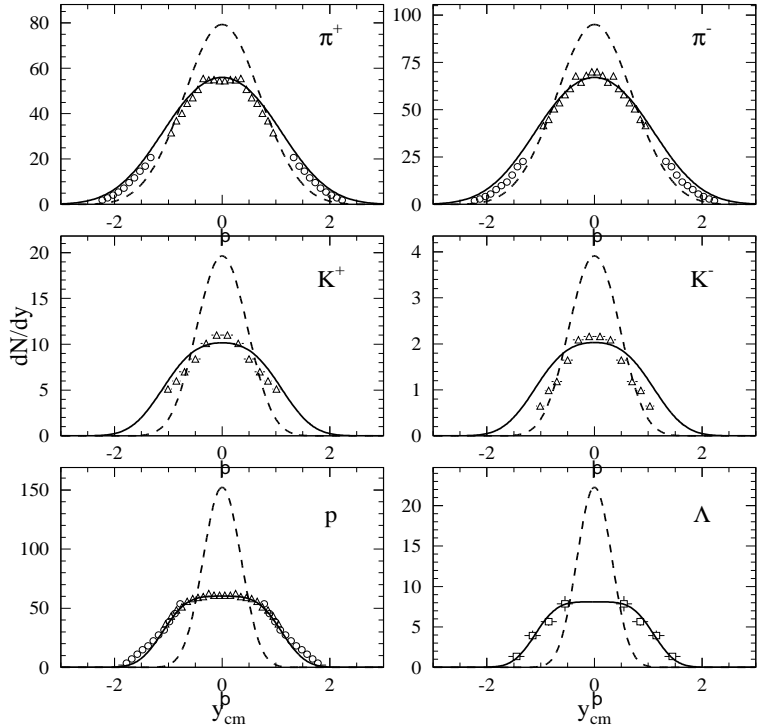,width=11.cm,height=10.5cm}}
\end{minipage}
\begin{minipage}[t]{4cm}
\vspace{-7cm}
\caption{Rapidity distributions of various particles as measured
in central Au+Au collisions at  10 AGeV.  
Dashed lines: isotropic thermal distribution for T = 0.13 GeV.  
Solid lines: distributions for a source at the same temperature expanding
with $<\beta_l>$= 0.5. The picture is taken from 
\protect\cite{stachel96}, the data are from  
\protect\cite{e802,e814,lacasse,Ahmad}.
}
\label{dndy_ags}
\end{minipage}
\end{figure}

Fig.~\ref{dndy_ags} 
shows the rapidity distributions of pions, kaons, protons and Lambdas  
measured in central Au+Au collisions  
at 10 AGeV  \cite{stachel96,e802,e814,lacasse,Ahmad}.
All distributions have a full width 
at half maximum around 2.2 units of rapidity which translates 
into a width of $\delta$y$^{(0)}$=1.4, 
except for the negatively charged kaons for 
which the width comes out smaller by 20\% (see Table \ref{dndywid}).
The widths of the rapidity distributions are larger than expected 
for an isotropic thermal source with a temperature of T=0.13 GeV 
(dashed lines in Fig.~\ref{dndy_ags}). 
The data (except for K$^-$) can be reproduced   
assuming a longitudinally expanding source with an 
average expansion velocity of $<\beta_l>$=0.5 and a temperature of T=0.13 GeV
\cite{stachel96}. This corresponds to  local isotropic sources
distributed over a rapidity interval of $\pm$ 1.1 units around midrapidity.
It is interesting to note that  
the widths of the pion and kaon rapidity distributions  
are  independent of the system size (see Table \ref{dndywid}). The values
found for  Si+Al collisions at 13.7 AGeV \cite{abbot} and even for
p+p collisions at 12 AGeV \cite{Blobel} are within errors the same.

\begin{figure}[hpt]
%\vspace{-0.5cm}
\hspace{ 2.cm}\mbox{\epsfig{file=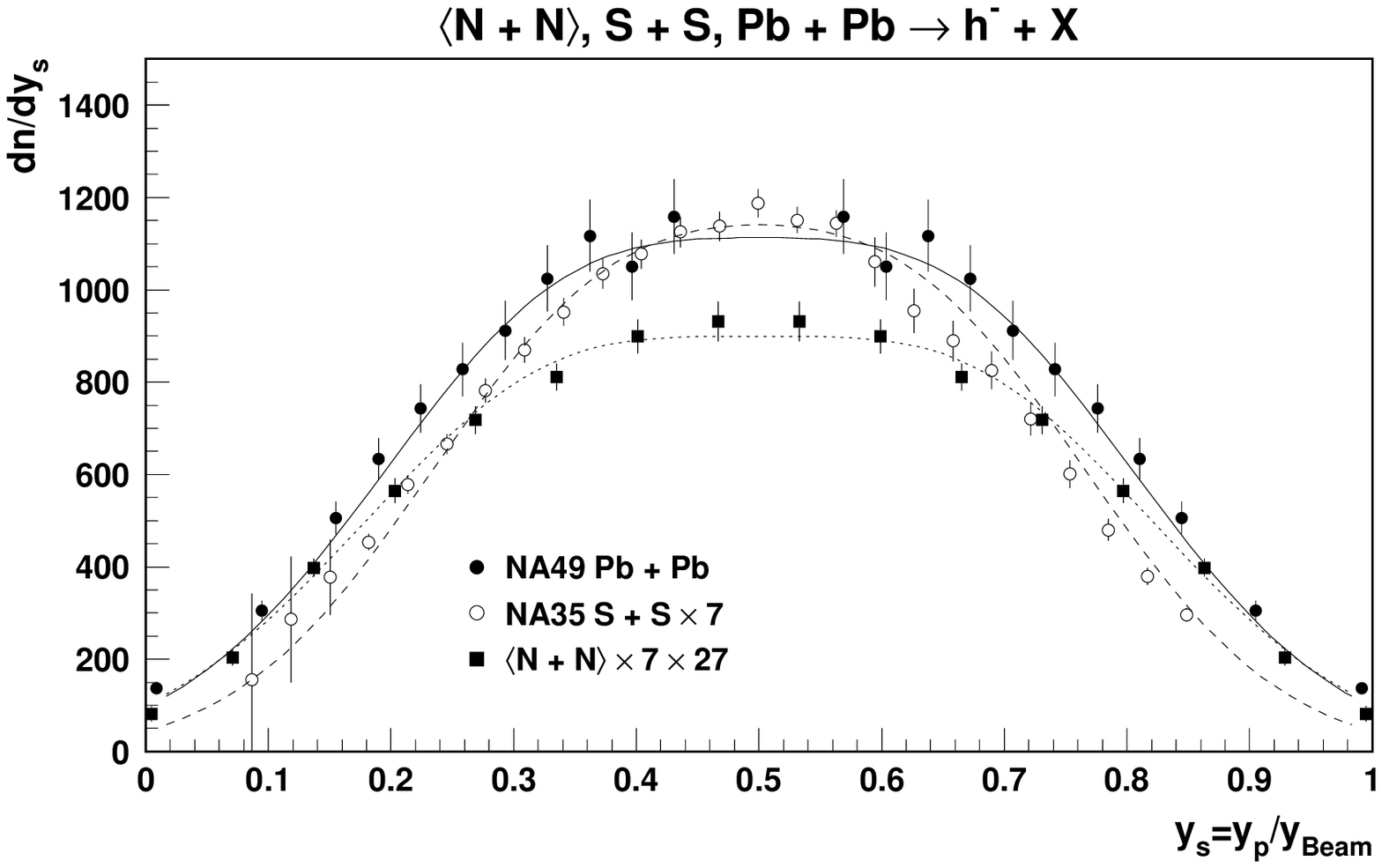,width=12.cm}}
\caption{
Rapidity density distributions of negatively charged hadrons from
central Pb+Pb  at 158 AGeV (full dots), S+S at 200 AGeV collisions 
(open dots) together with minimum bias nucleon-nucleon interactions 
(full squares). The Pb+Pb and N+N data are reflected around
midrapidity, the S+S data are not. The curves (gaussian shape for S+S, 
double gaussian for Pb+Pb and N+N) are meant to guide the eyes.
The picture is taken from \protect\cite{guen97}.
}
\label{dndy_na49pi}
\end{figure}

\begin{figure}
\vspace{0.cm}
%\begin{center}
\begin{minipage}[t]{11cm}
\hspace{ 1.5cm}\mbox{\epsfig{file=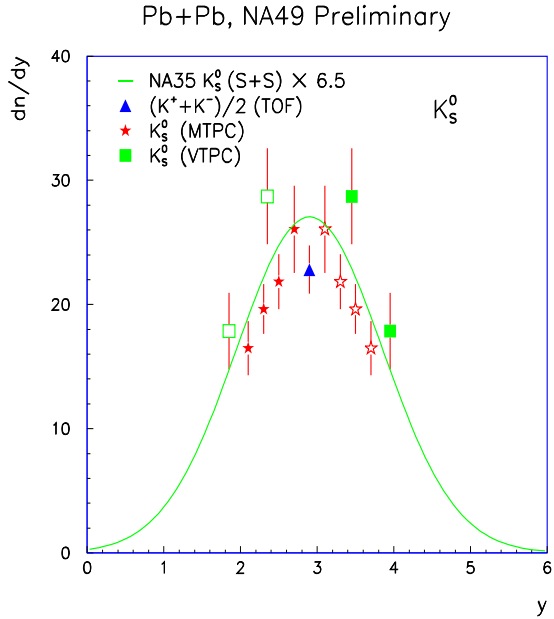,width=7.cm,height=8.5cm}}
\end{minipage}
\begin{minipage}[t]{5cm}
%\vspace{1.5cm}
%\end{center}
\vspace{-7cm}
\caption{
Rapidity density distributions
of kaons from  central Pb+Pb collisions at 158 AGeV (symbols, preliminary) 
compared to the kaons  measured in S+S collisions 
at 200 AGeV (line). The data are reflected around midrapidity. 
The figure is taken from \protect\cite{seyboth97}.
}
\label{dndy_na49k}
\end{minipage}
\end{figure}

\vspace{.4cm}
The situation is similar at SPS energies. Fig.~\ref{dndy_na49pi} shows the 
rapidity distributions of negatively charged particles (mostly pions
with an admixture of  8\% K$^-$ and 2\% $\overline{p}$)  
measured in central Pb+Pb (S+S) collisions at 
158 (200) AGeV by the NA49 (NA35) collaboration 
together with data from nucleon-nucleon collisions \cite{guen97}. 
The agreement of  nucleon-nucleon and nucleus-nucleus data excludes
collective flow to be the reason for the large width of the rapidity
distribution.  

Fig.~\ref{dndy_na49k} shows the rapidity distributions for kaons measured 
in central  Pb+Pb collisions at 158 AGeV compared to the kaon distributions
from S+S at 200 AGeV \cite{seyboth97}.
The kaon data from Pb+Pb (which have large systematical 
errors) fall out of the general systematics which says that all widths of 
the kaon distributions are only slightly smaller than those of the pions and 
that the widths are independent of the size of the colliding system.

\begin{figure}
%\vspace{0.cm}
\begin{minipage}[t]{11cm}
\hspace{ 0.cm}\mbox{\epsfig{file=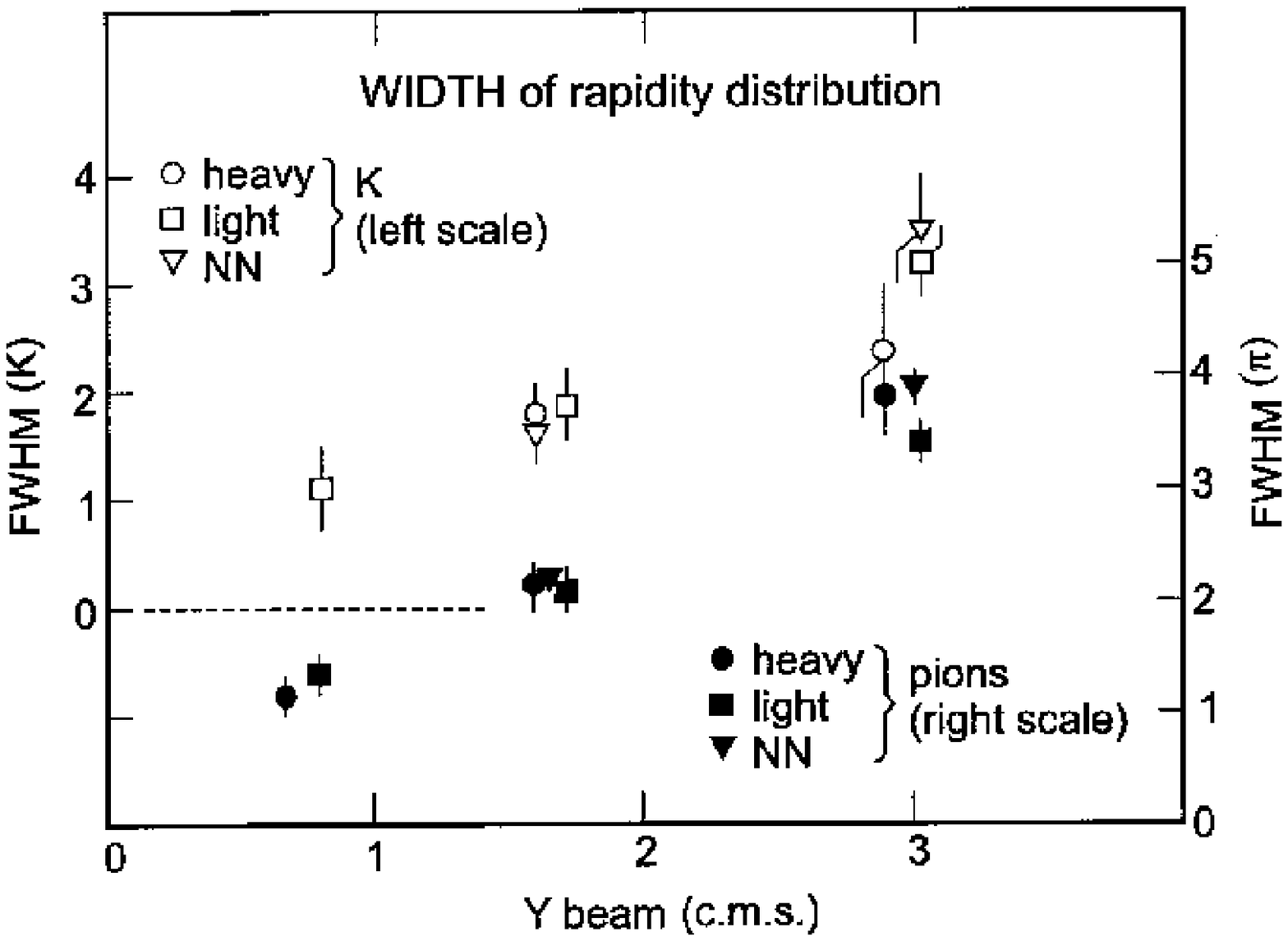,width=10.5cm}}
\end{minipage}
\begin{minipage}[t]{4.5cm}
\vspace{-7cm}
\caption{ Full width at half maximum of pion and kaon rapidity distributions 
in nucleon-nucleon and (heavy and light) 
nucleus-nucleus collisions as a function of 
the c.m. beam rapidity.}
\label{delta_y}
\end{minipage}
\end{figure}

\begin{table}
\caption{Widths (FWHM) of rapidity distributions of $\pi$- and K-Mesons in
central nucleus-nucleus and nucleon-nucleon collisions. The 158 AGeV Pb+Pb
data are preliminary; N+N data at 200 GeV are K$^0_s$ only.
$\delta$y$^{(0)}$ is the width in units of the beam rapidity in the c.m.
system.}
\vspace{0.5cm}
\begin{center}
\begin{tabular}{|c|c|c|c|c|c|c|}
\hline
beam energy& system & $\pi$& $\pi$& K & K &  Ref.\\
&&$\delta$y$^{(0)}$&$\delta$y&$\delta$y$^{(0)}$&$\delta$y&\\
\hline
\hline
 1 AGeV& Au+Au& 1.8& 1.2&&&\cite{pelte1}\\
 1.93 AGeV& Ni+Ni& 1.8& 1.6& 1.2& 1.07&\cite{pelte2,best}\\
\hline
 10 AGeV& Au+Au& 1.37 & 2.2 & 1.12 & 1.8$\pm$0.3&\cite{e802,lacasse}\\
 12 AGeV& p+p& 1.37 & 2.2$\pm$0.1 & 1.02 & 1.6$\pm$0.2&\cite{Blobel}\\
 13.7. AGeV& Si+Al& 1.22 & 2.1$\pm$0.1 & 1.1 & 1.9$\pm$0.3&(K$^0$)\cite{abbot}\\
 13.7 AGeV& Si+Al&  & &  & 2.0$\pm$0.3&(K$^+$)\cite{abbot}\\
 13.7 AGeV& Si+Al&  & &  & 1.6$\pm$0.3&(K$^-$)\cite{abbot}\\
\hline
158 AGeV&Pb+Pb& 1.3 & 3.8$\pm$0.4& 0.76& 2.4$\pm$0.6& \cite{bormann}\\
200 AGeV&S+S& 1.13 & 3.4$\pm$0.2& 1.06& 3.0$\pm$0.3& \cite{Bae93,
Alb94}\\
200  GeV&N+N& 1.3 & 3.9$\pm$0.2& 1.12& 3.4$\pm$0.5& \cite{Eisenberg,
Jae75}\\
\hline
T=160MeV&thermal&&1.75&&1.1&\\
\hline
\end{tabular}
\end{center}
\label{dndywid}
\end{table}

The results for the widths of the rapidity distributions are
summarized in table \ref{dndywid} and Fig.~\ref{delta_y} 
in which for comparison the 
data from nucleon-nucleon (p+p) interactions have been added.
Except for the lowest (SIS) energies the width of all rapidity 
distributions are wider than expected for an isotropically emitting  
thermal system. From Fig.~\ref{delta_y} one concludes that the widths
of the rapidity density distributions scale with beam rapidity both for
pions and kaons. The spread of the rapidity distribution measured 
in units of c.m. beam rapidity shows a small but continuous decline from
1.8 at low (SIS) energies to 1.2 at high (SPS) energies 
(see Table \ref{dndywid}).
The corresponding number for the high energy data
from the microscopic transport model UrQMD is 1.1 \cite{SBass}. It will be 
interesting to see whether this model also finds the same trend as a 
function of energy.

\vspace{.4cm}
At SPS energies the distribution of the produced particles 
is twice as wide as expected from a thermal fireball.
This longitudinal boost of the pions and kaons from A+A collisions is 
very similar to what is observed  in p+p interactions. In fact at the 
highest energy the latter exhibits even a slightly larger width than 
what is obtained from nucleus-nucleus collisions. 

The longitudinal boost in p+p interactions was explained by the formation
and fragmentation of strings which are the dominant source of new
particles. The fact that a very similar boost is observed in nuclear
collisions at the same energy suggests that the longitudinal
distribution of the strings is the same in A+A and p+p collisions.
Any additional decompressional flow will have only a minor impact
on the widths of the rapidity distributions. It will, however, modify
significantly the transverse momentum distributions as will be discussed
in the following section.

\section{Transverse momentum distributions}

Transverse momenta of particles produced in hadronic interactions 
are one of the measures of the internal energy of the system. 
However, quantitative comparisons must pay attention to the proper
definition of the transverse directions. Normally they are
defined with respect to the beam direction. This may not always be 
appropriate \footnote{ For example in hadronic interactions involving 
hard parton 
scattering the source of the hadrons may be the jet in the wake of the 
leading quark. The transverse momenta of those hadrons  are then the components 
perpendicular to the jet axis}. In the case of a single source emitting 
particles according to 
a thermal spectrum the kinetic energies are Boltzmann distributed. 
This scenario seems to be adequate for low energy nuclear collisions.

For high-energy  nucleon-nucleon and nucleus-nucleus  
collisions Bjorken proposed a longitudinal expanding boost-invariant reaction
zone \cite{Bjorken}. In this case the transverse momentum spectrum is 
independent of rapidity and the shape of the invariant cross section
as a function of transverse momentum  coincides with the one obtained from
a single fireball (at the same temperature) integrated over all rapidities.
If the kinetic energy (in the transverse direction) of the hadrons
is given not only by the  thermal motion but also by collective
effects like flow, the magnitude and the shape of the 
transverse momentum distribution may change
but  still reflect the total kinetic 
(transverse ) energy. In such a case and for transverse momenta between
approximately 0.5$\times$m$_0\times$c and 5$\times$m$_0\times$c 
a single slope is in general a good 
approximation but it will be different for 
different particle masses (m$_0$ = rest mass, c = velocity of light).

\vspace{.4cm}
In the following paragraph we avoid the 
problem of nonthermal slopes  by
using  the average momentum as a measure of the energy in the transverse 
degrees of freedom.
For heavy particles this quantity is not always available because
of inclomplete coverage of p$_T$. In such cases 
the average p$_T$ is derived from the slopes, 
since for heavy particles  a single slope seems to 
describe well the transverse mass spectra.
   
\begin{figure}[ht]
\vspace{0.cm}
\begin{minipage}[t]{11.5cm}
\hspace{ 0.cm}\mbox{\epsfig{file=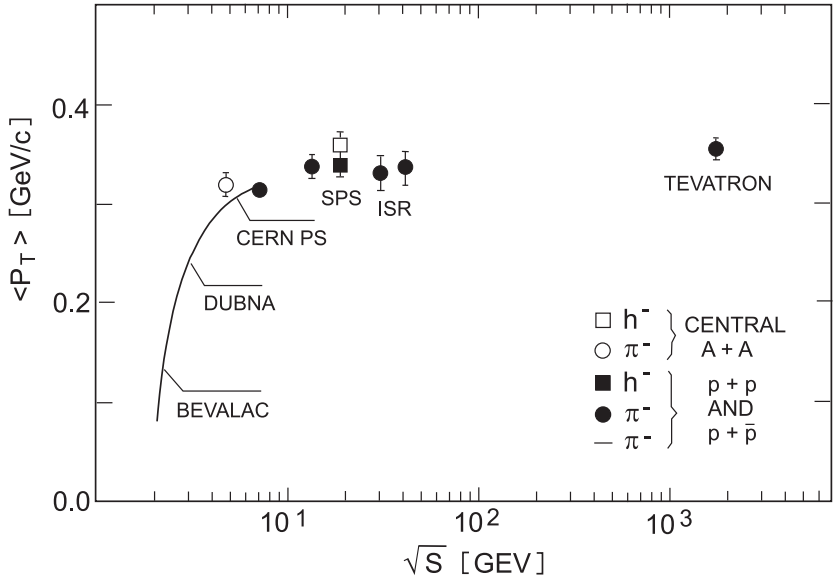,width=11.cm}}
\end{minipage}
\begin{minipage}[t]{4cm}
\vspace{-7.cm}
\caption{ Average transverse momenta of negatively charged
hadrons produced in p+p, p+$\protect\overline p$ and A+A collisons 
as function of $\protect\sqrt s$.
}
\label{pt_pp}
\end{minipage}
\end{figure}

Fig.~\ref{pt_pp}
summarizes the average transverse momenta of negatively charged
hadrons measured in hadronic
interactions. The average transverse momenta of $\pi^-$ produced in
p+p (p+$\overline{p}$) interactions rise steeply with $\sqrt{s}$ 
from low (BEVALAC/SIS) to medium (AGS) energies with a saturation
beyond. In fact the mean transverse momentum of negatively charged
pions remains constant within some ten MeV/c above $\sqrt{s}$=10 GeV 
as depicted in Fig.~\ref{pt_pp}.
This value corresponds to a temperature of 140-150 MeV
in a thermal picture and is consistent with Hagedorn's temperature limit for 
hadronic fireballs \cite{Hag2}. The comparison with $A+A$ collisions 
is straight forward 
at AGS energies for which the average transverse momenta of pions have been 
determined in full phase space \cite{e802}. The value of 320$\pm$5 MeV/c
for $<p_T>$ of all pions in central 
$^{197}Au+$$^{197}Au$ collisions at $\sqrt{s}$=5 GeV falls
slightly above the universal curve in Fig.~\ref{pt_pp}
which passes through 300 MeV/c
at $\sqrt{s}$= 5 GeV. At SPS energies results on negatively charged 
particles in central $^{32}S+$$^{32}S$ collisions \cite{Baechler,alber98} 
give $<p_T>$=355$\pm$ 5MeV/c 
which is again somewhat higher than the universal curve
at $\sqrt{s}$=20 GeV (330 MeV/c). The increase is partly due to the 
admixture of
$K^-$ and $\overline{p}$. Recent measurements of 
negatively charged particle spectra in central
$^{208}Pb+$$^{208}Pb$ collisions at the SPS yield a similar value for $<p_T>$ 
of approximately 360 MeV/c \cite{Jones}. In a thermal picture the difference 
of 20 MeV/c in $<p_T>$ corresponds to a  difference in temperature of 
only 10 MeV 
or 7\%.  In terms of energy (stored in the pions) the higher $<p_T>$ in $A+A$ 
collisions amounts to an increase of the average transverse energy 
of only 5\%.  This means that  the
pion emission characteristics are similar in $p+p$ and $A+A$ collisions. 
At AGS and SPS energies, the observable $<p_T>$ seems to be insensitive 
to the environment the pions are emitted from.

\begin{figure}
\vspace{0.cm}
\begin{minipage}[t]{11cm}
\hspace{ 0.cm}\mbox{\epsfig{file=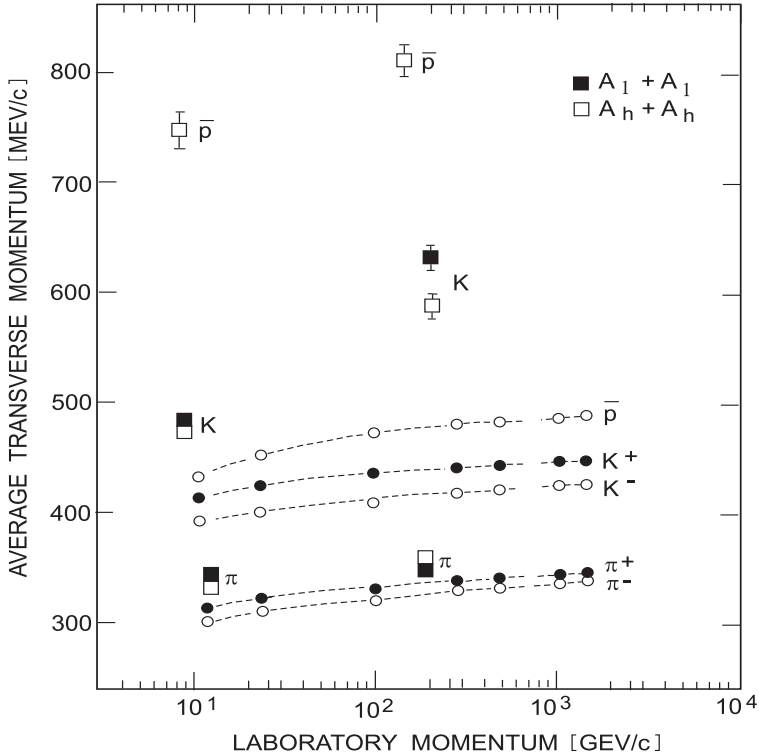,width=10.cm,height=10cm}}
\end{minipage}
\begin{minipage}[t]{4cm}
\vspace{-8.cm}
\caption{Average transverse momenta of different particle
species as function of beam energy for p+p interactions (circles) 
\protect\cite{Rossi} and nucleus-nucleus collisions (squares, l=light, 
h=heavy).
}
\label{pt_ppaa}
\end{minipage}
\end{figure}

\vspace{.4cm}
The systematics of the average transverse momenta of different particle
species as function of beam energy (above 10 GeV/c beam momentum)
is shown in Fig.~\ref{pt_ppaa} for p+p interactions
\cite{Rossi} and nucleus-nucleus collisions. The entries in the figure for 
the latter were derived from measurements near central rapidity. 
It is evident that the mean transverse momenta of the pions 
show only a slight 
increase with beam energy. However, the heavier particles experience
a boost in transverse direction when going from p+p to A+A collisions. The 
mean transverse momenta increase with particle mass, size of the colliding 
system and beam energy. 

\begin{figure}[hpt]
\vspace{.cm}
\begin{minipage}[t]{11cm}
\mbox{\epsfig{file=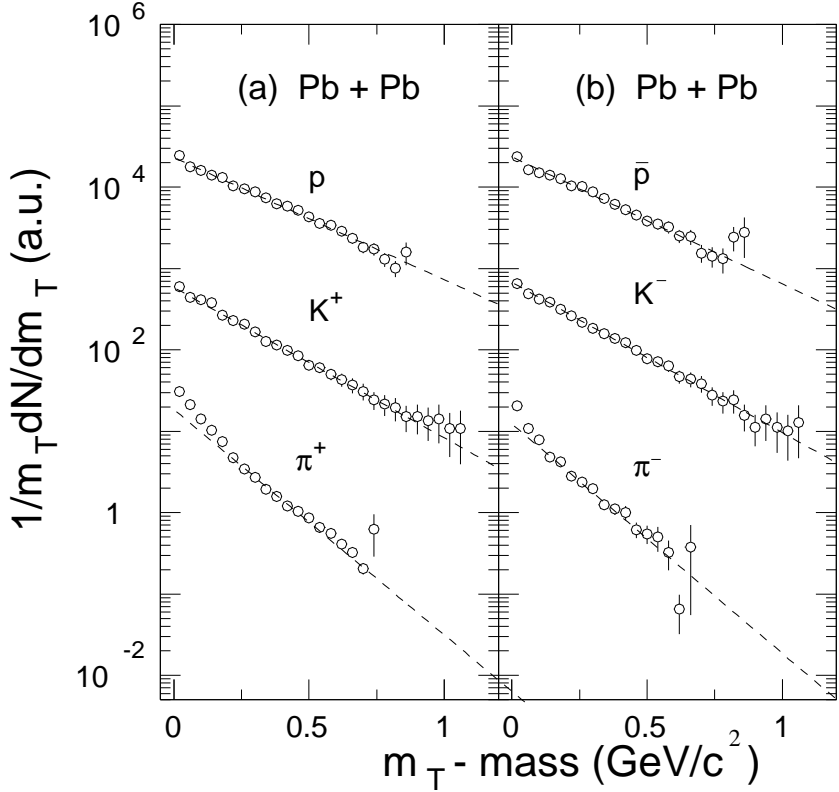,width=11.cm}}
\end{minipage}
\hspace{0.5cm}
\begin{minipage}[t]{4cm}
\vspace{-7cm}
\caption{Transverse mass distributions for positive (left)
and negative particles (right)  from Pb+Pb central collisions at
158 AGeV bombarding energy \protect\cite{na44}.
The spectra are described by
exponential functions (dashed lines).
}
\label{mt_na44}
\end{minipage}
\end{figure}

\begin{figure}[hbt]
\vspace{0.cm}
\hspace{ 1.cm}\mbox{\epsfig{file=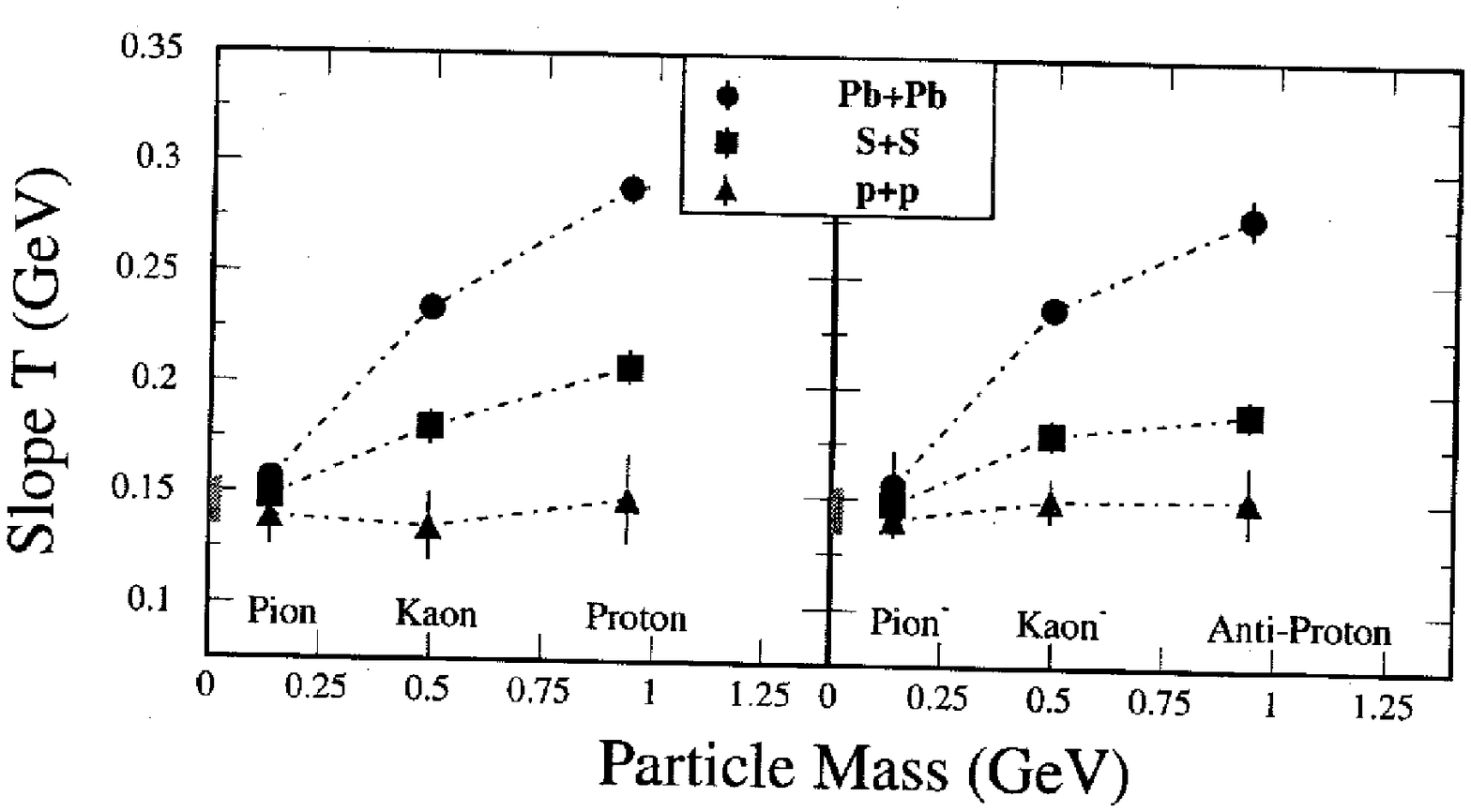,width=13.cm,angle=1 }}
\vspace{0.cm}
\caption{Inverse slope parameters  for positive (left)
and negative particles (right) 
from Pb+Pb collisions
($\protect\sqrt{s}$=17.3 AGeV) in comparison to those from
S+S ($\protect\sqrt{s}$=19.4 AGeV) and p+p collisions 
($\protect\sqrt{s}\approx$23 AGeV)
\protect\cite{na44}.
}
\label{t_na44}
\end{figure}

\vspace{.5cm}
The same picture emerges if the transverse mass spectra 
(m$_T=\sqrt{m_0^2+p_T^2}$)      are analysed.
Fig.~\ref{mt_na44} 
shows transverse mass (m$_T$-m$_0$)  distributions for charged pions, 
kaons and antiprotons from Pb+Pb central collisions at 
158 AGeV bombarding energy \cite{na44}.
The spectra can be parameterized by  
exponential functions (dashed lines) except for the low m$_T$ pions.  
In this kinematical region, resonance decays are expexted to contribute
to the pion yield. 
Therefore, the pion spectra were fitted above m$_T$ = 0.2 GeV only.  
The resulting  inverse slope parameters increase with increasing 
particle mass.

In Fig.~\ref{t_na44} 
the inverse slope parameters of pions, kaons and antiprotons 
from Pb+Pb collisions 
($\sqrt{s}$=17.3 AGeV) are compared to those from 
S+S ($\sqrt{s}$=19.4 AGeV) and p+p collisions ($\sqrt{s}\approx$23 AGeV)
\cite{na44}. 
In the case of p+p the inverse slope parameter T is constant for the 
different particle species whereas for nucleus-nucleus collisions T increases
as the mass of the particle and the mass of the collision system increases
\footnote{we use T for the inverse slope parameter because  
sometimes it is identical to the temperature. Whenever T appears as a subscript
it denotes the transverse direction}.  
This characteristic behavior  suggests that the transverse motion 
has a component which behaves like a velocity rather than a momentum.

\vspace{.4cm}
There are two scenarios in which the observed effect is present: (i) initial
state scattering which broadens the effective distribution of beam
directions thus introducing a mass-dependent transverse momentum 
component of the produced particles
and (ii) the build-up of pressure with its decompressional
flow  which induces a velocity field. In this hydrodynamical 
description the reaction volume expands after compression
and the nuclear matter, i.e. all particles, flow with the same 
collective velocity (on top of their thermal motion). 
Hence particles with higher masses will have
higher energies. In collisions between very heavy nuclei, 
multiple scattering becomes
more important which results in a higher nuclear stopping and a stronger
collective flow in the expanding system. 
Therefore, the collective motion will be more pronounced in the Pb+Pb 
than in the S+S system and will vanish for p+p collisions. 

It has been shown for CERN-SPS energies that initial scattering may   
also account for the observed transverse momentum spectra of pions and 
nucleons from 
proton-nucleus and nucleus-nucleus collisions \cite{satz}.  On the other hand,
the hydrodynamical picture describes well  
the inverse slope parameters of particles and light fragments from 
nucleus-nucleus collisions at LBL/SIS and AGS all the way up to
 CERN-SPS energies. Therefore we analyze in the following the transverse
momentum distributions of A+A collisions in terms of the  hydrodynamical model.
 
Using the model of Ref. \cite{schned}, the transverse momentum
distributions are calculated according to 
\begin{equation}
\frac{dN}{m_T dm_T} \propto \int_{0}^{R} r dr m_T I_o 
\left[\frac{p_T cosh(\rho)}{T_o}\right] \times K_1
\left[\frac{m_T cosh(\rho)}{T_o}\right] 
\label{eq:dndm}
\end{equation}
with $\rho = tanh^{-1}\beta_T,$ the transverse velocity profile 
$\beta_T(r)=\beta_T^{max}(r/R)$ and 
Bessel functions $I_o, K_1$. 
The two free parameters $T_0$ and $\beta_T^{max}$ determine the shape of the 
$m_T$ spectra. 
In general, the fits to the experimental data
show an anticorrelation of these two parameters.
Temperatures in the range from 100 to 200 MeV together with appropriate 
flow parameters describe well the data for pions, kaons and protons.
Assuming a value of $T_0$=140 MeV,
the maximum transverse velocities are found to be 
$\beta_T^{max}$=0.41c and 0.6c for S+S and Pb+Pb collisions, respectively.
These values correspond to average transverse flow velocities of 0.27c and 
0.4c \cite{na44}.

\begin{figure}
%\vspace{0.cm}
\begin{minipage}[t]{11cm}
\mbox{\epsfig{file=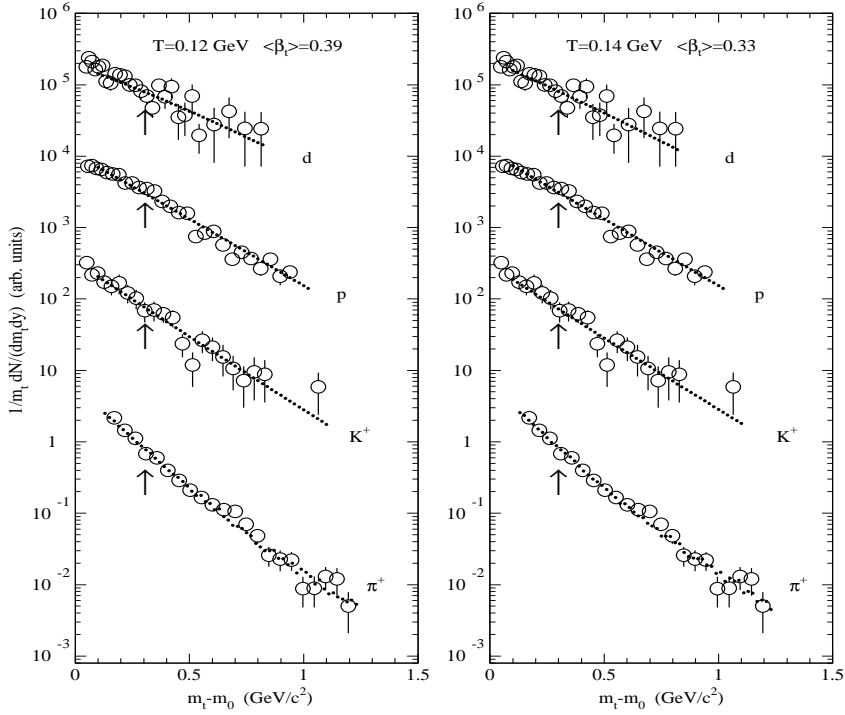,width=11.cm,height=10.cm}}
\end{minipage}
\hspace{0.5cm}
\begin{minipage}[t]{4cm}
\vspace{-6 cm}
\caption{Transverse mass spectra of $\pi^+$, K$^+$, p and d measured
in Si+Au collisions at 14.6 AGeV/c \protect\cite{pbm1,abbot}.
The solid lines are results of calculations using equ. 
\protect\ref{eq:dndm} with 
T = 0.12 GeV and $<\beta_l>$ = 0.39 (left) and T = 0.14 GeV and 
$<\beta_l>$ = 0.33 (right). For details see text.
}
\label{mt_ags}
\end{minipage}
\end{figure}

\begin{figure}
\vspace{1.5cm}
\begin{minipage}[t]{7.5cm}
\mbox{\epsfig{file=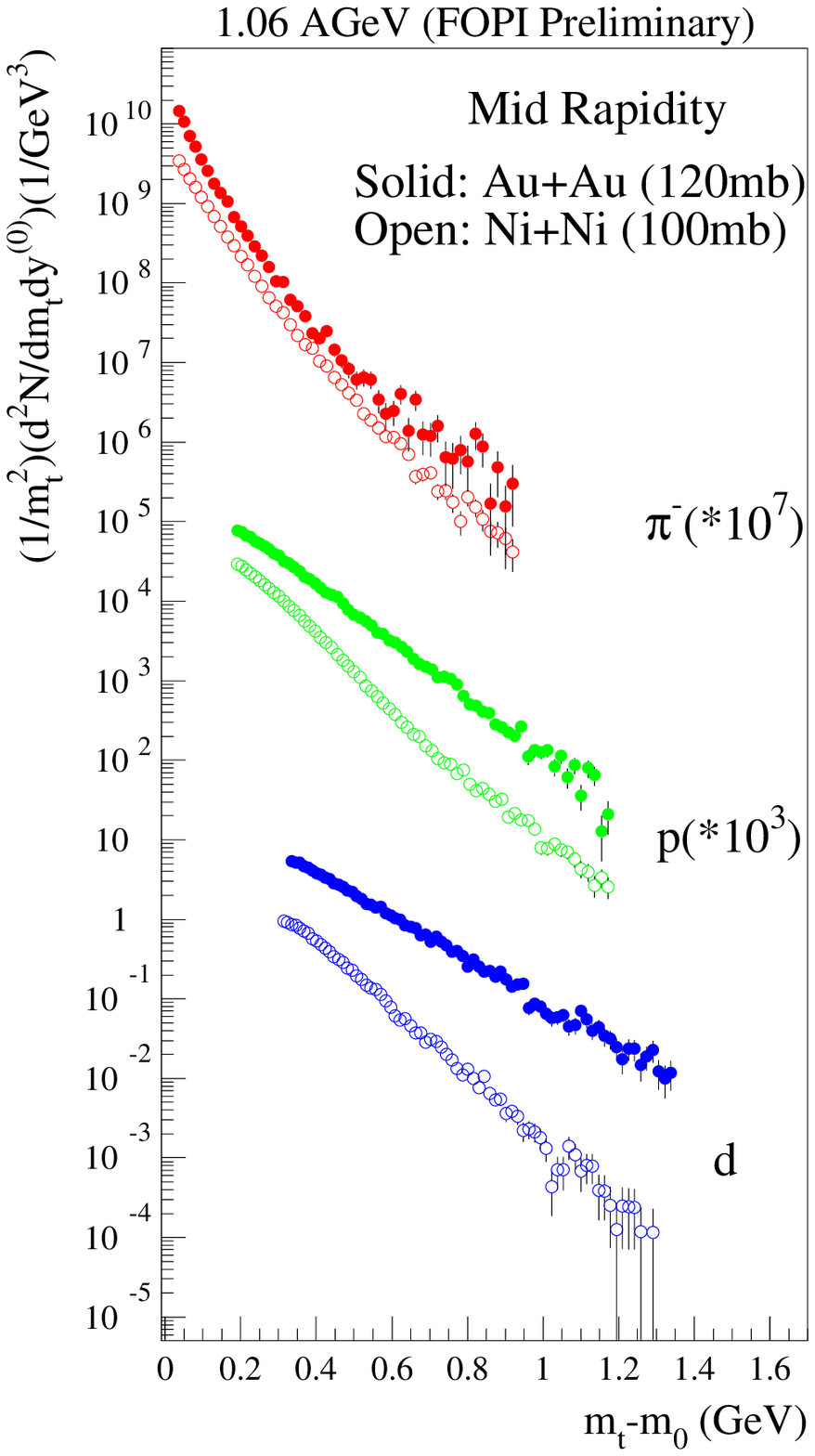,width=7.cm,height=9.cm}}
%\vspace{0.5cm}
\caption{Transverse mass spectra of $\pi^-$, p and d measured in
central Ni+Ni (open points) and Au+Au collisions (full points) 
at 1.06 AGeV (preliminary)
\protect\cite{pelte1,pelte2,hong,herrmann}.
}
\label{mt_sis}
\end{minipage}
\hspace{0.5cm}
\begin{minipage}[t]{7.5cm}
\vspace{-9.cm}
\mbox{\epsfig{file=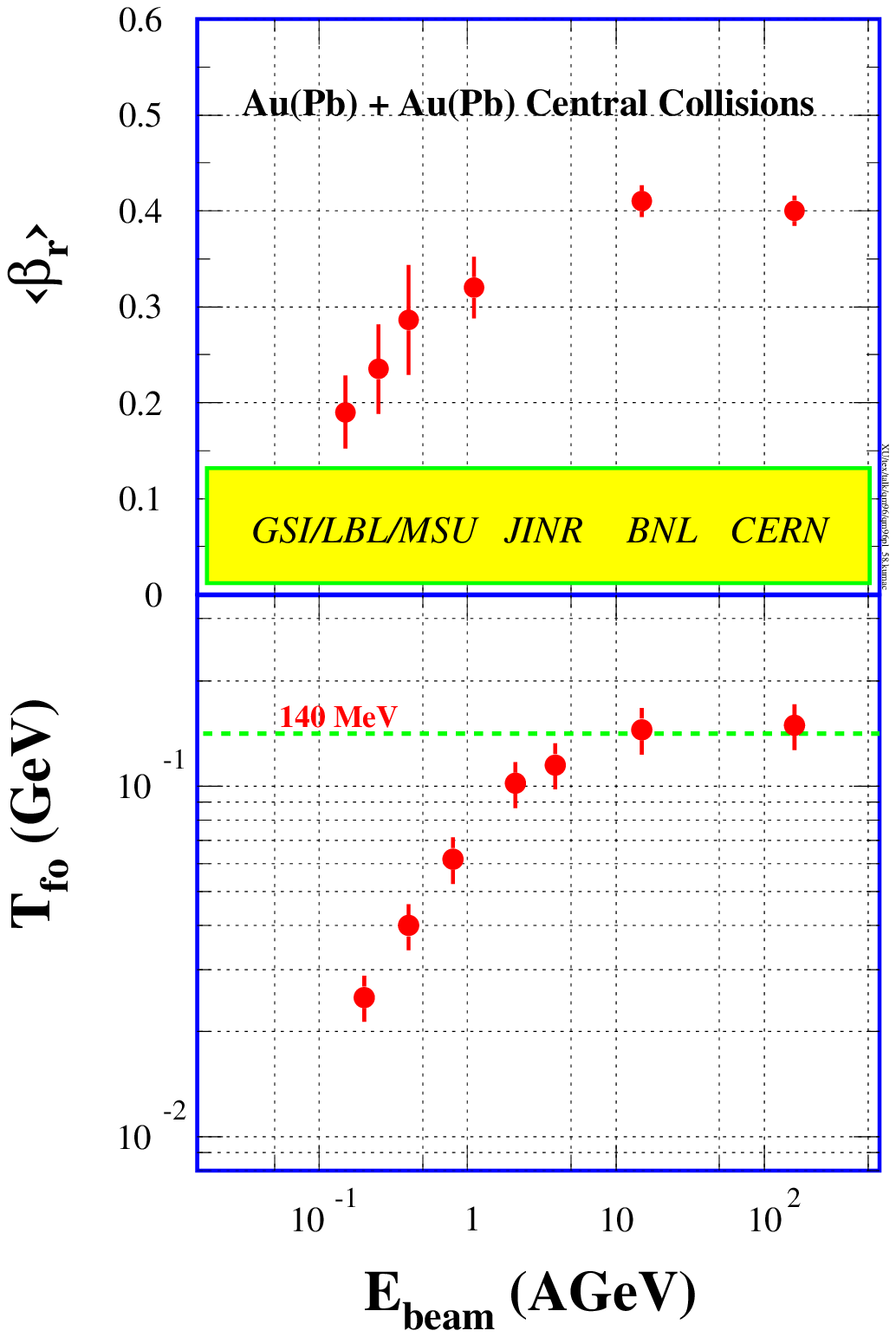,width=7.cm,height=9.cm}}
\caption{Compilation of average 
flow velocities and inverse slope parameters
as a function of bombarding energy (see text). The picture is taken 
from \protect\cite{na44}.
}
\label{t_beta}
\end{minipage}
\end{figure}

Similar values both for the intrinsic temperature and the transverse 
particle velocity
have been found in nucleus-nucleus collisions at AGS energies.
Fig.~\ref{mt_ags} 
shows transverse mass spectra of $\pi^+$, K$^+$, p and d measured
in Si+Au collisions at 14.6 AGeV/c \cite{pbm1,abbot}. 
The solid lines are results 
of calculations using equ.\ref{eq:dndm}. 
In order to avoid complications due to resonance decays, the fit was
restricted to $(m_T-m_o)>$0.3 GeV/c$^2$.  Fig.~\ref{mt_ags} demonstrates that 
a good agreement between data and calculations
is obtained for $T_o$=0.12 (0.14) and $\beta_T^{max}$=0.58 (0.50) corresponding
to average velocities of $<\beta_T>$=0.39 (0.33).

\vspace{.4cm}
At LBL/SIS energies, a transverse (azimuthally symmetric) flow component
was observed by measuring 
light fragment spectra in very central Au+Au collisions 
\cite{lisa,jeong}. 
It was found that the collective expansion uses
about 50\% of the available energy \cite{reisdorf,lisa}.
The analysis of transverse mass spectra of light fragments 
using a blast wave model \cite{siemens} yields a temperature of
$T_o=81\pm24$ MeV and a constant radial velocity of 
$\beta=0.32\pm$0.05 for central 
Au+Au collisions at 1.0 AGeV \cite{lisa}.   

Fig.~\ref{mt_sis} shows transverse mass spectra of $\pi^-$, p and d measured in
central Ni+Ni and Au+Au collisions at 1.06 AGeV 
\cite{hong,pelte1,pelte2,herrmann}.
It can be seen that (i) the slope parameter increases with particle mass
and (ii) this increase is more pronounced for the heavy system. Both 
observations indicate a collective motion of matter.

In summary, the slope of the transverse mass spectra of emitted particles 
in a wide range of masses 
can be successfully explained within a hydrodynamical picture by a 
freeze-out temperature and a transverse flow velocity
($\beta_T$= constant or $\beta_T\propto$ r/R).
Fig.~\ref{t_beta} presents the excitation function of these parameters 
with respect to the bombarding energy \cite{na44}. The effect of
resonance formation and decay is neglected in this picture. For its 
influence see for example  reference \cite{SBass}.

Both the temperature and the transverse flow velocity exhibit a saturation 
above  AGS energies. the limiting temperature is about  140 MeV  again 
in line with the limit stated by Hagedorn. The average transverse flow velocity
increases up to 0.4 c which corresponds to a maximum velocity around
0.6c - 0.7c.

At SIS energies, particles heavier than pions (such as etas, 
kaons and antiprotons) can hardly be produced in a first chance
nucleon-nucleon collision.  
The threshold energies for producing a $\eta$, K$^+$, K$^-$ and a
$\bar p$ in a free nucleon-nucleon collision are 1.25 GeV, 1.58 GeV, 2.5 GeV and
5.6 GeV, respectively. 
Therefore, their production in nucleus-nucleus collisions at beam energies 
below these thresholds requires collective and/or medium effects. According 
to transport models, subthreshold particle production proceeds predominantly 
via multiple hadron-hadron interactions which occur more frequently in the    
dense phase of the  collision. At this stage, the collective
expansion has not developed yet. Those particles which freeze out in this phase 
should not be affected by the radial motion.
Indeed it was found in nucleus-nucleus collisions at 1 - 2 AGeV 
that the emitted high-energy pions, kaons, etas and
antiprotons exhibit similar slope parameters  \cite{miskowiec,ahner,
averbeck,shor1,schroeter}. These common slope parameters depend 
on the beam energy and on the mass of the  colliding nuclei.

\section{Energy in produced particles}
Particle production in hadronic collisions can be studied in various 
observables. One of them is the multiplicity of 
different particle species (see chapters 3 and 4). 
An alternative and more global approach to study regularities in 
particle production makes use of the inelasticity of a hadronic interaction.
Inelasticity is here defined as that fraction of the center-of-mass kinetic 
energy of the incident nucleons which is deposited in produced particles.
There are two ways to measure the corresponding observable:

\begin{itemize}
\item The fraction of the energy  (measured in the center-of-mass frame) lost
by the colliding particles is determined from the difference between the 
average kinetic 
energies before and after the collision. This implies the measurement of
the kinetic energies of the participating nucleons in the final state of the
collision.
\item The second, equivalent, measurement compares the total energy of all 
produced particles with the sum of the kinetic energies of all participating 
nucleons before the collision. This choice requires identification and the 
momentum or energy measurement of all produced particles in the final state. 
With present experiments this is an ambitious task. However, symmetries and
regularities often 
allow to determine the energy carried by all produced particles
to a precision of better than 10\%.
\end{itemize}

In the following we will use both prescriptions to determine the inelasticity. 
Furthermore we will differentiate between the energy transferred to strange
and nonstrange particles.

\subsection{Inelasticity}
In order to compare 
nucleon-nucleon and nucleus-nucleus collisions 
we need to define the particle multiplicity per N+N interaction.   
However, the definition of a N+N interaction is ambiguous: 
should one take into account all (elastic and inelastic) N+N collisions or 
inelastic collisions only?
In studies of hadronic particle production usually  only inelastic 
interactions are used. 
On the other hand it is obvious that even in central nucleus-nucleus
collisions some of the nucleons suffer only elastic 
interactions. 
Furthermore, the study of particle multiplicities as function of beam
energy is more plausible if it is related to all interactions: 
close to the threshold of pion
production in $N+N$ interactions, the pion multiplicity is equal to 
unity if only
inelastic collisions are considered whereas it tends to zero if the total
cross section is taken as the reference. Therefore, in the following we will 
use for the evaluation  of inelasticity in $N+N$ interaction 
the fraction of energy lost by the incident nucleons averaged over all 
elastic and inelastic strong interactions.

At low energies for which single pion production dominates the inelastic 
cross section, the inelasticity of $N+N$ collisions is simply given by
\begin{equation}
<E_{\pi}>\times\sigma_{inel}/\sigma_{tot}/(2\times E_N^{kin})
\end{equation}
where $<E_\pi>$ is the average c.m. energy of 
the pion, $\sigma_{inel}$ the inelastic cross section, $\sigma_{tot}$ 
the total (elastic and inelastic) cross section and $E_N^{kin}$
the c.m. kinetic energy of the incident nucleon.   
For $<E_\pi>$
we have used a value of 0.25 GeV (0.30 GeV) at  1.5 GeV/c
(2.5 GeV/c) laboratory beam momentum \cite{ABD}.

At AGS  and SPS  beam energies (i.e. 12 AGeV and 200 AGeV, 
respectively) we 
computed the energy in the produced particles from the particle multiplicities
in $p+p$ collisions as compiled by Rossi et al. \cite{Rossi}. 
The numbers were rescaled with the factor $\sigma_{inel}/\sigma_{tot}$ for
consistency with the low energy data (see beginning of this subsection).
The mean particle energies are computed as the sum of the expression
\begin{equation}
<E>_{y_i}=<m_T>_{y_i} \cdot cosh(y_i) 
\label{eq:eyi}
\end{equation}
evaluated in each rapidity bin. The result for the energy in the
produced particles was checked against the energy loss of the incident
protons and the two results were consistent to within 10\%. No correction
was applied for possible effects of the difference in isospin between
$p+p$ and $N+N$ interactions.

\begin{figure}[ht]
\vspace{.cm}\begin{minipage}[t]{9cm}
\hspace{ 0.cm}\mbox{\epsfig{file=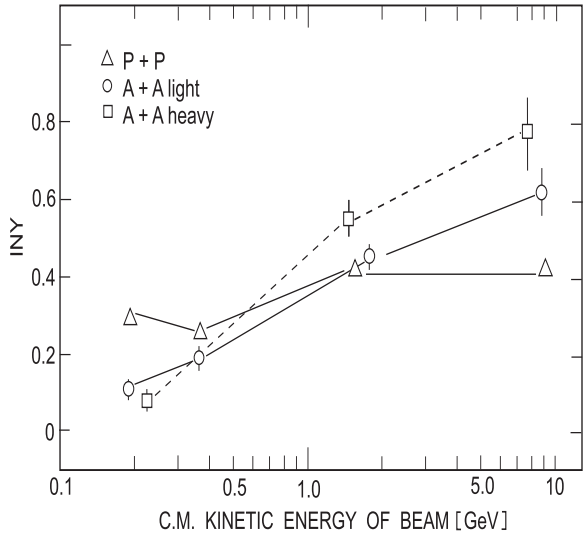,width=7.7cm,height=7.1cm}}
\end{minipage}
\begin{minipage}[t]{6cm}
\vspace{-5cm}
\caption{Inelasticity of 
nucleon-nucleon and nucleus-nucleus collisions
as function of c.m. kinetic energy of the projectile nucleons.}
\label{iny}
\end{minipage}
\end{figure}

The same procedure was applied to central $^{28}Si+$$^{27}Al$ and 
$^{197}Au+$$^{197}Au$ collisions at 
AGS energies \cite{e802} and central $^{32}S+$$^{32}S$ collisions at the
SPS \cite{Baechler}. For central $^{197}Au+$$^{197}Au$ \cite{pelte1} and 
$^{40}Ar+KCl$ \cite{harris1} 
collisions from
SIS/BEVALAC, the energy in the produced particles was calculated from the
pion multiplicity and their c.m. energy which was taken as 230 (280) MeV 
at 0.83 GeV (1.73 GeV) kinetic beam energy. Central $^{208}Pb+$$^{208}Pb$ 
collisions at SPS 
energies were characterized only by the rapidity and $p_T$ distributions 
of the participant protons \cite{PJacobs}. The result of all these calculations 
is shown Fig.~\ref{iny} 
(an earlier version of this compilation was presented in \cite{str96}). 
The inelasticity is plotted as a function of the 
center-of-mass beam energy per nucleon for $p+p$, 
central $A_l+A_l$ and $A_h+A_h$ collisions (l=light, h=heavy).
In nucleon-nucleon interactions
the inelasticity stays remarkably constant with beam energy
from a few 100 MeV to a few GeV, a well established 
behavior already since a long 
time: in each interaction the nucleons lose on the average half of their
energy. The inelasticity in $p+p$ collisions in Fig.~\ref{iny}
is lower than the expected value of 1/2 because the total instead
of the inelastic cross section was employed. In contrast to the
trend in $N+N$ interactions, nuclear collisions show a significant increase of 
the inelasticity as a function of the collision energy. 
This could mean that in a high-energy nucleus-nucleus collision the 
participating nucleons scatter inelastically more than once. On the other hand,
one should keep in mind that 
the inelasticity is determined mainly by pion  production and in chapter 3
we will see that  pion reabsorption may be more important at SIS energies than
at SPS energies. This effect would also decrease the inelasticity   
with decreasing beam energy. An observable less affected by reabsorption 
and thus directly proportional to 
the inelasticity would be the yield of  K$^+$ mesons (see chapter 4).

Pion reabsorption could also be the origin of the 
projectile/target mass dependence of the inelasticity.
Experiments  on pion production in 
$^{197}$Au+$^{197}$Au collisions at 1 AGeV
find that the number of pions per 
participating nucleon is about 25\% smaller than in $^{40}Ar+KCl$ collisions
(see section 3.1). 
Preliminary results from $^{208}Pb+$$^{208}Pb$ 
collisions at SPS energies indicate that
the inelasticity is only slightly higher in the heavy system than in the light 
$^{32}S+$$^{32}S$ configuration.
This behaviour suggests that pion absorption is correlated with the number
of pions per participating nucleon which is 30 times smaller at SIS than at SPS
energies. 

\subsection{Fraction of energy in strange particles}

\begin{figure}[ht]
\begin{minipage}[t]{9cm}
\vspace{-0.1cm}
\hspace{ 0.cm}\mbox{\epsfig{file=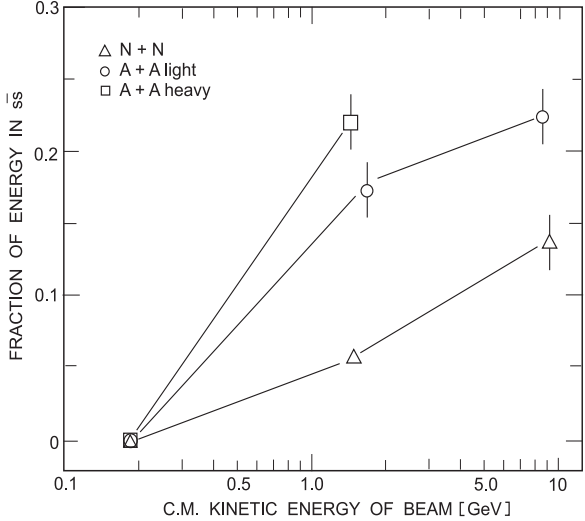,width=7.7cm,height=6.9cm}}
\end{minipage}
\begin{minipage}[t]{6cm}
%\hspace{0.5cm}
\caption{Fraction of energy deposited 
in all strange particles produced 
in p+p and A+A (l=light, h=heavy) collisions as a function of the c.m.
kinetic energy of the projectile nucleons.}
\label{fes}
\end{minipage}
\end{figure}
In this subsection we introduce a new observable which is motivated by 
two experimental findings: (i)  
in section 2.2 we have seen that the average transverse momenta of heavy
particles (in contrast to pions) 
increase drastically with system size and  (ii) there
is the well established phenomenon of the strangeness enhancement
in nuclear collisions \cite{GR2}  (see section 4). 
Both phenomena may be correlated since all strange mesons contain 
a heavy strange quark. 
We have computed the energy deposited in the
strange particles relative to the total energy in produced particles
in order to further quantify the
 different trends in $N+N$ and $A+A$ collisions.  
Here again the energy has been  
computed from the rapidity and $m_T$ distributions
according to \ref{eq:eyi}.
K$^0_S$, K$^+$ and K$^-$ mesons have been used whenever sufficient spectral 
information is available. For the hyperons from associated production
only a fraction of the total energy has been considered. 
It has been calculated
using the ratio of the difference between the hyperon and the nucleon mass to
the nucleon mass. At SPS energies also hyperon pair production has been taken
into account. The total energy in the hyperons has been obtained from
$\Lambda$ and $\overline{\Lambda}$ particles, which include already 
$\Sigma^0$, by multiplication with a factor of 1.6 to cover also charged 
hyperons\cite{wrob}. Fig.~\ref{fes}
shows the fraction of energy deposited in all strange particles
(FES). As expected
this fraction increases with beam energy in $p+p$ interactions in
accordance with the trend in the particle multiplicities. 
In nucleus-nucleus collisions, FES is systematically higher than in 
$p+p$ interactions. At the lowest energy  this is not
visible in Fig.~\ref{fes} due to the linear scale of the vertical axis. Around
$\sqrt{s}$=5 GeV, FES increases by a factor 2.4  from $p+p$ to the
light collisions system $^{28}Si+$$^{27}Al$. For central 
$^{197}Au+$$^{197}Au$ collisions an additional 
increase of 20\% occurs. Such a dependence of FES on the size of the system
is weaker than the increase of the number of participant nucleons.
Above AGS energies, FES increases with energy less in 
light nucleus-nucleus than in nucleon-nucleon collisions.
This behaviour is mainly due to the fact that the pion yield 
(i.e. the energy stored in nonstrange particles) increases with beam energy.
This increase  is stronger for A+A than for
p+p collisions (see Fig.~\ref{pimult}).

\vspace{.3cm}
It is an open question whether FES in A+A
and N+N collisions converge to the same universal (hadronic) value.
Alternatively, the onset of partonic effects in A+A collisions 
perhaps may lead to a more effective deposition of energy in strange particles. 
So far the number of strange quarks was used to quantify a possible strangeness
enhancement. The relative abundance of 
s-quarks is the appropriate observable in a situation in which chemical 
equilibrium governs particle multiplicities whereas we expect the 
fraction of energy in strange particles to be
more sensitive to the dynamics of particle
production in the non-equlibrium case.

\chapter{Pion Production}

Pion production is the dominating inelastic process in nucleus-nucleus 
collisions. 
At beam energies of 1-2 AGeV the pion-to-participant ratio is about 10-20\%.  
In this energy range pions are predominantly created via the 
excitation and the decay of the $\Delta_{33}$  resonance 
\cite{verwest,huber,bass1,teis1}. 
At higher bombarding energies, heavier baryonic resonances are excited which
also decay predominantly into pions.
At beam energies of 5 - 10 AGeV, 
the pion-to-participant ratio reaches  values of  0.7 - 1.
At ultrarelativistic beam energies  (above 100 AGeV), pions have become  
the most abundant particle species'
with a 10-20\% "contamination" of baryons and K-mesons. 
The high center-of-mass energy 
leads to the formation of vector mesons which become an important source of 
the pions observed in the final state.

\vspace{.4cm}
In a baryonic environment pions are reabsorbed with a certain probability
via the sequential process $\pi$N$\rightarrow\Delta$ and 
$\Delta$N$\rightarrow$NN. Experiments on ''true pion absorption'' 
found a cross section for this
process of about 1 barn for pions 
with kinetic energies of 125 and 187 MeV impinging on heavy nuclei
\cite{navon,nakai}. Such an absorption cross section  corresponds
to a mean free path of about  5 fm at normal nuclear matter density. 
The mean free path decreases with increasing baryon density.
Hence pion reabsorption is enhanced at SIS energies as
compared to SPS energies and also in heavy collision systems as compared to 
light ones. 
In the framework of transport models, pion production, propagation and 
reabsorption  depends sensitively on the in-medium properties of the
$\Delta$ resonance (i.e. lifetime, spectral function, 
interaction cross sections)
which are not well known. This might be the reason that transport models 
still have difficulties to reproduce the pion yields and momentum 
distributions measured in A+A collisions \cite{teis1}.

\vspace{.4cm}
Due to their large cross section, most of the pions are 
trapped in the dense reaction zone and have a chance to leave only when
the hadrons cease to interact. Therefore, the measured pion 
phase space distributions mainly reflect the fireball properties at 
freeze-out. However, at low beam energies where the pion to nucleon ratio is 
small,   
high-energy pions freeze-out much earlier
as they are beyond the kinematical region of the $\Delta$ resonance.
These pions offer the possibility to study the dense phase of the collision
and therefore are important messengers of the reaction dynamics.

In this chapter we review the data on pion 
multiplicites, their transverse momentum and angular distributions
(rapidity distributions have already been discussed in chapter 2).
Moreover, we discuss data on the correlation between pion emission and
the reaction plane which provide indications for a time scale in 
pion emission. Finally we present data on the pion isospin ratios 
which can give access to the size of the pion emitting source
\cite{barz}.

\vspace{.5cm}

\section{Multiplicities}

Early experiments performed 
with the streamer chamber at the LBL found that the 
$\pi^-$ multiplicity per event increases linearly
with the number of participants. In Ar+KCl collisions
the ratio of $\pi^-$ mesons to participating
protons is about 0.08 for 1 AGeV beam energy 
and about 0.2 for 1.8 AGeV \cite{harris1}. 

Very similar results were obtained for La+La collisions
\cite{harris2}. 
Taking into account the isospin 
asymmetry of the La-nucleus, the total pion multiplicity n$_{\pi}$ was
calculated from the measured $\pi^-$ multiplicity. The number of 
participating nucleons A$_{part}$ was determined from the charge sum of 
projectile spectator fragments measured  with a plastic scintillator  
hodoscope at forward angles.
It was found, that for La+La collisions 
the ratio of the total pion multiplicity 
to the number of participant nucleons varies from 0.035 at a beam energy of 
0.53 AGeV  to about 0.18 at 1.35 AGeV.  
At a beam energy of 1 AGeV, for example, both the  
Ar+KCl and the La+La measurements yield a ratio of 
n$_{\pi}$/A$_{part}$ $\approx$ 0.12.

\begin{table}[htb]
\caption{Total pion multiplicity per number of participating 
nucleons for A+A collisions at 1 AGeV}  
\begin{center}
\begin{tabular}{|c|c|c|c|c|}
\hline
A+A &  n$_{\pi}$/A$_{part}$ & $\pi$ acceptance & measured data &  Ref.\\
\hline
C+C& 0.168$\pm$0.012 & midrapidity &  $\pi^0$& \cite{averbeck}\\ 
\hline
Ar+Ca& 0.09$\pm$0.009 & midrapidity &  $\pi^0$& \cite{berg,schwalb}\\ 
\hline
Ar+KCl & 0.114$\pm$0.015 & 4$\pi$ &  $ \pi^-$& \cite{harris1}\\
\hline
Ni+Ni& 0.12$\pm$0.01 &  $40^o<\Theta_{lab}<48^o$&  $\pi^+, \pi^-$& \cite{wagner1}\\
\hline
Ni+Ni& 0.13.62$\pm$0.139 & $32^o<\Theta_{lab}<150^o$ &  $\pi^+, \pi^-$& \cite{pelte2}\\
\hline
Kr+Zr& 0.075$\pm$0.021 &  midrapidity & $\pi^0$& \cite{berg,schwalb}\\ 
\hline
La+La & 0.119$\pm$0.013 & 4$\pi$ &  $ \pi^-$& \cite{harris2}\\
\hline
Au+Au& 0.069$\pm$0.015 & midrapidity & $\pi^o$ & \cite{schwalb,averbeck2}\\
\hline
Au+Au& 0.074$\pm$0.012 & $30^o<\Theta_{lab}<150^o$&  $\pi^+, \pi^-$& \cite{pelte1}\\
\hline
Au+Au& 0.09$\pm$0.01 & $40^o<\Theta_{lab}<48^o$ &  $\pi^+, \pi^-$& \cite{wagner2}\\
\hline
\end{tabular}
\end{center}
\label{pimul}
\end{table}
Table \ref{pimul} presents a 
compilation of average pion multiplicities  n$_{\pi}$
as a function of A$_{part}$  for A+A collisions at 1 AGeV.
The table includes data from Au+Au collisions measured recently  by 
three different experiments at SIS
\cite{wagner2,pelte1,schwalb}. The KaoS \cite{wagner2} and TAPS 
\cite{schwalb} data
were taken around midrapidity and extrapolated to 4$\pi$ assuming isotropic
pion emission in the fireball frame. The FOPI data \cite{pelte1} 
were measured within $30^o<\Theta_{lab}<150^o$ and are 
somewhat lower than the KaoS data for central Au+Au collisions. 
The differences in  the data might result
from different methods in the determination of A$_{part}$ 
(the inclusive differential cross-sections as a function
of transverse momentum as measured by the two experiments agree within 10\%, 
see Fig.~\ref{pi_au_sis}).

The pion-to-participant ratios in Table \ref{pimul} 
are not constant when changing
the  mass of the collision system. There seems to be a trend of decreasing 
n$_{\pi}$/A$_{part}$ values with increasing A:
the largest value is found in C+C, the smallest in  Au+Au collisons whereas
in the mass range from Ar+KCl to La+La the values stay constant.
It is interesting to note that the corresponding 
value for nucleon-nucleon collisions 
at 1 GeV beam energy is even larger than the one for C+C 
(see Table \ref{pimule}).  

\begin{figure}[hpt]
%\vspace{-2.cm}
\hspace{ 1.5cm}\mbox{\epsfig{file=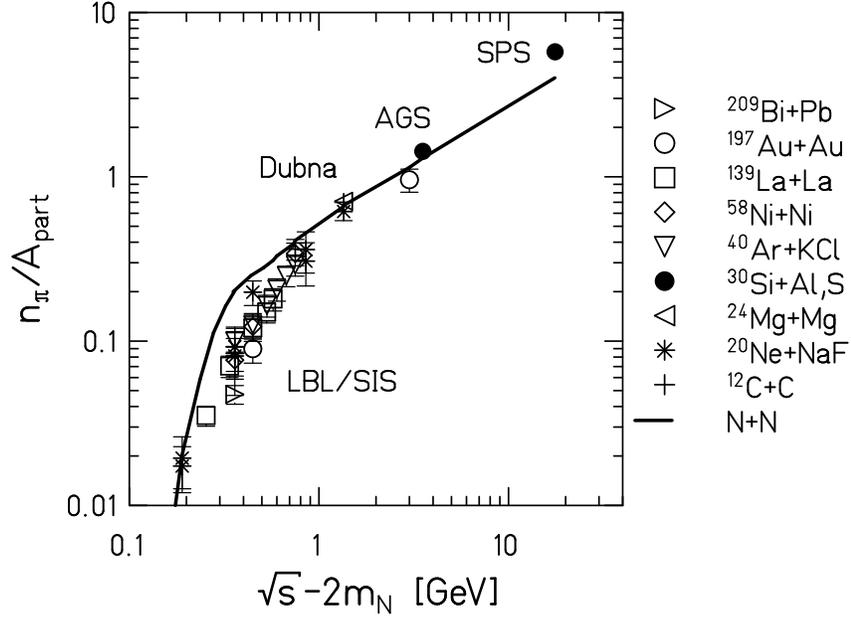,width=14.cm}}
\vspace{0.cm}
\caption{Compilation of the pion multiplicity per participating nucleon
as a function of the kinetic energy of two colliding nucleons 
in the center-of-mass
frame. The A+A data are taken from \protect\cite{GR1,muentz1,muentz2,nagamiya,
harris1,harris2,averbeck}. The N+N data (solid line) 
are taken from \protect\cite{GR1} but 
rescaled by $\sigma_{inel}/\sigma_{tot}$ (see text).
}
\label{pimult}
\end{figure}

In Table \ref{pimule} and Fig.~\ref{pimult}  
we address the question how the pion multiplicity 
per participant evolves with beam energy. The findings from symmetric
nuclear systems are compared with the corresponding numbers from 
nucleon-nucleon collisions as compiled in reference \cite{GR1}. 
Following the line of arguments presented in Chapter 2 we have rescaled
the values of the number of pions per nucleon from reference \cite{GR1}  
by the ratio $\sigma_{inel}/\sigma_{tot}$ for a consistent comparison
between N+N and A+A collisions.

For A+A collisions the values of
n$_{\pi}$/A$_{part}$ increase 
from n$_{\pi}$/A$_{part}$=0.1-0.2 at LBL/SIS energies   
to $\approx 0.7$ at Dubna energies
and to about 1 for AGS energies.  
At  SPS energies, the pion-to-participant ratio  reaches a value of 
n$_{\pi}$/A$_{part}$ $\approx$ 5 which means that   
a new ''state'' of meson dominated matter has been created.

\begin{table}
\caption{Total pion multiplicity per participating 
nucleon for different beam energies. The value for Au+Au at 1 AGeV
is averaged over the experimental results listed in Table 
$\ref{pimul}$.
}
\vspace{0.5cm}
\begin{center}
\begin{tabular}{|c|c|c|c|c|}
\hline
A$_1$+A$_2$ &E$_{beam}$ (AGeV)& $\sqrt{s_{NN}}-2m_N$ (GeV)
& n$_{\pi}$/A$_{part}$&Ref.\\
\hline
La+La& 0.53 & 0.25 & 0.035$\pm$0.005& \cite{harris2} \\
\hline
N+N& 1.0 & 0.45 & 0.24$\pm$0.008& \cite{GR1} \\
La+La&  &  & 0.12$\pm$0.017& \cite{harris2} \\
Au+Au&  &  & 0.086$\pm$0.1 & \cite{pelte1,wagner2,schwalb} \\
\hline
N+N& 1.8 & 0.75 & 0.41$\pm$0.01& \cite{GR1} \\
Ar+KCl&  &  & 0.18$\pm$0.03& \cite{harris1} \\
\hline
N+N& 3.66 & 1.35 & 0.67$\pm$0.08& \cite{GR1} \\
Mg+Mg&  & & 0.7$\pm$0.04& \cite{GR1}\\
\hline
N+N& 10.7& 3.0 & 1.16$\pm$0.1& \cite{GR1} \\
Si+Al& 13.7 & 3.5 & 1.4$\pm$0.14&  \cite{GR1}\\
Au+Au&10.7  &3.0  &1.11$\pm$0.18 & \cite{GR1}\\
\hline
N+N& 200 & 17.6 & 4.0$\pm$0.1& \cite{GR1} \\
S+S& 200 & 17.6 & 5.4$\pm$0.2& \cite{GR1}\\
Pb+Pb& 158  & 15.4 &5.1$\pm$0.2 & \cite{PJacobs} \\
\hline
\end{tabular}
\end{center}
\label{pimule}
\end{table}

The pion-to-participant ratio measured in N+N collisions clearly deviates
from the A+A data for low and high beam energies. 
At BEVALAC/SIS energies, the ratio n$_{\pi}$/A$_{part}$ for A+A collisions is 
in average a factor of about 2 smaller than for N+N collisions. 
This difference increases with increasing A.      
At SPS energies, n$_{\pi}$/A$_{part}$ is about 35\%
larger in  A+A than in N+N interactions.
The discrepancy at SIS energies
may be due to true pion absorption which proceeds via
sequential processes occuring in A+A collisions 
($\pi$N$\rightarrow\Delta$ and $\Delta$N$\rightarrow$NN). Pion reabsorption
requires high nucleon densities
i.e. a small value of n$_{\pi}$/A$_{part}$ which is realized at low beam 
energies. At SPS energies, multiple  pion production is possible and 
pion reabsorption is suppressed because of the low baryon density.

\section{Transverse momentum distributions}

%\vspace{1.5cm}
\begin{table}
\caption{Pion inverse slope parameters from a fit to the spectra 
according to d$^3\sigma$/dp$^3$ = C$_1$ exp(-E/T$_1$) +  C$_2$ exp(-E/T$_2$) 
The data were taken at SIS energies.}
\vspace{0.5cm}
\begin{center}
\begin{tabular}{|c|c|c|c|c|c|}
\hline
A+A & E [AGeV]& $\pi$ & T$_1$ [MeV] & T$_2$ [MeV]&  Ref.\\
\hline
Au+Au & 1.0& $\pi^+$& 45$\pm$3 &76$\pm$3 & \cite{muentz2}\\
\hline
Au+Au & 1.0& $\pi^+$& 49$\pm$2 &85$\pm$3 & \cite{wagner2}\\
\hline
Au+Au & 1.0& $\pi^-$& 41$\pm$3 &76$\pm$3 & \cite{wagner2}\\
\hline
Au+Au & 1.0& $\pi^+$& 49$\pm$4 &96$\pm$10 & \cite{pelte1}\\
\hline
Au+Au & 1.0& $\pi^-$& 42$\pm$3 &96$\pm$10 & \cite{pelte1}\\
\hline
Au+Au & 1.0& $\pi^o$& 38$\pm$4 &78$\pm$4 & \cite{schwalb}\\
\hline
Ni+Ni & 1.0& $\pi^+$& 45$\pm$3 &75$\pm$3 & \cite{muentz2}\\
\hline
Ni+Ni & 1.0& $\pi^+$& 47$\pm$3 &77$\pm$3 & \cite{wagner1}\\
\hline
Ni+Ni & 1.06& $\pi^+$& 49.4$\pm$3.7 &96.4$\pm$5.1 & \cite{pelte2}\\
\hline
Ni+Ni & 1.06& $\pi^-$& 42.2$\pm$2.7 &96.4$\pm$5.1 & \cite{pelte2}\\
\hline
Kr+Zr & 1.0& $\pi^o$& 48$\pm$8 &76$\pm$10 & \cite{schwalb}\\
\hline
Ne+NaF & 1.0& $\pi^+$& 43$\pm$3 &63$\pm$3 & \cite{muentz2}\\
\hline
Ni+Ni & 0.8& $\pi^+$& 40$\pm$3 &68$\pm$3 & \cite{muentz2}\\
\hline
Ni+Ni & 1.45& $\pi^+$& 54.3$\pm$3.3 &103.8$\pm$4.9 & \cite{pelte2}\\
\hline
Ni+Ni & 1.8& $\pi^+$& 50$\pm$3 &95$\pm$3 & \cite{muentz2}\\
\hline
Ni+Ni & 1.8& $\pi^+$& 36$\pm$3 &90$\pm$3 & \cite{wagner1}\\
\hline
Ni+Ni & 1.8& $\pi^-$& 40$\pm$3 &91$\pm$3 & \cite{wagner1}\\
\hline
Ni+Ni & 1.93& $\pi^+$& 56.6$\pm$4.4 &111.7$\pm$9.5 & \cite{pelte2}\\
\hline
Ar+KCl & 1.8& $\pi^-$& 59 &102 & \cite{brockmann}\\
\hline
C+C & 0.8& $\pi^o$& 50$\pm$4 & - & \cite{averbeck}\\
\hline
C+C & 1.0& $\pi^o$& 54$\pm$3 & - & \cite{averbeck}\\
\hline
C+C & 2.0& $\pi^o$& 83$\pm$2 & - & \cite{averbeck}\\
\hline
C+C & 1.0& $\pi^+, \pi^-$& 45$\pm$3 & 62$\pm$3 & \cite{foerster}\\
\hline
C+C & 2.0& $\pi^+, \pi^-$& 40$\pm$3 & 86$\pm$3 & \cite{foerster}\\
\hline
\end{tabular}
\end{center}
\label{slopep}
\end{table}

\begin{figure}
%\vspace{-2.5cm}
\hspace{1.5cm}\mbox{\epsfig{file=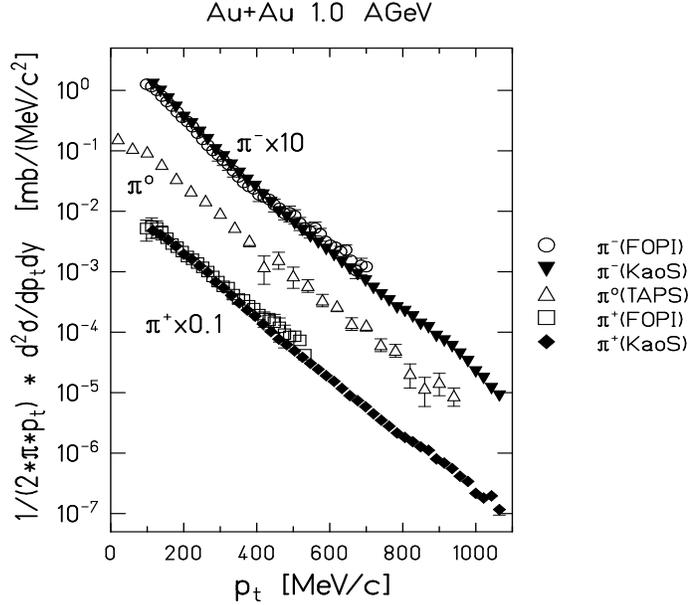,width=11.cm,height=9.cm}}
%\vspace{0.cm}
\caption{Invariant pion production cross sections as a function of 
transverse momentum for Au+Au at 1 AGeV measured around midrapidity by 
FOPI ($\pi^-$: open circles, $\pi^+$: open squares \protect\cite{pelte1}, 
KaoS (full symbols \protect\cite{wagner2}) and TAPS 
(open triangles \protect\cite{schwalb}).
}
\label{pi_au_sis}
\end{figure}

Pioneering experiments at the LBL BEVALAC found that the 
transverse momentum spectra 
of negatively charged pions measured in central Ar+KCl  collisions 
at 1.8 AGeV exhibit a nearly exponential shape with a characteristic 
enhancement at low transverse momenta \cite{brockmann}.
Recent experiments performed at SIS/GSI have extended the study of  
charged and neutral pion production to very heavy collision systems and found 
similar effects. Fig.~\ref{pi_au_sis} 
shows the pion invariant cross sections versus 
transverse momentum for Au+Au at 1 AGeV measured around midrapidity by 
three different setups. The pion spectra deviate from a single exponential
and have been parameterized   by a superposition of two 
Maxwell-Boltzmann distributions:

\begin{equation}
d^3\sigma/dp^3 = C_1 exp(-E/T_1) +  C_2 exp(-E/T_2) 
\label{eq:invx}
\end{equation}

Table \ref{slopep} presents a compilation of the inverse
slope parameters as obtained by a fit to the measured 
spectra according to equ.~\ref{eq:invx}.   
The values of the inverse slope parameters T$_1$ and T$_2$ 
are correlated and depend somewhat on the measured momentum range. 
Nevertheless, the fit results have some features in common: 
T$_1$ has values between 40 and 50 MeV for most of the systems
and T$_2$ increases with both the mass number of the colliding nuclei
and the bombarding energy.  
For Au+Au collisions at 1 AGeV the pion inverse slope parameters are 
found to be constant for different 
center-of-mass polar emission angles \cite{pelte1}.
Any interpretation of the slope parameters in Table \ref{slopep} 
should take into account the correlations 
between the different fit parameters.
What would be needed in addition to the numbers presented
in Table \ref{slopep} are the cross correlations between the two slope factors
and relative yields in the two components.

\vspace{.4cm}
The enhancement at low pion momenta  (described by the T$_1$ component)
was explained by the contribution of pions from delta resonance decay   
\cite{brockmann}. In this case 
the pion transverse momentum is strongly influenced by the Q-value 
of the decay and does not reflect the ''temperature'' of the nuclear fireball.
This interpretation of the pion spectra was motivated by the fact that 
the pion production cross section in proton-proton collisions  
up to about 2 GeV bombarding energy is dominated by the $\Delta$   
resonance \cite{verwest,huber}. 
Recent data on pion production in Au+Au collisions at 1 AGeV 
\cite{muentz2} and Ni+Ni collisions at 1.06, 1.45 and 1.93 AGeV \cite{hong}
were analyzed in terms of thermal models including pions from
baryonic resonances which are embedded in a radial flow. In Ref.
\cite{muentz2} it was assumed that all pions stem from resonance decays
and the measured pion momentum distribution could be reproduced by varying
the in-medium spectral function of the resonances. 
In the analysis of Ref. \cite{hong}, the pion transverse 
momentum spectra were
explained by a (high-energy) thermal component and a contribution
from $\Delta$ decays at low transverse momenta. In this approach,
the $\Delta$ mass distribution is parameterized by a relativistic 
Breit-Wigner form with  the maximum fixed to the free mass of 1232 MeV and 
a momentum dependent width. When increasing the beam energy from 
1.06 to 1.93 AGeV the ratio 
$\Delta$/baryons at freeze-out increases from 10$\%$ to 18$\%$ whereas the  
ratio of pions from $\Delta$ decays to the total pion multiplicity             
decreases from 77$\%$ to 66$\%$ \cite{hong}.

\vspace{.4cm} 
In a recent invariant mass analysis of p$\pi^{\pm}$ pairs in Ni+Cu 
collisions at 1.97 AGeV correlations from the decays of the 
$\Delta$ resonance have been observed \cite{hjort}. It was found, that the  
maxima of the $\Delta^{++}$ and $\Delta^0$ mass distributions  
are shifted to lower values as compared 
to the free spectral function. This mass reduction is maximized in central
collisions and has been interpreted as a simple kinematic phase space effect.

\begin{figure}
\begin{minipage}[t]{10cm}
%\vspace{-1.cm}
\hspace{1cm}
\mbox{\epsfig{file=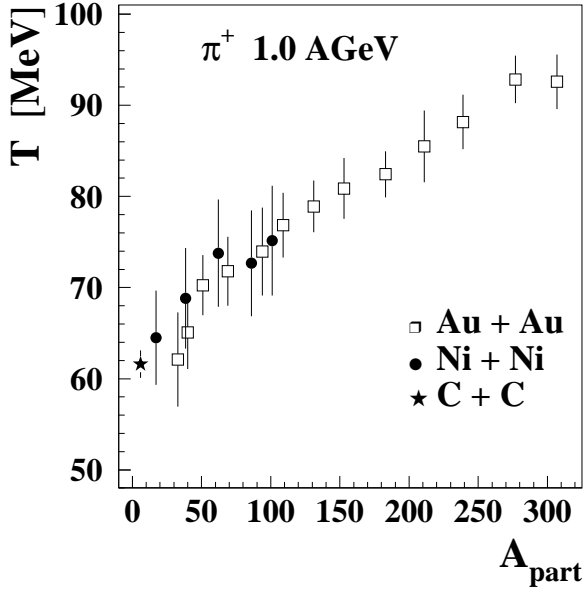,width=8.5cm}}
\end{minipage}
\begin{minipage}[t]{5cm}
\vspace{-5cm}
\hspace{2.cm}\caption{Inverse slope parameter
of the high-energy pions (E$_{cm}^{kin} >$ 300 MeV) 
for  Ni+Ni and Au+Au collisions at 1 AGeV as a function of the number of
participant nucleons \protect\cite{wagner1,foerster}.
}
\label{t_sis}
\end{minipage}
\end{figure}

\begin{figure}
%\vspace{-2.5cm}
\hspace{ 1.cm}\mbox{\epsfig{file=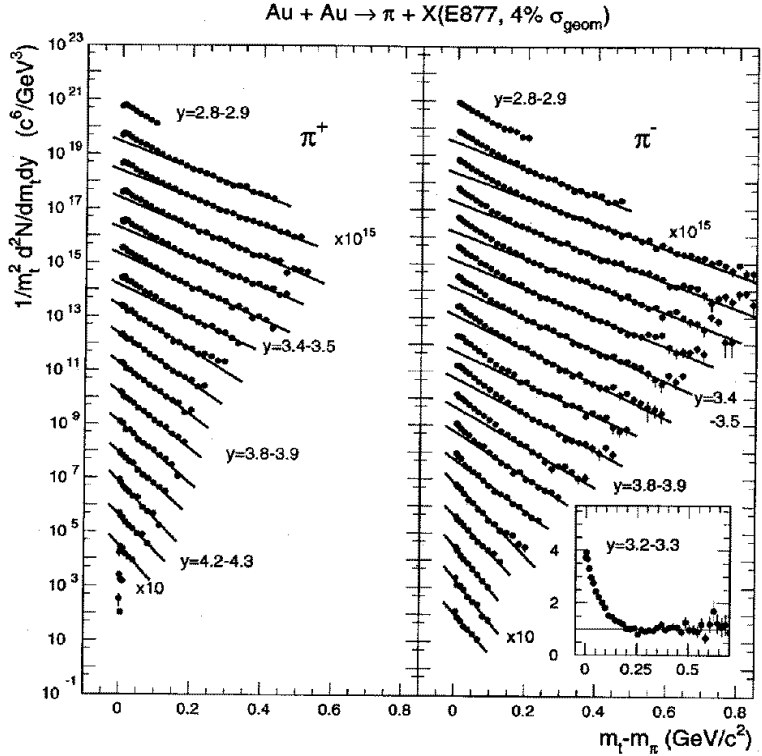,width=10.cm,height=10.cm}}
\vspace{0.cm}
\caption{Transverse mass distributions of $\pi^+$ and $\pi^-$ mesons 
measured at forward rapidities in central Au + Au collisions at 10 AGeV
beam energy \protect\cite{lacasse}.
}
\label{pi_mt_e877}
\end{figure}

\begin{figure}
\vspace{-0.5cm}
\hspace{1.cm}
%\begin{turn}{-90}
\mbox{\epsfig{file=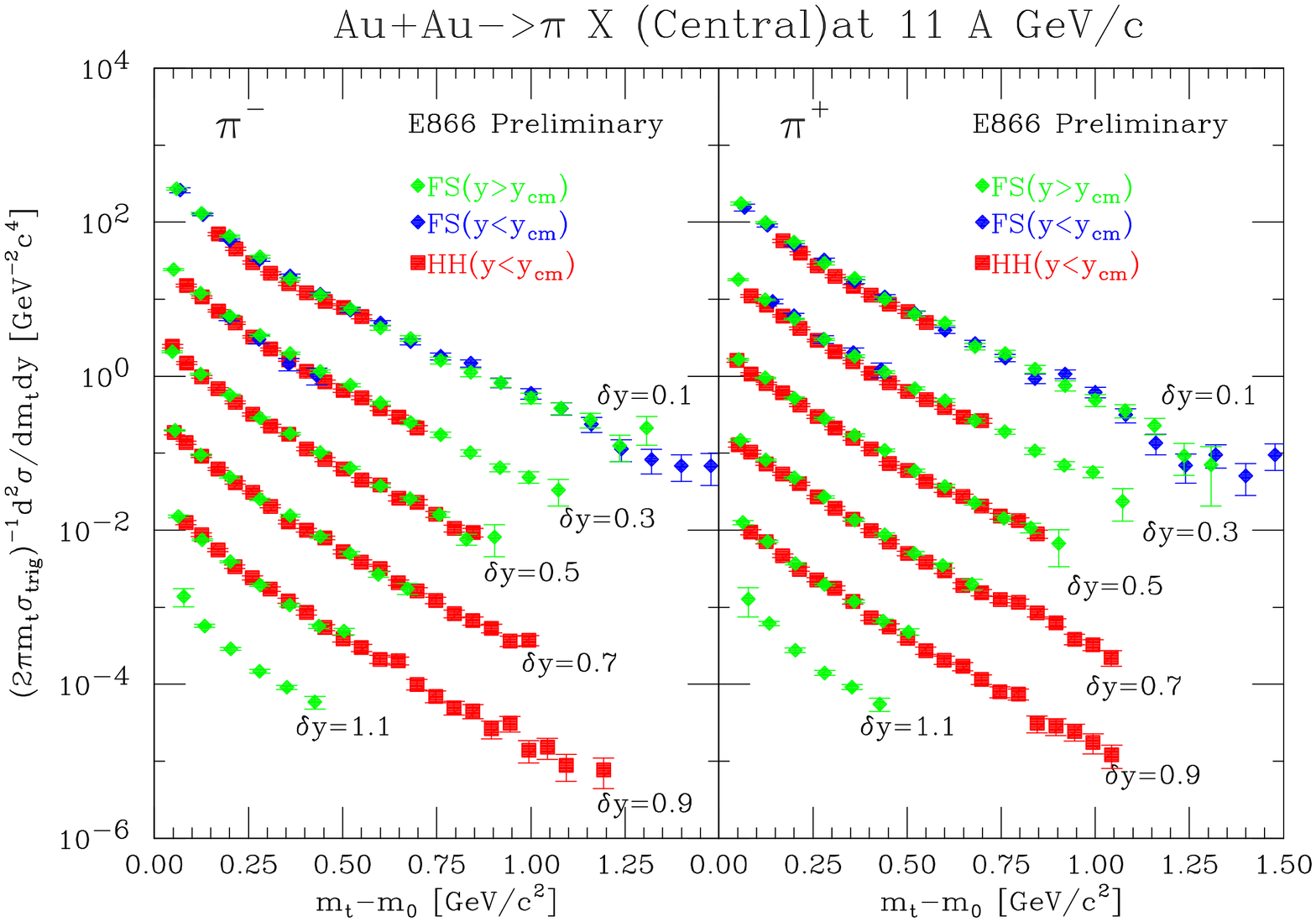,height=12cm,width=10cm,angle=90}}
%\end{turn}
%\vspace{-1.cm}
\caption{Transverse mass distributions of $\pi^+$ and $\pi^-$ mesons
measured around midrapidity in central Au + Au collisions at 10 AGeV
beam energy \protect\cite{e802}.
}
\label{pi_mt_e866}
\end{figure}
 
It is still under debate, whether the high-energy component of the 
pion spectra also 
reflects resonance decay kinematics \cite{pelte1,teis1,bass2} or represents  
the thermal pions \cite{weinhold}. 
Fig.~\ref{t_sis} shows the inverse slope parameter 
of the high-energy (T$_2$: E$_{cm}^{kin} >$ 300 MeV) pions 
for  Ni+Ni and Au+Au collisions at 1 AGeV as a function of the number of 
participants. The inverse slope parameter of the high-energy pions 
increases with increasing  of A$_{part}$ from T$_2$=60 MeV to 80 MeV
(for A$_{part}$=300). Furthermore, its value measured in 
central Ni+Ni collisions agrees with the one from peripheral Au+Au collisions.
These findings show that the pion spectral slope at high p$_T$    
depends on the size of the fireball and is not influenced by the
comoving spectator fragments. It should be noted that the  data 
shown in Fig.~\ref{t_sis} are derived from pion spectra 
which are integrated over the azimuthal angle.

\vspace{.5cm}

The high-energy pion yield was studied as a function of 
collision centrality for heavy collision systems at SIS energies
\cite{muentz2,muentz1}. In contrast to low-energy pions (which dominate the 
pion yield)  the multiplicity of high-energy pions (E$^{tot}_{c.m.} >$671 MeV)  
increases more than linearly with the number of participating nucleons
(the total pion c.m.energy of 671 MeV corresponds to the threshold for
K$^+$ production in a NN collision): assuming M $\propto$ A$_{part}^{\alpha}$,
values of $\alpha$ = 1.86$\pm$0.19 and 1.63$\pm$0.19 were determined
for Bi+Pb collisions at 0.8 AGeV and Au+Au at 1 AGeV, respectively.   
The production of these high-energy pions is 
''subthreshold'' and therefore requires collective effects like multiple
baryon-baryon interactions. As these processes 
are strongly enhanced in the dense medium,    
high-energy pions seem to probe the high-density stage of the collision. 
This picture is supported by the pion azimuthal emission pattern 
and the $\pi^+/\pi^-$ ratio as discussed in the next sections.

\vspace{.5cm}       
The ''low-p$_T$ enhancement'' of the pion yield  is also found at 
higher bombarding energies.
Fig.~\ref{pi_mt_e877} shows $\pi^+$ and $\pi^-$ transverse mass distributions
measured at forward rapidities in central Au + Au collisions at 10 AGeV  
beam energy \cite{lacasse}. At low transverse masses
a clear enhancement above the pure exponential is visible. 
The insert in Fig.~\ref{pi_mt_e877} shows the ratio of the pion yield 
to an exponential (fitted to the spectra above m$_T$-m$_{\pi}$$>$ 0.2 GeV/c) 
for a given rapidity bin.  The effect is most pronounced at
midrapidity (see also Fig.~\ref{pi_mt_e866}, \cite{e802}).
The low-p$_T$ enhancement of the pions 
was again explained by the contribution of pions from $\Delta$ decays
\cite{e814}.

This effect is also clearly visible in Pb+Pb collisions 
at SPS energies (see Fig.~\ref{mt_na44}, \cite{na44}).  
However, the large pion multiplicity  per participant of 
n$_{\pi}$/A$_{part} \approx$ 5 (see Table \ref{pimule} and Fig.~\ref{pimult}) 
indicates that the
pion yield at low  p$_T$ is due to a new source of pions other than 
the decay of baryonic resonances.
Relativistic transport calculations consider string fragmentation and 
$\rho$ and $\omega$ decays
to be the dominant source of low  p$_T$ pions. 

Earlier studies of light projectiles on heavy targets
revealed a stronger pion enhancement near target rapidity
\cite{alber98,Roehr94}.
It remains to be seen whether this trend persists
also for the heavy Pb projectile and what differences are found
between  central and semi-central collisions. 

\section{Angular distributions}

\vspace{.5cm}
Pions produced in proton-proton collisions exhibit a strong forward-backward
peaked angular distribution which is a consequence of the dominant
p-wave production channel \cite{wolf}. 
Pions emitted from a thermalized fireball formed in nucleus-nucleus 
collisions, however,
should exhibit an isotropic angular distribution
in the center-of-mass system.
It was found in Ne+NaF and Ar+KCl collisions that the pion 
emission pattern still contains a forward-backward peaked contribution 
\cite{wolf,nagamiya}. The anisotropy is larger for peripheral   
than for central collisions. 
For central Ar+KCl  collisions at 1.8 AGeV,
for example, about 90\% of the pions are emitted isotropically \cite{stock}.
This effect is caused by pion reabsorption 
and reemission which occurs more freqently in central events \cite{teis1}
and can be also considered to be a higher degree of thermalization.

\begin{figure}
\begin{minipage}[t]{9.cm}
\mbox{\epsfig{file=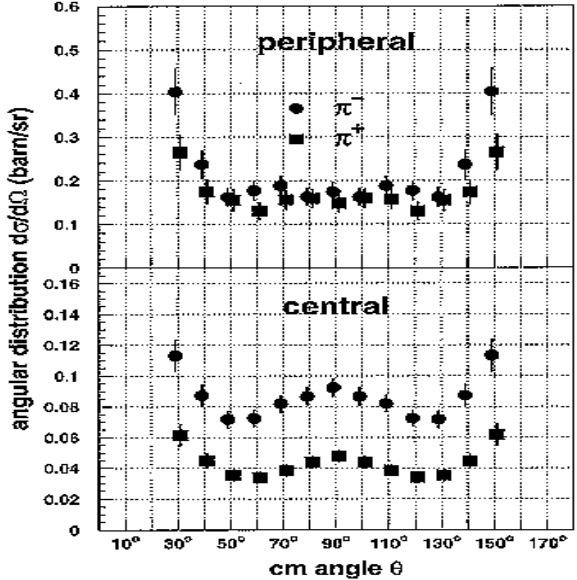,width=8.cm,height=8.cm}}
\end{minipage}
%\hspace*{1.cm}
\begin{minipage}[t]{6.cm}
\vspace{-8.cm}
\hspace{1cm}\caption{Pion angular distributions from peripheral (top) and 
central (bottom) Au+Au collisions at 1.06 AGeV \protect\cite{pelte1}.}
\label{pi_angle_fopi}
\end{minipage}
\end{figure}

Fig.~\ref{pi_angle_fopi} 
presents recent results of the FOPI Collaboration which 
measured the pion polar angular distribution
in Au+Au collisions at 1 AGeV \cite{pelte1}. A forward-backward
enhancement is clearly visible both for peripheral and central collisions.
In central collisions, however, an additional structure appears at 
$\Theta_{cm}$ = 90$^o$: the emission of $\pi^+$ and $\pi^-$ 
is enhanced at midrapidity. This effect is also observed for 'minimum bias'
collisions if pion kinetic energies above 365 MeV are selected \cite{pelte1}.

\vspace{.4cm}
Experiments at LBL/SIS have established the azimuthally anisotropic
emission of particles from the reaction zone of two colliding 
nuclei: the directed flow of nucleons in the reaction plane 
(''side splash'') and the preferred emission of nucleons perpendicular 
to the reaction plane (''squeeze-out'') 
\cite{gutbrod1,demoulins,gosset1,gutbrod2,leifels,lambrecht,brill2}.
A collective out-of-plane emission was predicted by early hydrodynamical 
calculations and interpreted as a dynamical squeeze out of matter due to the 
build-up of pressure in the interaction zone \cite{stoecker}.

First measurements of the emission pattern of pions in symmetric as well 
in asymmetric heavy-ion collisions were done with the streamer chamber
at the LBL BEVALAC \cite{keane,dani}. These studies showed evidence for a 
weak in-plane
correlation in the $\pi^-$ emission pattern. Similar findings for asymmetric 
systems were reported by the DIOGENE group \cite{gosset2}. These
measurements indicated a preferential in-plane emission of charged pions
on the projectile side. This behaviour was attributed to a stronger pion 
absorption by the heavier target spectator remnant on the side opposite 
to the projectile.

\begin{figure}
\vspace{0.7cm}
\begin{minipage}[t]{10cm}
\centerline{\hspace{0.cm}\mbox{\epsfig{file=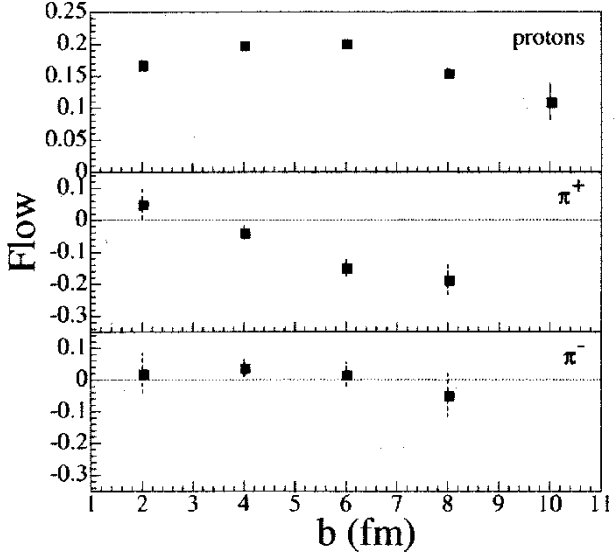,width=8.cm}}}
\end{minipage}
\begin{minipage}[t]{5cm}
\vspace{-6.5cm}
\hspace{5.5cm}\caption{Sidewards flow as a function of impact parameter for protons (top),
$\pi^+$ (middle) and $\pi^-$ (bottom) in Au+Au collisions at 1.15 AGeV
\protect\cite{kintner}.
}
\label{pi_flow_eos}
\end{minipage}
\end{figure}

%\vspace{.5cm}
Recently, the directed in-plane flow of pions and protons 
was measured in Au+Au collsions at 1.15 AGeV 
as a function of impact parameter 
by the EOS collaboration \cite{kintner}.
They found that  
the average in-plane transverse velocity  $<p_x/m>$ of pions and protons 
exhibits a S shape as a function of rapidity. The magnitude of the flow
is defined as the slope of the S curve at midrapidity. This slope is
shown in Fig.~\ref{pi_flow_eos} 
for protons, $\pi^+$ and $\pi^-$ as a function of impact 
parameter. The protons clearly show a flow signal for all
impact parameters. The positively charged pions show little flow
for central collisions and a large antiflow for peripheral collisions.
The negatively charged pions do not exhibit a clear trend.

Transport calculations predicted a  
weak pionic flow in central collisions due to the flow of baryon
resonances from which they are produced but an antiflow behaviour
in peripheral collisions as a result of the shadowing by spectators 
\cite{bass1,baoli}.
A weak pionic antiflow was also found in central collisions 
of Au+Au at 10.8 AGeV both for $\pi^+$ and $\pi^-$ \cite{lacasse}
and explained in a similar way \cite{baoli_ko}.

\begin{figure}
%\vspace{-1.5cm}
\begin{minipage}[t]{9cm}
\hspace{ 0.cm}\mbox{\epsfig{file=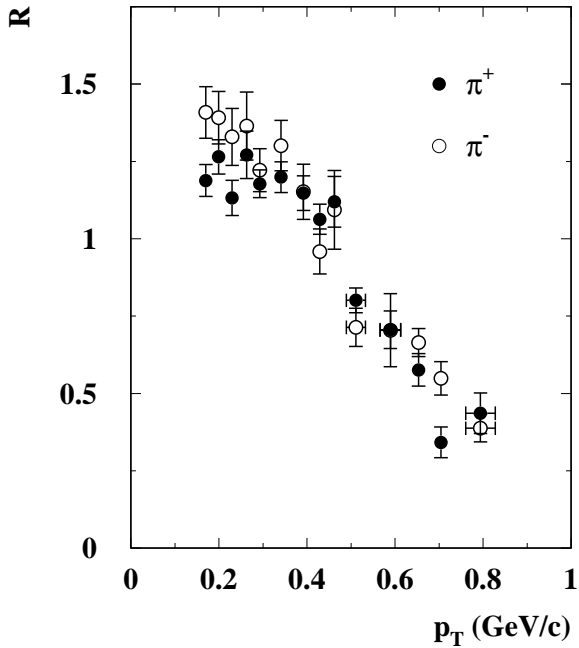,width=9.cm}}
\end{minipage}
\begin{minipage}[t]{5cm}
\vspace{-7.cm}
\caption{Yield ratio of pions  emitted to opposite directions in
the reaction plane (R = N$_p$/N$_t$) as a function 
of transverse momentum. N$_p$ and N$_t$ correspond to the number of pions
emitted within $\phi = \pm45^0$ (with respect to the reaction plane)
towards the projectile and target side, respectively. 
 The pions are measured at \protect$\Theta_{lab}$=84$^0$
(corresponding to normalized rapidities of 0.01 $< y/y_{proj}<$ 0.1)
in peripheral Au+Au collisions at 1 AGeV   \protect\cite{wagner2}.
}
\label{pi_flow_kaos_84}
\end{minipage}
\end{figure}

%\vspace{.5cm}
Very recently a measurement of pionic flow
was performed  in Au+Au collisions at 1 AGeV as a function of the 
impact parameter, rapidity and pion energy
by the KaoS Collaboration \cite{wagner2}.  
In peripheral collisions,  a directed in-plane antiflow was observed for 
low p$_T$ pions in accordance with the $\pi^+$ data of the 
EOS Collaboration (their integrated pion yields
are dominated by low-energy pions).
For high p$_T$ pions, however,  a (positive) flow behaviour 
was measured. 
This is demonstrated in Fig.~\ref{pi_flow_kaos_84} 
which shows the spectral ratio of pions
emitted into the event plane
to the ''projectile side'' over the ones emitted to the 
''target side'' (R=N$_p$/N$_t$) for peripheral Au+Au collisions at 1 AGeV. 
As this measurement is performed at target rapidity
(0.01$<$y/y$_{beam}<0.1$) a value of N$_p$/N$_t$ smaller (larger) 
than unity means flow (antiflow). 

The data presented in Fig.~\ref{pi_flow_kaos_84} show that the flow behaviour
of negative and positive pions is similar. This is in contradiction 
to the data presented in 
Fig.~\ref{pi_flow_eos} 
where the positively charged pions exhibit a clear antiflow signal
at b=8 fm whereas the antiflow of the negatively charged pions is consistent 
with zero.     

\begin{figure}
%\vspace{-7.cm}
%\begin{turn}{90}
%\vspace{0.cm}
\hspace{1.cm}\mbox{\epsfig{file=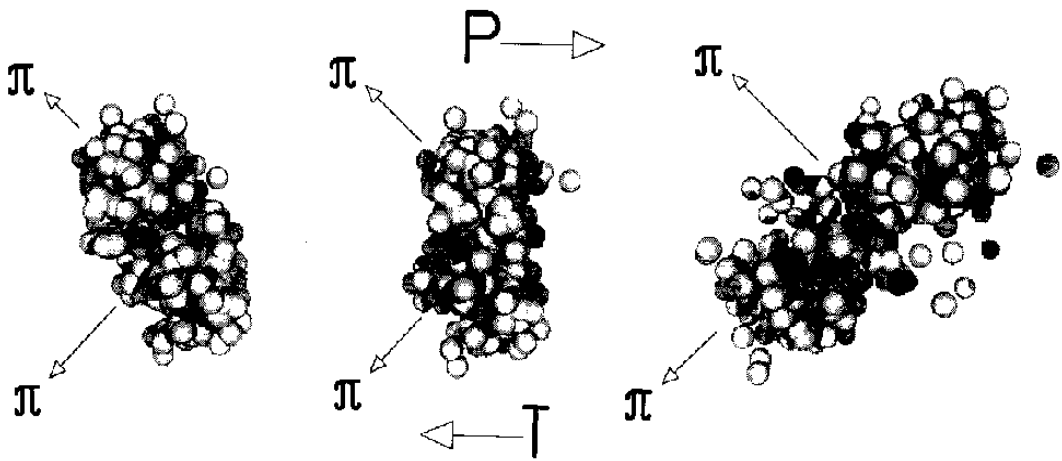,width=6.cm,height=14.cm,angle=270}}
%\vspace{6.cm}
\caption{Sketch of an Au+Au collision at 1 AGeV with an impact parameter 
of b = 7 fm as calculated by the QMD transport code \protect\cite{hartnack2}. 
The snapshots are taken at 4 fm/c (left), 8 fm/c (middle) and 20 fm.c (right). 
The pions are emitted in the reaction plane at backward  angles  
corresponding to the detector position.   
}
\label{pi_clock}
\end{figure}

The KaoS data (Fig.~\ref{pi_flow_kaos_84})  
can be  explained by the  shadowing of high p$_T$ pions
by the spectator fragments in the early phase of the collision. This effect
produces (i) a pronounced ''squeeze-out'' signal (see below) and 
(ii) a directed in-plane flow. The low p$_T$ pions, however, freeze out at 
a later time and the rescattering at the spectator fragments produces
an antiflow signal.      
Fig.~\ref{pi_clock} sketches the matter configuration    
at 4, 8 and 20 fm/c after the first touch of the nuclei  according to 
a QMD calculation for Au+Au at 1 AGeV and at an impact parameter of b=7 fm
\cite{hartnack2}.
The ''snapshots'' exemplify the effect of pion shadowing by spectator matter
at different stages of the collision.  
QMD calculations predict 
different freeze-out times and densities for low and high energy pions.
The model calculations find a correlation between high-energy
pions, early freeze-out times and high freeze-out densities \cite{bass2}.
This scenario is supported by 
the data shown in Fig.~\ref{pi_flow_kaos_84} 
which demonstrate for the first time experimentally
that pion emission is correlated to the reaction dynamics.

\vspace{.5cm}
First observations of a preferred out-of-plane emission of charged 
and neutral pions 
(''squeeze-out'') were reported by the KaoS and TAPS 
collaborations \cite{brill1,venema}. The azimuthal anisotropy was most 
pronounced for pions with high transverse momenta in
semicentral collisions. The experiments    
investigated Au+Au collisions at 1 AGeV.

Recently, the pion azimuthal emission pattern has been studied in detail for 
Bi+Pb collisions as a function of beam energy and impact parameter
by the KaoS-Collaboration \cite{brill3}. The experiment was focused 
on  the study of the pion emission perpendicular to the reaction plane 
and therefore the acceptance of the spectrometer covered midrapidity only.

\begin{figure}[ht]
%\vspace{-1.cm}
\begin{minipage}[t]{9cm}
\hspace{0.cm}\mbox{\epsfig{file=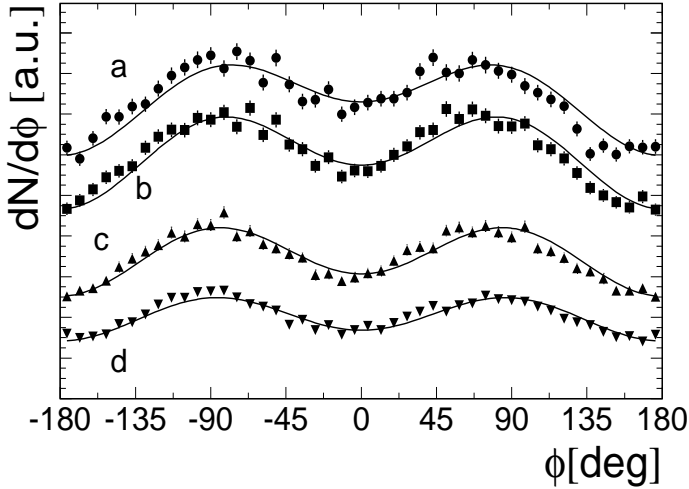,width=9.cm,height=7.cm}}
\end{minipage}
\begin{minipage}[t]{6.5cm}
\vspace{-7.cm}
\caption{Azimuthal distributions of $\pi^+$ measured
for semi-central Bi+Pb collisions at E = 700 AMeV \protect\cite{brill3}. 
The different symbols
correspond to different bins in transverse momentum: 
(a) 240 -300 MeV/c, (b) 300 -360 MeV/c, (c) 360 -420 MeV/c and 
(d) 420 -480 MeV/c. The linear ordinate starts at zero, the data are not 
corrected for the accuracy of the reaction plane determination.    
The solid lines are fits to the data (see text).
}
\label{pi_phi_bi}
\end{minipage}
\end{figure}

Fig.~\ref{pi_phi_bi} shows the azimuthal distributions of $\pi^+$ measured 
for semi-central Bi+Pb collisions at E = 700 AMeV. 
The different
symbols correspond to different bins in transverse momentum.
The distributions  show maxima
at azimuthal angles of $\phi = \pm 90^0$ which is  perpendicular to 
the reaction plane. 
The solid lines in Fig.~\ref{pi_phi_bi} correspond to the  
parameterization 
\begin{equation}
N(\phi) \propto 1+ a_1cos\phi + a_2cos2\phi
\end{equation}
The parameter $a_1$ quantifies the in-plane emission of the particles
parallel ($a_1>$1) or antiparallel ($a_1<$1) to the impact parameter vector,
whereas  $a_2$ stands for an elliptic emission pattern which may be
aligned with the event plane ($a_2>$0) or oriented perpendicular to the
event plane  ($a_2<$0).
The parameters  were determined by a fit to the data and corrected for the
uncertainty in the reaction plane reconstruction by
$a'_{1,2}$ = $a_{1,2}$/$<cos2\Delta\phi>$. The values of
$<cos2\Delta\phi>$ have been determined by a Monte Carlo simulation
and vary between  0.3 for peripheral and central collision and
0.5 for semi-central collisions (for details see \cite{brill3}).

The strength of the azimuthal anisotropy 
is given by the
ratio $R_N^{corr}$ which is the number of particles emitted perpendicular
to the event plane divided by
the number of particles emitted parallel to the event plane
(for $a_2<$0):
\begin{equation}
R_N^{corr} = \frac{N(+90^0) + N(-90^0)}{N(0^0) + N(180^0)} 
= \frac{1-a'_2}{1+a'_2}
\label{eq:r_n}
\end{equation}

\begin{figure}
%\vspace{-1.5cm}
\begin{minipage}[t]{9cm}
\centerline{\mbox{\epsfig{file=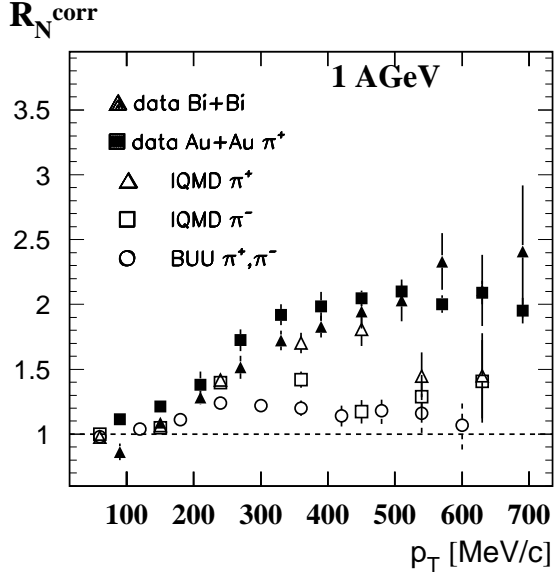,width=9.cm}}}
\end{minipage}
\begin{minipage}[t]{6.2cm}
\vspace{-8.cm}
\hspace{1.5cm}\caption{The anisotropy ratio for midrapidity pions from 
semi-central 
( b$\approx$7.5 fm) Au+Au and Bi+Bi
collisions at 1 AGeV (full symbols) as a function of transverse momentum 
\protect\cite{brill3}. The model calculations (open symbols) are performed at
5$\leq$b$\leq$10 (IQMD \protect\cite{bass1}) and 7$\leq$b$\leq$9
(BUU \protect\cite{baoli}).
}
\label{r_azimut_bi}
\end{minipage}
\end{figure}

The asymmetry parameter increases with 
pion transverse momentum. 
This is demonstrated in Fig.~\ref{r_azimut_bi} which shows the 
$R_N^{corr}$ as a function of transverse momentum
for Au+Au and Bi+Bi collisions at 1 AGeV.
The strength of the pion azimuthal anisotropy is almost constant
for impact parameters between 2.5 fm and 10 fm \cite{brill3}.
In contrast, the proton azimuthal anisotropy increases with increasing
impact parameter whereas the dependence of $R_N^{corr}$ on 
transverse momentum and  beam energy is similar for protons and pions
\cite{brill2}. According to transport calculations, 
the preferential emission perpendicular to the reaction plane is caused by 
rescattering and absorption in the spectator fragments \cite{bass3}.
Although the models predict or reproduce the out-of-plane emission of pions 
qualitatively, a quantitative agreement has not been reached up to now. 
This means that the emission of pions from a hot nuclear medium is   
not yet well understood within the framework of transport model calculations.

\vspace{0.5cm}

The  enhancement at 90$^0$ 
in the polar angle distribution
of the pions found in central collisions (see Fig.~\ref{pi_angle_fopi}) 
may be another 
manifestation of the same phenomenon. 
It is a challenge for transport models to describe both the 
polar and azimuthal distribution of pions quantitatively and then
answer the following questions: 
Is the pion azimuthal asymmetry
due to an enhancement perpendicular the event plane  or due to a suppression
in the event plane? Are the pions absorbed or rescattered? 
Is absorption the reason for the reduction of 
the mean pion multiplicity per nucleon
in A+A with respect to N+N collisions?

\vspace{.5cm}
The experimental and theoretical studies of the pion 
azimuthal emission pattern at LBL/SIS energies can be summarized as follows:
The anisotropies are probably caused mainly by the pion final-state interaction 
(rescattering and reabsorption).  
Low energy pions (which dominate the total pion yield) show 
a weak ``squeeze-out'' signal and weak antiflow for peripheral collisions. 
The pronounced 
out-of-plane emission of high-energy pions 
indicates that they freeze-out while the spectators are still close to the 
partizipant zone. The in-plane flow of high-energy pions is an additional 
signature for an early 
emission time. This finding is supported by an analysis of the 
$\pi^-/\pi^+$ ratio as discussed in the next section.  

Indications of azimuthal anisotropies in the pion emission pattern
have also been found in Au+Au collisions at AGS \cite{Barrette97}.
Data on the elliptic and directed flow of pions  and 
protons measured in Pb+Pb collisions at 158 AGeV 
have been very recently published by 
Na49 \cite{Appel98}.

\section{Pion isospin ratios}

\begin{figure}
%\vspace{-2cm}
\begin{minipage}[t]{9cm}
\mbox{\epsfig{file=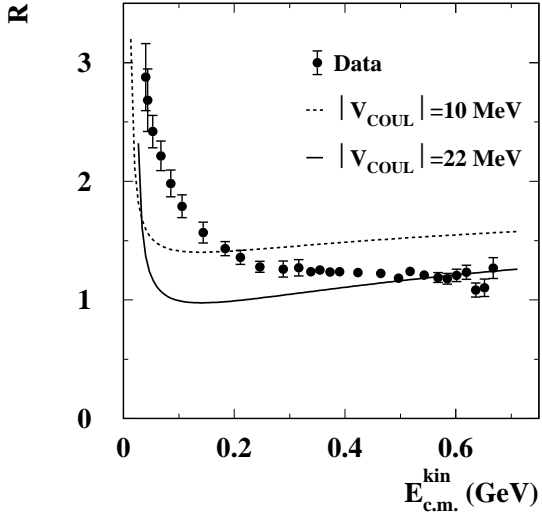,width=8.5cm}}
\end{minipage}
\begin{minipage}[t]{5.5cm}
\vspace{-5cm}
\caption{$\pi^-/\pi^+$ ratio for central  Au+Au collisions 
at 1 AGeV as a function of the pion kinetic energy. The solid (dotted)
curve shows the calculated ratio for a fixed Coulomb potential of
22 MeV (10 MeV) \protect\cite{wagner2}.  
}
\label{pi_iso_kaos}
\end{minipage}
\end{figure}

\begin{figure}
%\vspace{-2cm}
\begin{minipage}[t]{12.cm}
\mbox{\epsfig{file=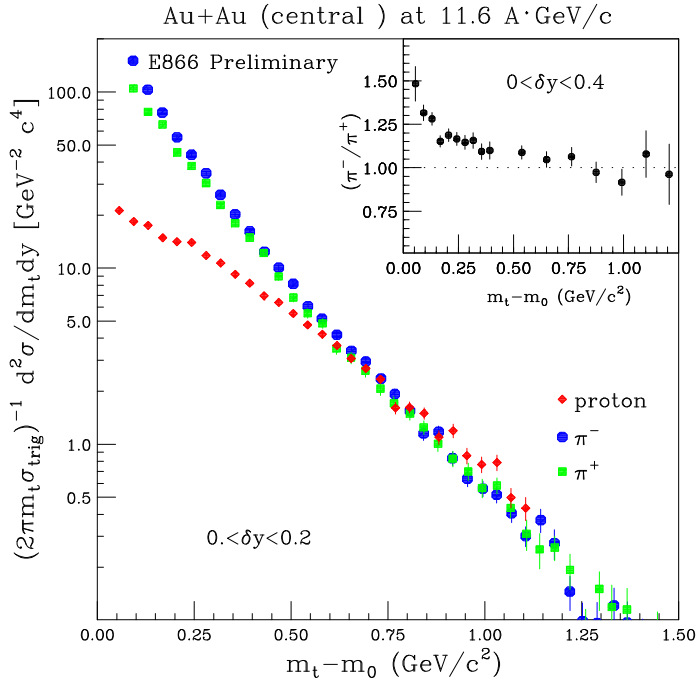,width=9.cm,angle=90}}
\end{minipage}
\begin{minipage}[t]{3.5cm}
\vspace{-5cm}
\caption{Transverse mass spectra of protons and pions measured in 
central Au+Au collisions at 10.7 AGeV around midrapidity \protect\cite{e802}.  
The insert shows the  corresponding
$\pi^-$/$\pi^+$ ratio.
}
\label{pi_iso_ags}
\end{minipage}
\end{figure}

\begin{figure}
%\vspace{-2cm}
\begin{minipage}[t]{10.cm}
\mbox{\epsfig{file=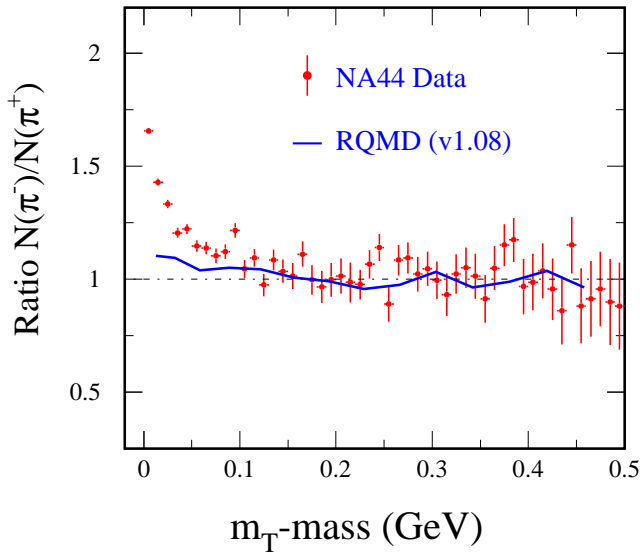,width=8.cm}}
\end{minipage}
\begin{minipage}[t]{5.5cm}
\vspace{-5cm}
\caption{$\pi^-$/$\pi^+$ ratio as a function of transverse mass
measured in central Pb+Pb collisions at 158 AGeV \protect\cite{nuxu}.
}
\label{pi_iso_sps}

\end{minipage}
\end{figure}

The multiplicity ratio of negatively to positively charged pions for 
Au+Au collisions at 1AGeV  
was found to be $n_{\pi^-}/n_{\pi^+}$=1.94$\pm$0.05 \cite{wagner2} and  
$n_{\pi^-}/n_{\pi^+}$=1.95$\pm$0.04$\pm$0.3  \cite{pelte1}; the latter value 
is obtained when assuming a linear dependence on A$_{part}$.   
According to the isospin decomposition
taking into account $\Delta_{33}$(1232) and N$^*$(1440) resonances the expected
ratio is $n_{\pi^-}/n_{\pi^+}$ = 1.90 \cite{verwest}. 
However, the $\pi^-/\pi^+$ ratio is found to be 
not constant as a function of the pion transverse momentum.       
The pion low-p$_T$ enhancement is much more pronounced for $\pi^-$ than 
for $\pi^+$ mesons.  This effect can be seen in 
Fig.~\ref{pi_iso_kaos}, Fig.~\ref{pi_iso_ags} and Fig.~\ref{pi_iso_sps} 
which show the $\pi^-$/$\pi^+$ ratios as a function 
of p$_T$ (or m$_T$) measured in central collisions
of very heavy nuclei at 1.0, 10.7 and 158 AGeV, respectively
\cite{wagner2,e802,na44,nuxu}.     
According to simulations with RQMD or RBUU transport codes, the increase
of the $\pi^-$/$\pi^+$ ratio towards smaller values of  transverse mass
is due to the 
Coulomb interaction between pions and  the net charge of the participants
\cite{bass1,teis2}.
It can be seen that the effect is largest for the lowest bombarding energy
(1 AGeV). At a beam energy of 158 AGeV the $\pi^-$/$\pi^+$ ratio 
exceeds  unity only for values of m$_T$-m$_0$ below 0.1 GeV \cite{nuxu}.

\vspace{.4cm}
The pronounced enhancement of the $\pi^-$/$\pi^+$ ratio at low pion 
energies as measured for 1 AGeV 
has been exploited to determine the size of 
the pion emitting source \cite{wagner2}. The procedure is based on 
the picture, that the primordial  $\pi^-$ and the $\pi^+$ spectra 
are shifted against each other by the Coulomb field of   
the fireball which changes its size slowly compared to the velocity of 
the pions. Hence, the effective Coulomb potential of the fireball 
is determined from the shift in energy of the charged pions.
It turns out that the deduced Coulomb potential decreases with decreasing 
pion energy.  
The effective freeze-out radius r$_{eff}$ is derived from the Coulomb potential  
V(r$_{eff}$) = e$^2$Z/r$_{eff}$ with Z the measured  
number of participating protons (which are the charged 
constituents of the fireball). Within the framework of this analysis 
a value of r$_{eff}$ = 7.2$\pm$1.1 fm has been determined for 
the freeze-out radius of high-energy pions in 
central Au+Au collsions at 1 AGeV \cite{wagner2}.

\chapter{Strangeness Production}

The production of $strange$ particles in nucleus-nucleus collisions is of 
particular interest as the heavy strange (and antistrange) quarks 
do not exist prior to the collision.
The annihilation of a produced s$\bar s$ pair is unlikely as long as their 
abundance is low. Therefore, the strange (and antistrange) 
quarks ''live'' in hadrons until they disappear by weak decays. 
The K$^+$ meson, for example, leaves the  hot and dense reaction zone 
nearly undisturbed because of its long mean free path
in nuclear matter ($\lambda \approx$ 5 fm). The corresponding
K$^+$N cross section is almost entirely due to the  
elastic scattering process 
($\sigma_{K^+N\rightarrow K^+N}\approx$ 12 mb for 
p$_{K^+} <$ 1 GeV/c). 
The charge exchange process is less important 
($\sigma_{K^+n\rightarrow K^0p}$ = 1-7 mb for p$_{K^+}$ = 0.2-0.6 GeV/c)
\cite{dover}.

\vspace{.4cm}
In heavy ion collisions 
at BEVALAC/SIS energies, kaons are produced near or below the
kinematical thresholds which
correspond to kinetic
beam energies of 1.58 GeV (for NN$\rightarrow$K$^+\Lambda$N) and 2.5 GeV 
(for  NN$\rightarrow$K$^+$K$^-$N).  
For central nucleus-nucleus collisions at ''subthreshold'' beam energies,
experimental results suggest 
that a major  fraction of the kaons is created in processes which require
multiple interactions of the participant nucleons
\cite{miskowiec,cieslak,mang,barth}. 
Within the framework of microscopic transport models, 
these kaons are produced in
secondary collisions like $\Delta$N$\rightarrow$KYN and 
$\pi$N$\rightarrow$KYN with Y=$\Lambda,\Sigma$
\cite{ran_ko,randrup,ko,cugnon,maruyama,fuchs}. 
Sequential processes occur preferentially at  high baryonic densities and 
therefore the K$^+$ yield  is expected to be sensitive 
to nuclear matter properties \cite{aich_ko}.
This sensitivity is enhanced at  bombarding energies 
far below the K$^+$ threshold. 
For Au+Au at 1 AGeV, transport codes predict a measurable effect 
on the K$^+$ production cross section when 
considering different nuclear equation
of states \cite{aich_ko,maruyama,li_ko}.  
However, the reliability of the model calculations still suffers from an
insufficient understanding (and the simplified treatment) 
of the in-medium nucleon-nucleon interactions
and from the lack of information on 
$\Delta$N$\rightarrow$KYN reactions.
Nevertheless, nuclear matter properties can be explored by detailed 
comparisons of specific model predictions to experimental data as function
of energy and system size.   

\vspace{.4cm}
In the last years, the question of in-medium modifications of hadron
properties has attracted increasing interest. Calculations 
based on  chiral Lagrangians predict an in-medium kaon-nucleon
potential which is  weakly repulsive for kaons but strongly attractive
for antikaons \cite{kaplan,brown,weise}.  
Recent transport model calculations,
which incorporate these medium effects, find  a  considerable  impact on 
the K$^+$ and K$^-$ yields and on their azimuthal angular distributions
for nucleus-nucleus collisions at SIS  energies 
\cite{fang_ko,cassing,brat,li_flow,li2}.  
In particular, the  yield of antikaons is 
expected to be significantly enhanced in dense nuclear matter.
Indeed, experimental results on K$^-$ production and on K$^+$ directed flow 
in Ni+Ni collisions were interpreted as indication for
in-medium modifications of kaon properties \cite{schroeter,barth,ritman}.

\vspace{.4cm}
At ultrarelativistic energies, strangeness production is considered as a 
signal from the quark-gluon plasma (QGP) phase. 
An extraordinary increase of the strange particle
yield is expected if the fireball has spent a non negligible amount of time 
in the deconfined phase. During this period, chemical equilibrium abundances
of s,u,d quarks are reached faster than in an hadronic environment: 
in the QGP phase the temperature is comparable to the strange quark mass
whereas in the hadronic phase the production of strange hadrons is  
suppressed by their large masses. 
This is, however, not the only scenario for an enhanced production of 
strange (with respect to nonstrange) particles. We have mentioned 
for example the effective lowering of the production threshold
by secondary interactions. 
Experimentally the strangeness enhancement has been established 
in nucleus-nucleus collisions  at SIS, AGS and SPS energies. 
It is a challenge to unravel the different underlying dynamics 
at the different energies. 

\vspace{.4cm} 
In the following sections we review the existing experimental data on the 
production cross sections  
of kaons and on their spectral distributions.
The low energy data and their impact on the determination of nuclear
matter properties will be considered first (section 4.1 through 4.5).
In section 4.6 the production of K-mesons at AGS and SPS 
energies will be reviewed. 
Multistrange hadrons will be covered together with  
production of antibaryons in a separate
chapter.     

\section{Kaon production probabilities}

First experiments on strangeness production in nucleus-nucleus 
collisions were performed in the early 1980's at the LBL in Berkeley 
\cite{schnetzer,harris3,shor1} and at the JINR in Dubna 
\cite{anikina}.
Inclusive  K$^+$ production cross sections 
$d^2\sigma/dpd\Omega$ were measured 
in Ne+NaF and Ne+Pb collisions at 2.1 AGeV between $15^o < \Theta_{lab} < 80^o$
\cite{schnetzer}. 

The second generation experiments at SIS/GSI have extended the study of
K$^+$  production to very heavy collision systems and to beam energies 
far below the threshold for free nucleon-nucleon collisions which is 1.58 GeV 
\cite{best,miskowiec,ahner,barth,cieslak,mang,ritman}.
Table \ref{kxsec} presents a compilation of total K$^+$ cross sections measured 
in nucleus-nucleus collisions at energies between 0.8 and 2 GeV per nucleon.
The data were taken mostly within a restricted range of emission angles
and momenta. The cited total K$^+$ cross sections were calculated by 
extrapolating 
to  full phase space assuming a Maxwell-Boltzmann distribution and
isotropic emission (the validity of the second assumption will be discussed 
in section 4.3).

\begin{figure}[hpt]
\vspace{0.cm}
\begin{minipage}[t]{9cm}
\centerline{\hspace{ -1.cm}\mbox{\epsfig{file=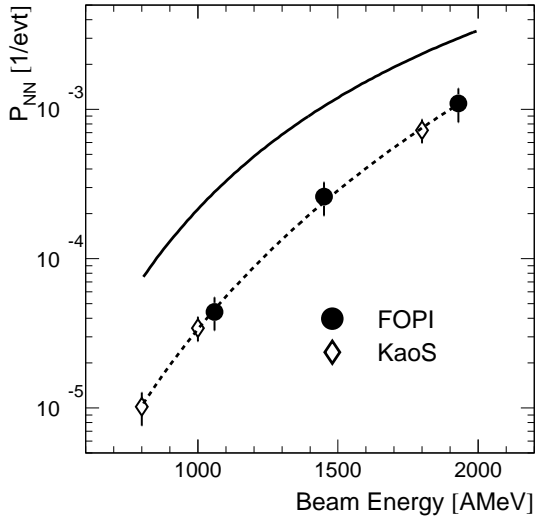,width=7.cm}}}
\end{minipage}
\begin{minipage}[t]{6.5cm}
\vspace{-7.cm}
\caption{K$^+$ production probability 
(P$_{NN}$ = multiplicity per participating 
nucleon) measured in Ni+Ni collisions as a function of beam energy
(full symbols: \protect\cite{best}, open symbols: \protect\cite{barth}). 
Dashed line: P$_{NN} \propto$ E$_{beam}^{5.3\pm0.3}$. The solid line
represents the production probability expected for pions and eta-mesons 
(when corrected for the threshold energies) 
according to an empirical scaling by Metag (\protect\cite{metag}, 
see Fig.~\protect\ref{eta_pi_exci}). Taken from
\protect\cite{best}.
}
\label{kp_excit}
\end{minipage}
\end{figure}

\vspace{.4cm}
For Ni+Ni collisions, the excitation function of K$^+$ production 
at beam energies below and near threshold can be parameterized 
by P$_{K^+} \propto$ E$_{beam}^{\alpha_E}$ with $\alpha_E$ = 5.3$\pm$0.3 
(see Fig.~\ref{kp_excit}). 
P$_{K^+}$ is the K$^+$ production probability per participating nucleon. 
The data are taken from \cite{best,barth}. 
The corresponding exponent $\alpha_E$
in p+p$\rightarrow{K^+}$+X 
reactions near threshold is much larger ($\approx$ 18). 
In fact, a value of  $\alpha_E$=5.3 is found for the p+p reaction 
only at beam energies  between 2.0 and 2.5 GeV i.e.
more than 400 MeV above threshold.

\begin{table}[htb]
\caption{Inclusive K$^+$ production cross sections and inverse 
slope parameters T from symmetric 
nucleus-nucleus collisions. The values are determined by fitting the data with
a Maxwell Boltzmann distribution d$^3\sigma/$dp$^3\propto$ exp(-E/T) 
 and integrating the fitted curve over momentum.
The total cross section is calculated assuming isotropic emission in 4$\pi$.
The Au+Au cross sections 
agree within the error bars, the slopes do not. However, the 
dataset of Ref. \protect\cite{miskowiec} consists of about 100 K$^+$ whereas 
the results of Ref. \protect\cite{mang} are based on about 10$^4$ K$^+$ mesons.
The data marked with (*) are preliminary.}
\vspace{0.5cm}
\begin{center}
\begin{tabular}{|c|c|c|c|c|c|}
\hline
A+A & E (AGeV) & $\Theta_{lab}$ (deg)& 4$\pi\times$d$\sigma/d\Omega$ (mb)&T (MeV)&Ref.\\
\hline
C+C& 1.0 (*)& 44& 0.1$\pm0.03$  & 61$\pm$5  &\cite{sturm}\\
\hline
C+C& 1.8 (*)& 40& 3.3$\pm1$  & 73$\pm$6  &\cite{sturm}\\
\hline
Ne+NaF& 1.0& 44 & 0.33$\pm0.1$  & 61$\pm$6  &\cite{ahner}\\
\hline
Ne+NaF& 2.0& 44 & 20$\pm10$  & 79$\pm$6  &\cite{ahner}\\
\hline
Ne+NaF& 2.1& 15 - 80 & 23$\pm8$  & 122  &\cite{schnetzer}\\
\hline
Ni+Ni& 0.8& 44 & 0.85$\pm0.2$  & 59$\pm$6  &\cite{barth}\\
\hline
Ni+Ni& 1.0& 44 & 2.7$\pm0.5$  & 75$\pm$6  &\cite{barth}\\
\hline
Ni+Ni& 1.8& 44 & 57$\pm15$  & 88$\pm$7  &\cite{barth}\\
\hline
Au+Au& 1.0& 44 & 41$\pm11$  & 67$\pm$5  &\cite{miskowiec}\\
\hline
Au+Au& 1.0 (*)& 34, 44, 54, 84& 26$\pm5$  & 85$\pm$6  &\cite{mang}\\
\hline
Bi+Pb& 0.8& 44 & 9.4$\pm4.9$  & 64$\pm$11 &\cite{cieslak}\\
\hline
\end{tabular}
\end{center}
\label{kxsec}
\end{table}

\begin{figure}
%\vspace{-1.3cm}
\begin{minipage}[t]{7.5cm}
\centerline{\hspace{ 0.cm}\mbox{\epsfig{file=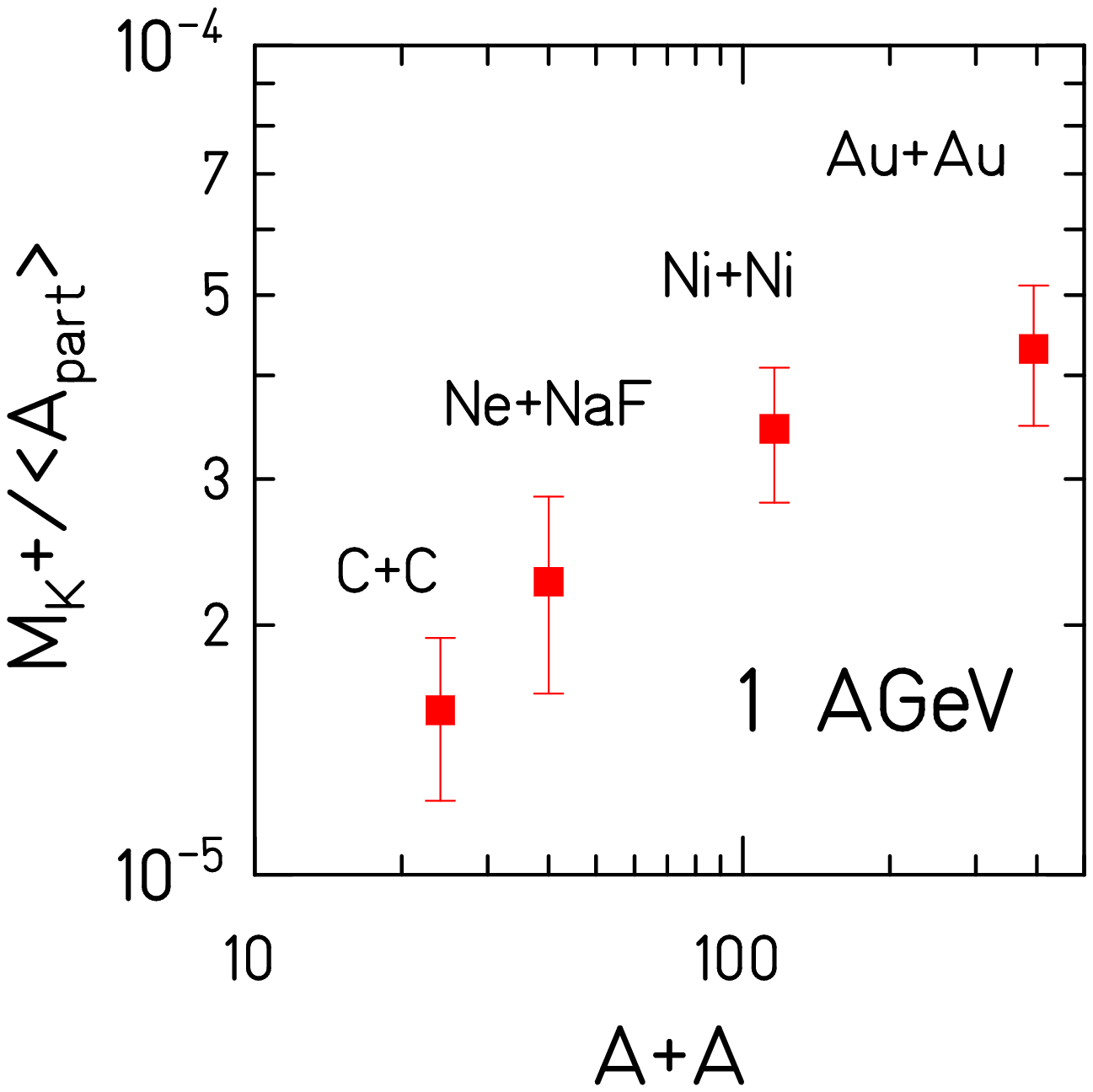,width=9cm}}}
\vspace{0.cm}
\caption{K$^+$ multiplicity per average number of participating nucleons
as a function of the mass of nuclei  colliding at 1 AGeV 
\protect\cite{ahner,mang,barth,sturm}.  
}
\label{kp_mass}
\end{minipage}
\hspace{0.5cm}
\begin{minipage}[t]{7.5cm}
\vspace{-8.cm}
\hspace{ -0.5cm}\mbox{\epsfig{file=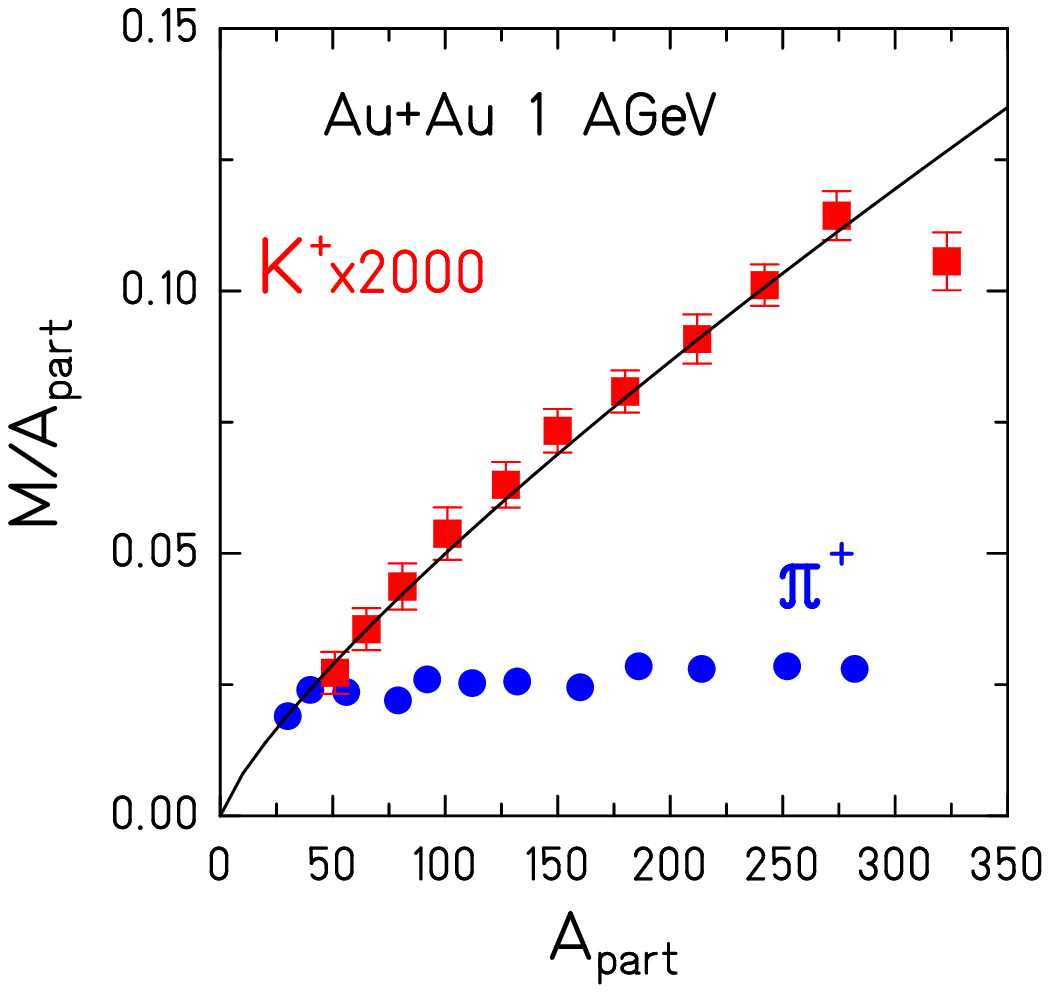,width=8.6cm}}
\vspace{-0.1cm}
\caption{K$^+$ and $\pi^+$ multiplicity per number participating nucleons
A$_{part}$ as a function of A$_{part}$ for Au+Au collisions at 1 AGeV
\protect\cite{wagner1,mang} (preliminary).
}
\label{kp_mult_au}
\end{minipage}
\end{figure}

\vspace{.4cm}
The average  K$^+$ multiplicity per participating nucleon 
was studied as a function of the mass of the colliding system 
for a beam energy of 1 AGeV.
It was calculated from the 
K$^+$ production cross section according to 
n$_{K^+}$(A) = $\sigma^{K^+}/\sigma_R$ 
with the reaction cross section 
$\sigma_R$=4$\pi$(r$_o$A$^{1/3}$)$^2$  (r$_o$=1.2 fm).
The K$^+$ production cross section was measured for C+C, Ne+NaF, 
Ni+Ni and Au+Au \cite{ahner,sturm,mang,barth}. 
Fig.~\ref{kp_mass} 
shows the K$^+$ multiplicity per participating nucleon 
n$_{K^+}/<$A$_{part}>$ with $<$A$_{part}>$= A/2 according to the geometrical
model for symmetric collisions.    
The increase of n$_{K^+}/<$A$_{part}>$ with the mass of the collision system  
is a clear experimental  signature for kaon production 
via collective processes.  
According to model calculations the kaons are produced predominantly
in $\Delta$-nucleon and pion-nucleon 
collisions \cite{ran_ko,fuchs,fang_ko,hartnack}.

\vspace{.4cm}
Kaon production was also studied as a function of impact parameter. 
It was found that the 
K$^+$ yield normalized to the number of participating nucleons
is enhanced in central
as compared to peripheral nucleus-nucleus collisions \cite{miskowiec}.  
Fig.~\ref{kp_mult_au} 
shows the meson  multiplicities n$_{K^+}$ and n$_{\pi^+}$
per participating nucleon (A$_{part}$)
as a function of A$_{part}$ measured by the KaoS Collaboration 
for Au + Au at 1 AGeV \cite{mang}.
The $\pi^+$ muliplicity per A$_{part}$ is roughly constant (circles) whereas
n$_{K^+}$ increases according to A$_{part}^{\alpha_A}$ with
$\alpha_A = 1.8\pm0.15$. Values of  $\alpha_A$ 
larger than unity indicate that kaons are produced
in multiple collisions of participants. The A$_{part}$-dependence 
of n$_{K^+}$(A$_{part}$) 
is more pronounced than the A-dependence of the 
average kaon multiplicity per participating nucleon 
for different  A+A collisions as shown in Fig.~\ref{kp_mass}.

\begin{figure}
%\vspace{-2cm}
\begin{minipage}[t]{10cm}
\hspace{ 0.cm}\mbox{\epsfig{file=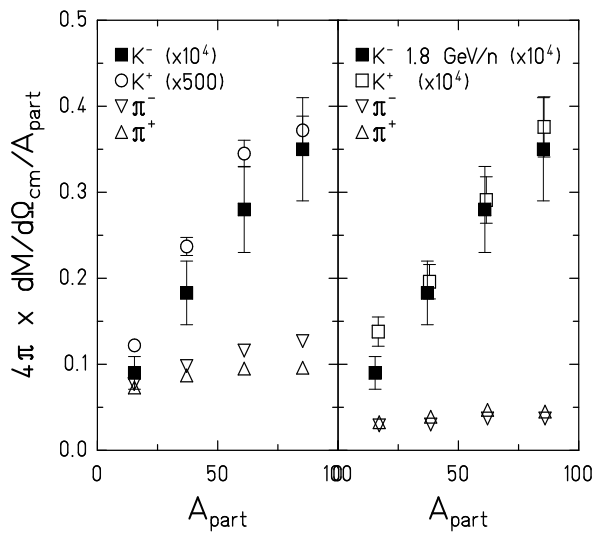,width=10.cm,height=8.5cm}}
\end{minipage}
\begin{minipage}[t]{5cm}
\vspace{-4.5cm}
\caption{Kaon and pion multiplicities per number participating nucleons
A$_{part}$ as a function of A$_{part}$ for Ni+Ni collisions.
Left panel: K$^-$, K$^+$, $\pi^+$, $\pi^-$ for a beam energy of 1.8 AGeV.
Right panel: K$^-$ at 1.8 AGeV and K$^+$, $\pi^+$, $\pi^-$  at 1.0 AGeV. 
\protect\cite{barth} 
}
\label{ka_mult_ni}
\end{minipage}
\end{figure}

The K$^+$ scaling factor $\alpha_A$   has been also measured  
for Ni+Ni collisions as a function of the bombarding energy 
(see Fig.~\ref{ka_mult_ni}) \cite{barth}. 
Values of $\alpha_A$ = 1.9$\pm$0.25, 1.7$\pm$0.25 and 1.65$\pm$0.15 
have been determined for 0.8, 1.0 and  1.8 AGeV, respectively.  
Thus $\alpha_A$ does not vary significantly with bombarding energy
between 0.8 and 1.8 AGeV for intermediate mass systems. 
Hovever, for K$^+$ production  in Bi+Pb collisions 
at 0.8 AGeV a value of $\alpha_A$=2.26$\pm$0.14 was found
\cite{cieslak}. This value is larger than the one obtained
for Au+Au at 1 AGeV (see above) indicating a weak increase 
of $\alpha_A$  with decreasing beam energy for very heavy collision systems.   

\begin{figure}
%\vspace{0.cm}
\hspace{ 1.cm}\mbox{\epsfig{file=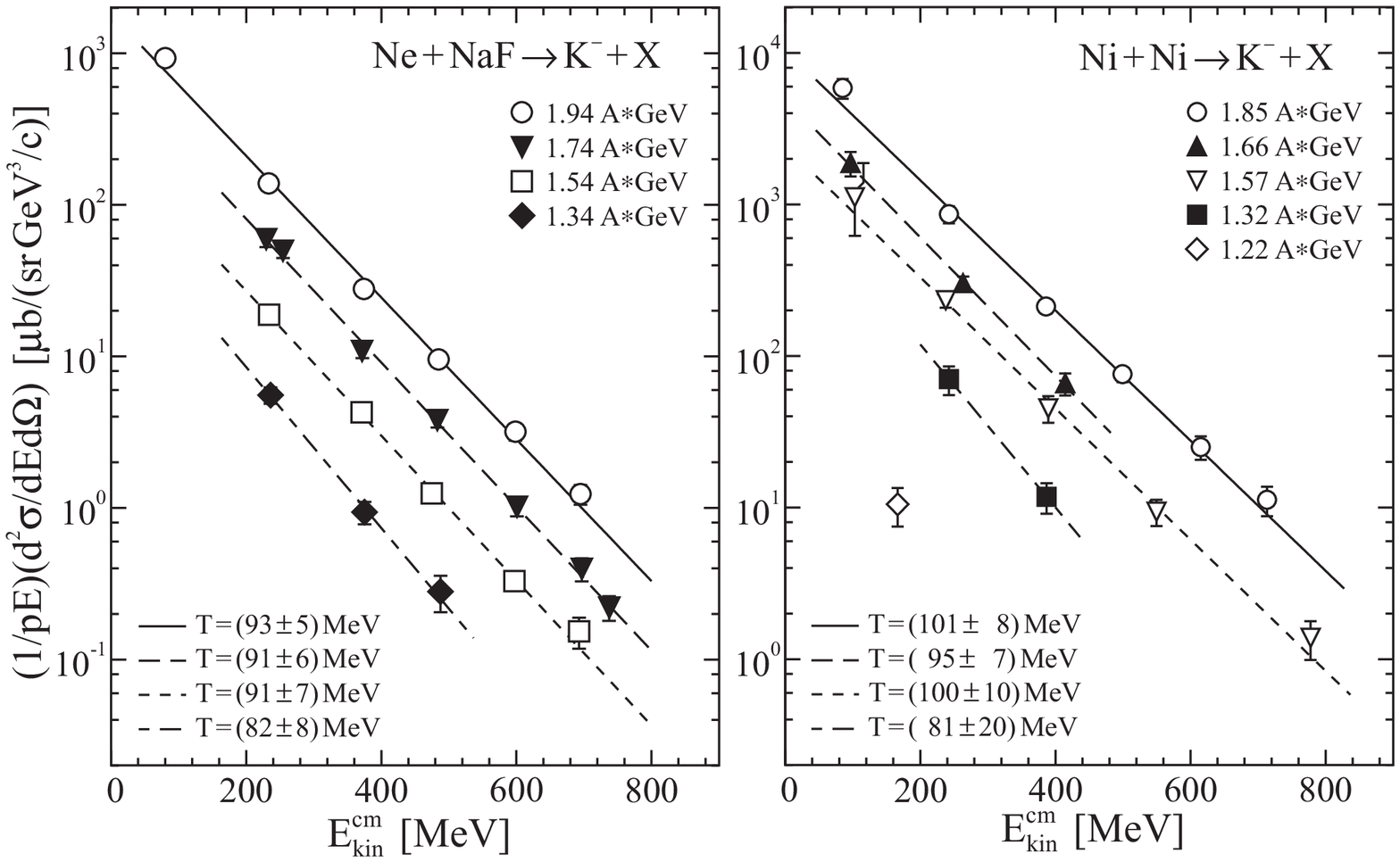,width=14.cm,height=9.cm}}
\vspace{0.cm}
\caption{Invariant K$^-$ production cross section as a function of
the K$^-$ c.m. kinetic energy for the reactions Ne+NaF and Ni+Ni at the
bombarding energies indicated \protect\cite{kienle}. The lines
are exponential fits to the data points with inverse slope parameters
T as indicated.  
}
\label{km_spec_frs}
\end{figure}

\vspace{.4cm}
The data on K$^-$ production in nucleus-nucleus collisions are rather
scarce. Pioneering experiments at the BEVALAC studied K$^-$ yields 
in the system Si+Si at beam energies of 1.16 - 2.0 AGeV  and in 
Ne+NaF at 2.0 AGeV at laboratory angles of 0$^o$ \cite{shor1,shor2}. 
Similar experiments (at $\Theta_{lab}=0^0$)  
were  performed with the Fragment Separator (FRS) 
at SIS/GSI \cite{schroeter,gillitzer,kienle}. 
Fig.~\ref{km_spec_frs} shows the K$^-$ invariant
production cross section for Ne+NaF (at 1.34 - 1.94 AGeV ) 
and Ni+Ni collisions (at 1.22 - 1.85 AGeV) \cite{kienle}.     
The spectral distributions fall off exponentially
as indicated by the dashed and solid lines. 
The values of the inverse slope parameters
are listed in Table \ref{invsl}. They seem to be independent of beam energy and
projectile mass (within the large error bars).
The data allow to extract an excitation function 
$\sigma^{K^-}\propto$E$_{beam}^{\alpha_E}$ with $\alpha_E \approx$ 10.

The mass dependence of K$^-$ production can also be estimated from 
the FRS data (some data given in \cite{schroeter} have been revised 
in \cite{kienle}). The ratio of the K$^-$ production cross sections
measured in Ni+Ni and Ne+NaF was found to be 
$\sigma_{Ni}^{K^-}/\sigma_{Ne}^{K^-}$ $\approx$10 \cite{kienle}.
When correcting for the geometrical reaction cross section
$\sigma_R$=4$\pi$(r$_o$A$^{1/3}$)$^2$ one obtains the ratio of multiplicities
M$_{Ni}^{K^-}$/M$_{Ne}^{K^-}$ $\approx$ 5 corresponding to a 
scaling of M$^{K^-}\propto$A$^{\approx 1.5}$. 

Recently, K$^-$ production cross sections have been measured 
also around midrapidity ($\Theta_{c.m.} \approx 90^o$) 
in Ni+Ni and C+C collisions at 1.8 AGeV by the KaoS Collaboration 
\cite{barth,laue}.         
From these data one can estimate a scaling according to 
M$^{K^-}\propto$A$^{1.6\pm0.2}$. 
This scaling behaviour is in agreement with the dependence of K$^-$ production 
on the number of participants as found in Ni+Ni collisions for different impact
parameters:
M$^{K^-}\propto$A$_{part}^{\alpha_A}$
with $\alpha_A$ = 1.8$\pm$0.3 \cite{barth}.  

\vspace{.4cm}
In Ni+Ni collisions at 1.8 AGeV, the K$^-$/K$^+$ ratio was found 
to be 4.8$\pm$2$\times 10^{-2}$  \cite{barth}. 
In C+C collisions at 1.8 AGeV, the preliminary value for the 
K$^-$/K$^+$ ratio is approximately
2.5$\times 10^{-2}$ \cite{laue}.  
When increasing the mass of the collision system from C+C to Ni+Ni, 
the K$^-$ yield increases by about   
a factor of 2 faster than the K$^+$ yield. 
Part of this K$^-$ enhancement may be due to the fact that at a 
beam energy of 1.8 AGeV  
the production of K$^-$ mesons is a subthreshold process which is 
more sensitive to the size of the collision system than 
K$^+$ production which is above threshold. 
Nevertheless, the experimental
result is quite intriguing since one expects  that  due to 
K$^-$ absorption the K$^-$ yield increases slower than the K$^+$ yield 
with increasing mass and size of the collision system \cite{zwer}.

A measurement of K$^+$ and K$^-$ at slightly higher (3.65 AGeV) energies has 
been made available in reference \cite{pantuev}. The authors report
ratios K$^+$/$\pi^+$=0.07  and  K$^-$/K$^+$=0.07 for C+C collisions 
at midrapidity. A very similar value for the K$^-$/K$^+$ ratio 
has been measured in proton-proton collisions at the same 
bombarding energy (see Fig.~\ref{ka_pp}).

\begin{table}[htb]
\caption{Inclusive K$^-$ production cross sections and inverse 
slope parameters T from symmetric 
nucleus-nucleus collisions. The values are determined by fitting the data with
a Maxwell Boltzmann distribution d$^3\sigma/$dp$^3
\propto$ exp(-E/T) and integrating the fitted curve over momentum.
The total cross section is calculated assuming isotropic emission in 4$\pi$.
The data marked with (*) are preliminary.}
\vspace{0.5cm}
\begin{center}
\begin{tabular}{|c|c|c|c|c|c|}
\hline
A+A & E (AGeV) & $\Theta_{lab}$ (deg)& 4$\pi\times$d$\sigma/d\Omega$ (mb)&T (MeV)&Ref.\\
\hline
C+C& 1.8 (*) & 40 & 0.075$\pm$0.02  & 60$\pm$5 &\cite{laue}\\
\hline
Ne+NaF& 1.34& 0 & - & 82$\pm$8  &\cite{kienle}\\
\hline
Ne+NaF& 1.54& 0 & - & 91$\pm$7  &\cite{kienle}\\
\hline
Ne+NaF& 1.74& 0 & - & 91$\pm$6  &\cite{kienle}\\
\hline
Ne+NaF& 1.94& 0 & - & 93$\pm$5  &\cite{kienle}\\
\hline
Si+Si& 2.0&  0 &  & 103$\pm$7  &\cite{shor1}\\
\hline
Ni+Ni& 1.32& 0 & -   & 81$\pm$20  &\cite{kienle}\\
\hline
Ni+Ni& 1.57& 0 & -   & 100$\pm$10  &\cite{kienle}\\
\hline
Ni+Ni& 1.66& 0 & -  & 95$\pm$7  &\cite{kienle}\\
\hline
Ni+Ni& 1.85& 0 & -  & 101$\pm$8  &\cite{kienle}\\
\hline
Ni+Ni& 1.8& 44 & 2.7$\pm$1.0  & 90$\pm$15  &\cite{barth}\\
\hline
\end{tabular}
\end{center}
\label{invsl}
\end{table}

\section{Kaon spectral distributions}
At sufficiently low center-of-mass energies,  
particle emission is more or less isotropic. 
Under this condition the appropriate kinematical 
observable is the kinetic energy. Its distribution is a straight exponential
in the case of a completely chaotic motion with different particles 
having the same slope factor which is simply the inverse temperature. 
Collective radial flow will 
modify the shape of such spectra. This effect increases with increasing 
mass of the particles with the consequence that particles with different masses 
will have different slope factors. 
Another complication in the interpretation of
particle spectra is the effect induced by resonance decays
which occur after freeze-out. At low (SIS) 
energies kaons are not affected by this phenomenon.
Therefore, kaons are especially suited to study the interplay of 
chaotic and ordered motion.

\vspace{.4cm}
The K$^+$ spectra measured at LBL/SIS energies were parameterized by 
Maxwell-Boltzmann distributions \cite{ahner,barth}. The resulting inverse slope parameters 
as derived from inclusive differential cross sections at midrapidity 
are listed in Table \ref{kxsec}. 

\begin{figure}
\vspace{-1.cm}
\begin{minipage}[t]{12.cm}
\centerline{\hspace{ 0.cm}\mbox{\epsfig{file=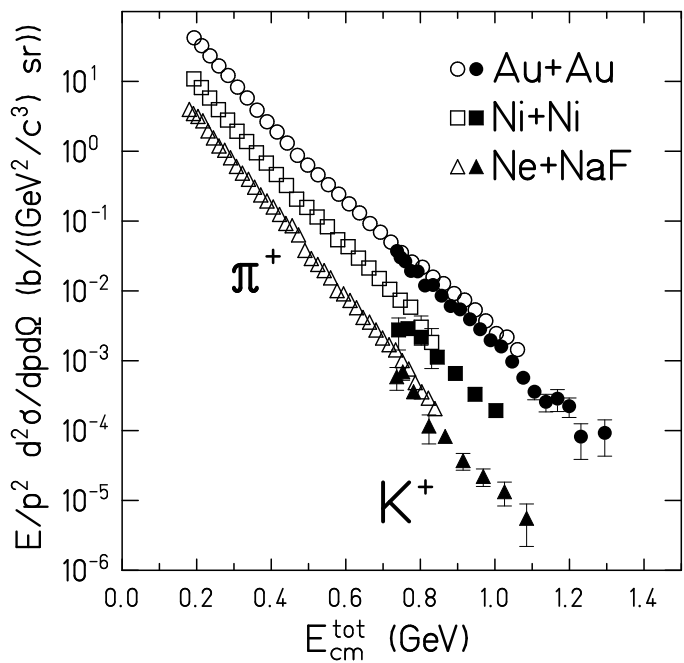,width=12.cm}}}
\end{minipage}
\begin{minipage}[t]{4.cm}
\vspace{-7.cm}
\caption{Invariant production cross sections for $\pi^+$ (open symbols)
and K$^+$ mesons (full symbols)  
as a function of the kinetic energy (plus the energy 
needed to produce the particle) in
Ne+NaF, Ni+Ni and Au+Au collisions at 1 AGeV (see text) 
\protect\cite{ahner,barth,mang}.  
}
\label{pi_ka_1agev}
\end{minipage}
\end{figure}

\begin{figure}
\begin{minipage}[t]{11cm}
\vspace{0.cm}
\centerline{\hspace{ 0.cm}\mbox{\epsfig{file=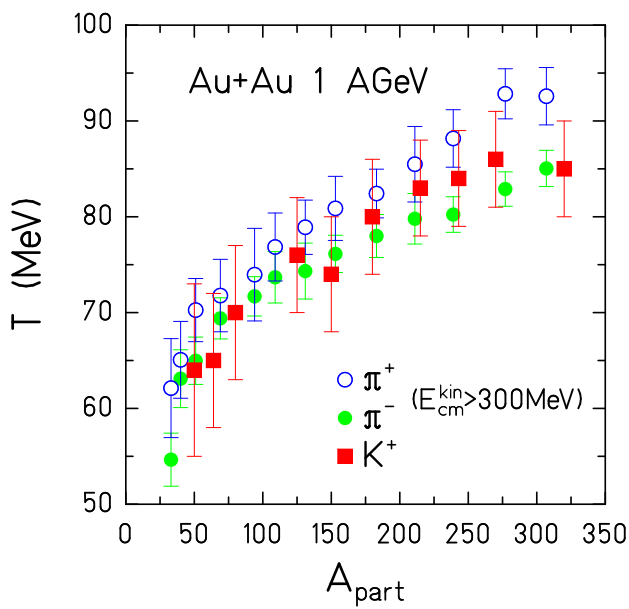,width=11.cm}}}
\end{minipage}
\begin{minipage}[t]{5.cm}
\vspace{2.cm}
\caption{Inverse slope parameter for high-energy pions 
(E$^{kin}_{cm}> 300 MeV$)
and K$^+$ mesons in Au+Au collisions at 1 AGeV (see text).  
The data are taken at $\Theta_{lab}$ = 44$^0$ \protect\cite{wagner1,mang}.
}
\label{t_pi_ka_au}
\end{minipage}
\end{figure}

It was found that the K$^+$ inverse slope parameters agree 
with the those of  high-energy pions (see tables \ref{slopep} and \ref{kxsec}).
This accordance is demonstrated   for  different collision systems (at 1 AGeV) 
in Fig.~\ref{pi_ka_1agev} which shows
the K$^+$ and $\pi^+$ invariant cross section as a function of the total
energy carried by the particle which includes the energy needed
to produce the particle. The total energy
is E$_{tot}$=E$_{kin}$+m$_{\pi}$ for pions and 
E$_{tot}$=E$_{kin}$+m$_K$+(m$_{\Lambda}$-m$_N$) for the associated production
of a K$^+$. The differential cross sections of the different particle types
coincide within a factor of about 2  
in the interval in which their total energies overlap.

Spectral distributions  of K$^+$ and pions were also measured as a function 
of collision centrality  in Au+Au collisions at a bombarding energy of 1 AGeV  
(see Fig.~\ref{t_pi_ka_au}). 
The inverse slope parameters both of high-energy pions and K$^+$ mesons 
increase with increasing centrality. The fact that particles
with different masses  exhibit very similar spectral slopes, indicates 
that the impact of collective flow and  resonance decays is small. 
It rather suggests that the ''temperature'' 
(i.e. the energy in chaotic motion)  of the system 
rises with the number of participant nucleons and thus with system size.  
Furthermore, these observations are a signature
for a common freeze-out temperature realized probably in the early phase 
of the  nuclear fireball with particle emission  
according to the phase space available.

\vspace{.4cm}
According to RBUU calculations
the K$^+$ mesons are produced in the high density phase ($\rho>2\rho_o$
in central Au+Au collisions at 1 AGeV) and the K$^+$ yield  reaches its
final value after about 17 fm/c \cite{fang_ko}. At this time the system starts 
to expand. The primordial
kaon spectrum is relatively soft due to the limited phase space but will
be modified  by the
K$^+$ propagation and rescattering in the nuclear medium. 
The kaons get (partially) thermalized by K$^+$N rescattering which makes 
the spectrum harder \cite{fang_ko}. 

\vspace{.4cm}
In Ne+NaF collisions at 2 AGeV, invariant cross sections 
for kaon and pions production have been measured \cite{ahner}. 
Inverse  slope parameter of about 80 MeV
were found both for K$^+$ mesons  and high-energy pions \cite{ahner}.
This result is at variance with the high  
K$^+$ ''temperature'' of T = 122 MeV reported  
for Ne+NaF collisions at 2.1 AGeV \cite{schnetzer} which were interpreted 
as a signature for freeze-out during an early high temperature phase 
\cite{nagamiya,nagamiya2}. 
The large value of the slope parameter may be an artefact of the averaging 
over c.m-angle in reference \cite{schnetzer} in the presence of a
nonisotropic K$^+$ angular distribution.  This observable will be 
discussed in the following section.

\section{K$^+$ angular distributions}

K$^+$ inclusive double differential
cross sections were measured at $\Theta_{lab}$ = 44$^o$ with a 
magnetic spectrometer \cite{senger} and at $\Theta_{lab}$ = 85$^o$ 
and 125$^o$ with 
plastic-scintillator range telescopes \cite{elmer} 
by the KaoS/CHIC Collaboration. The quoted K$^+$ yields  at 
different angles are inconsistent with the assumption of an 
isotropic emission in the fireball system and were explained 
by an anisotropy of the polar angle distribution 
caused by K$^+$ rescattering \cite{elmer}.

\begin{figure}
%\vspace{0.cm}
\begin{minipage}[t]{10cm}
\hspace{ 0.cm}\mbox{\epsfig{file=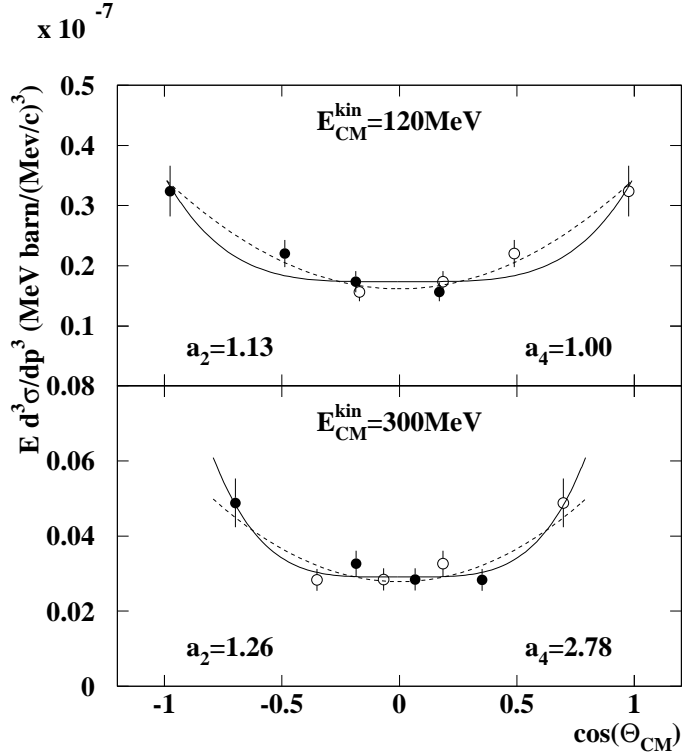,width=10cm}}
\end{minipage}
\hspace{-0.2cm}
\begin{minipage}[t]{5.6cm}
\vspace{-7cm}
\caption{Invariant K$^+$ production cross section as a function of the 
c.m. polar emission angle in Au+Au collisions at 1 AGeV \protect\cite{mang}
for a kaon kinetic energy of 120 MeV (top) and 300 MeV (bottom).
The full symbols represent measured data, the open ones are reflected
around 0. The parameterization 
$\protect{\sigma_{inv} \propto (1 + a_ncos^n\Theta_{cm})}$ 
is fitted to the data with n=2 (dashed lines) and n=4 (solid lines).
}
\label{kp_angle}
\end{minipage}
\end{figure}

\vspace{.4cm} 
Recently K$^+$ production has been studied in Au+Au collisions
at 1 AGeV as a function of impact parameter with a 
more complete phase space coverage. 
\cite{mang}. Fig.~\ref{kp_angle} presents the K$^+$ invariant cross section  
at E$_{cm}^{kin}$=120 MeV and 300 MeV as
a function of cos$\Theta_{cm}$ for inclusive collisions. 
The data are parameterized by
$\sigma_{inv} \propto (1 + a_ncos^n\Theta_{cm})$ with n=2 (dashed line)
and n=4 (solid line). The fitted parameters $a_2$ and $a_4$ are given
in the figure. 
When taking into account the  polar anisotropy
(which is assumed to be independent of E$_{cm}^{kin}$) the total  K$^+$ production
cross section in Au+Au collisions at 1 AGeV is found to be 26$\pm$5 mb
(preliminary value).
This value agrees (within the systematic errors) with the one obtained  
from the K$^+$ data measured around 
midrapidity and extrapolated  to 4$\pi$ assuming isotropic emission:
$\sigma_{K^+}$ = 24$\pm$5 mb.

Transport model calculations are in quantitive agreement 
with the measured K$^+$ polar angular distribution \cite{brat,wang}. 
These calculations claim that
the K$^+$ forward-backward anisotropy is caused by (i) kaon-nucleon 
rescattering and (ii) by 
the alignment of pion-nucleon collisions along the beam axis 
which are the dominant source of kaons 
(via secondary interactions $\pi$N$\rightarrow$KYN).  

\vspace{.4cm}
Another important feature of kaon emission in nucleus-nucleus collisions 
is the azimuthal angular distribution which is    affected by
kaon propagation  and hence by
the kaon-nucleon interaction in the nuclear medium\cite{fang_ko}.

In section 3.3 we have discussed the azimuthally anisotropic emission
of pions. Microscopic models explain this effect by rescattering
and/or reabsorption in the spectator matter. In contrast,
the enhanced emission of nucleons and light fragments in-plane 
(''sideward flow'') and out-of plane (''squeeze-out'')  
was explained by the hydrodynamical expansion of nuclear
matter \cite{stoecker}.
                       
\begin{figure}
\vspace{0.cm}
\begin{minipage}[t]{9.5cm}
\mbox{\epsfig{file=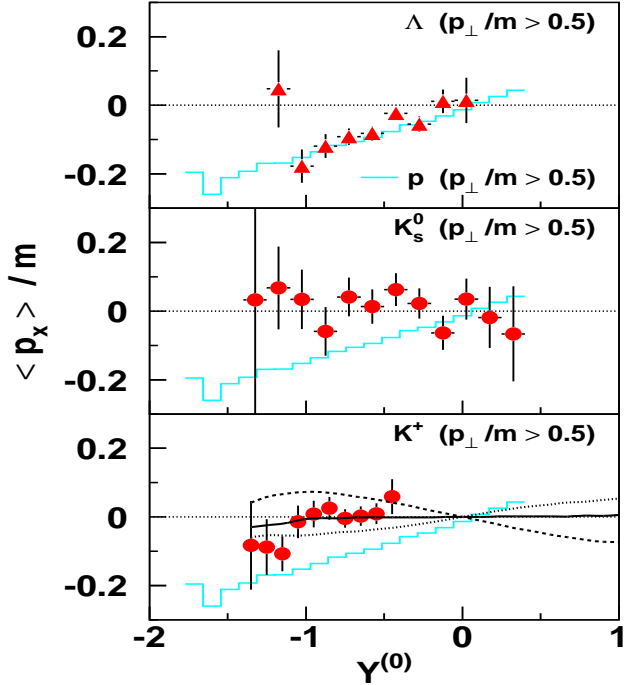,width=8.cm,height=9cm}}
\end{minipage}
%\hspace{2cm}
\begin{minipage}[t]{5.5cm}
\vspace{-8cm}
\caption{ Sideward flow of strange particles \protect\cite{ritman}. Average transverse momenta per mass projected onto the event plane as a 
function of the normalized rapidity for $\Lambda$ (top), K$^0$ (middle) and 
K$^+$ (bottom). Protons are shown for comparison (diamonds  and  histogramms).
The lines in the lower panel correspond to RBUU calculations
\protect\cite{li_flow}.   
}
\label{kp_flow_ni_fopi2}
\end{minipage}
\end{figure}

The in-plane flow  of $\Lambda$, K$^0_s$ and K$^+$  has been measured
for Ni+Ni collsions at 1.93 AGeV by the FOPI Collaboration \cite{ritman}.   
The results are shown in Fig.~\ref{kp_flow_ni_fopi2}. 
No  evidence for sidewards flow of kaons is seen whereas 
both the $\Lambda$'s and protons exhibit a clear flow signal.
The absence of kaon flow was interpreted as a signature for a weakly
repulsive in-medium kaon-nucleon potential \cite{li_flow}.
A more detailed discussion of in-medium effects on strange mesons 
is presented in the next subsection.

\begin{figure}[hpt]
%\vspace{0.cm}
\begin{minipage}[t]{9.5cm}
\mbox{\epsfig{file=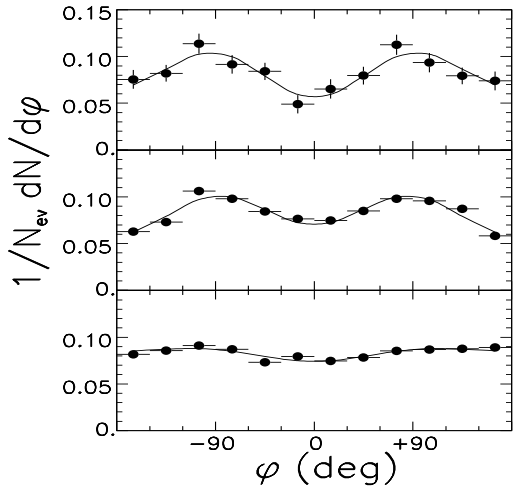,width=10.cm,height=8.cm}}
\end{minipage}
\hspace{1cm}
\begin{minipage}[t]{5cm}
\vspace{-7cm}
\caption{K$^+$ azimuthal angular distribution for peripheral 
(b$\geq$10 fm),
semi-central (b = 5 - 10 fm) and central (b$\leq$5 fm)
Au+Au collisions at 1 AGeV (from top to bottom).
The data cover normalized rapidities in the interval
0.2$\leq y/y_{proj}<$0.8 and transverse momenta in the
interval 0.2 GeV/c$\leq p_t<$0.8 GeV/c. The lines represent fits to the data
\protect\cite{shin} (see text)}
\label{kp_azimut_b}
\end{minipage}
\end{figure}

An anisotropic azimuthal emission pattern of 
K$^+$ mesons has been found recently by the KaoS Collaboration \cite{shin}.
The kaons were measured in Au+Au collisions at 1 AGeV
within the range of transverse momenta of $0.2 < p_T< 0.8$ GeV/c and 
normalized rapidities of $0.2 < y/y_{proj}< 0.8$. 
Fig.~\ref{kp_azimut_b} 
shows the azimuthal distribution dN/d$\phi$ as a function of the 
azimuthal emission angle $\phi$ for peripheral (top), semicentral (middle)
and central collisions (bottom). The data are corrected for the dispersion 
of the reaction plane measurement \cite{shin}.
The distributions peak at $\phi=\pm90^0$ which corresponds to the directions
perpendicular to the 
reaction plane. The anisotropy ratios (see equation \ref{eq:r_n}) are 
R$_N^{corr}$ = 1.68$\pm$0.18, 1.58$\pm$0.06 and 1.09$\pm$0.03 for peripheral,
semi-central and central collisions, respectively.

If the anisotropy is  caused by interaction of the kaons with the
spectator matter the emission pattern provides a time scale for kaon emission:
the K$^+$ mesons have to freeze-out within 15-20 fm/c after
the first nuclear contact. This duration corresponds to 
the time which a spectator fragment
needs to pass by the reaction zone at a beam energy of 1 AGeV. 
Transport calculations confirm this time span which coincides
with the lifetime of the dense nuclear fireball:  
In Au+Au collisions at 1 AGeV the K$^+$ mesons are produced within 
the time interval of 5 fm/c$<$t$<$17 fm/c. During this period  
the baryonic  density of the reaction zone is 
expected to be more than twice as high as the 
nuclear ground state density \cite{fang_ko}. 

\section{Probing the nuclear equation of state}
%\vspace{0.3cm}
\begin{figure}
%\vspace{0.cm}
\centerline{
\hspace{1.cm}\mbox{\epsfig{file=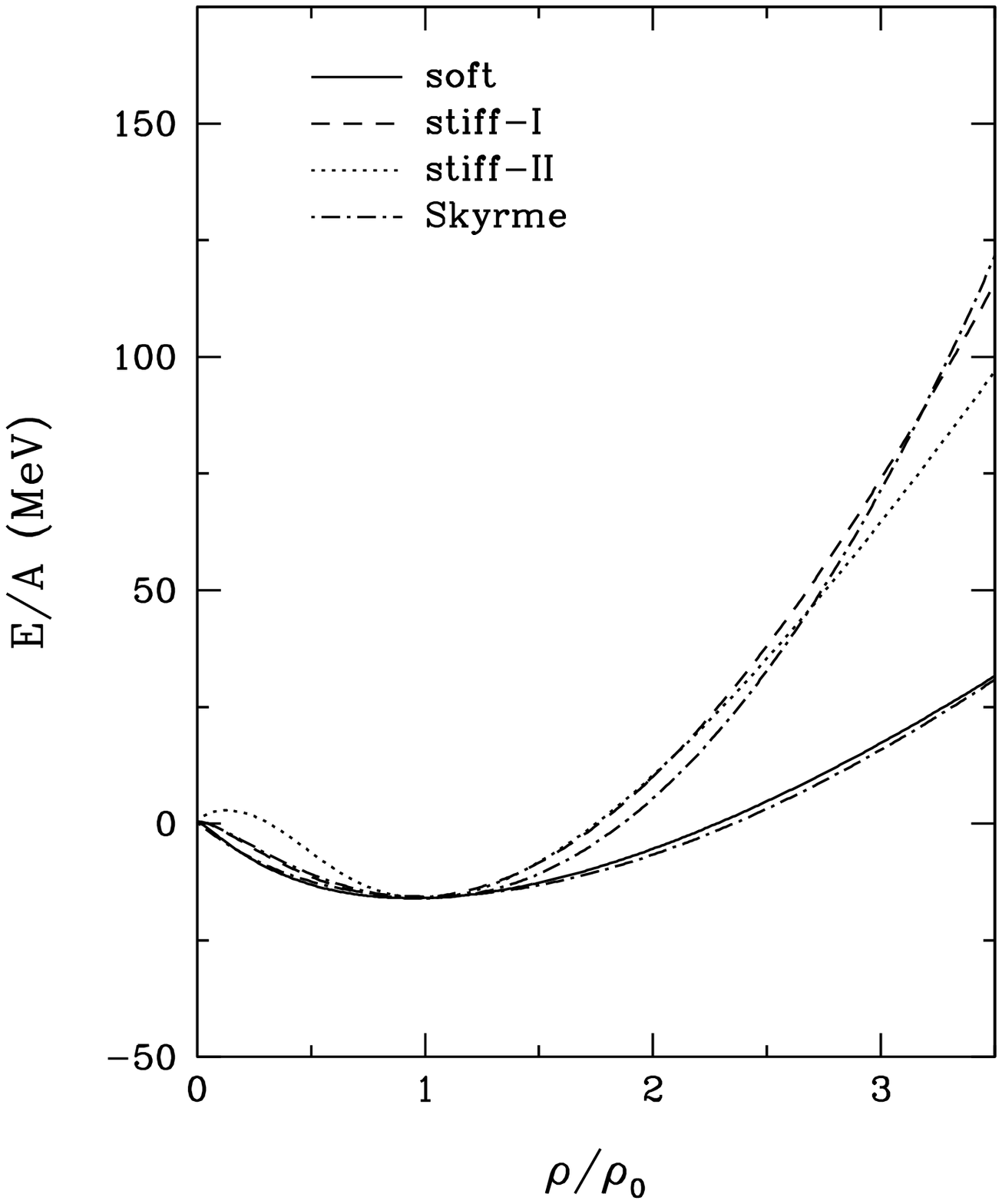,width=9.cm,height=9cm}}
\hspace{-1.cm}\mbox{\epsfig{file=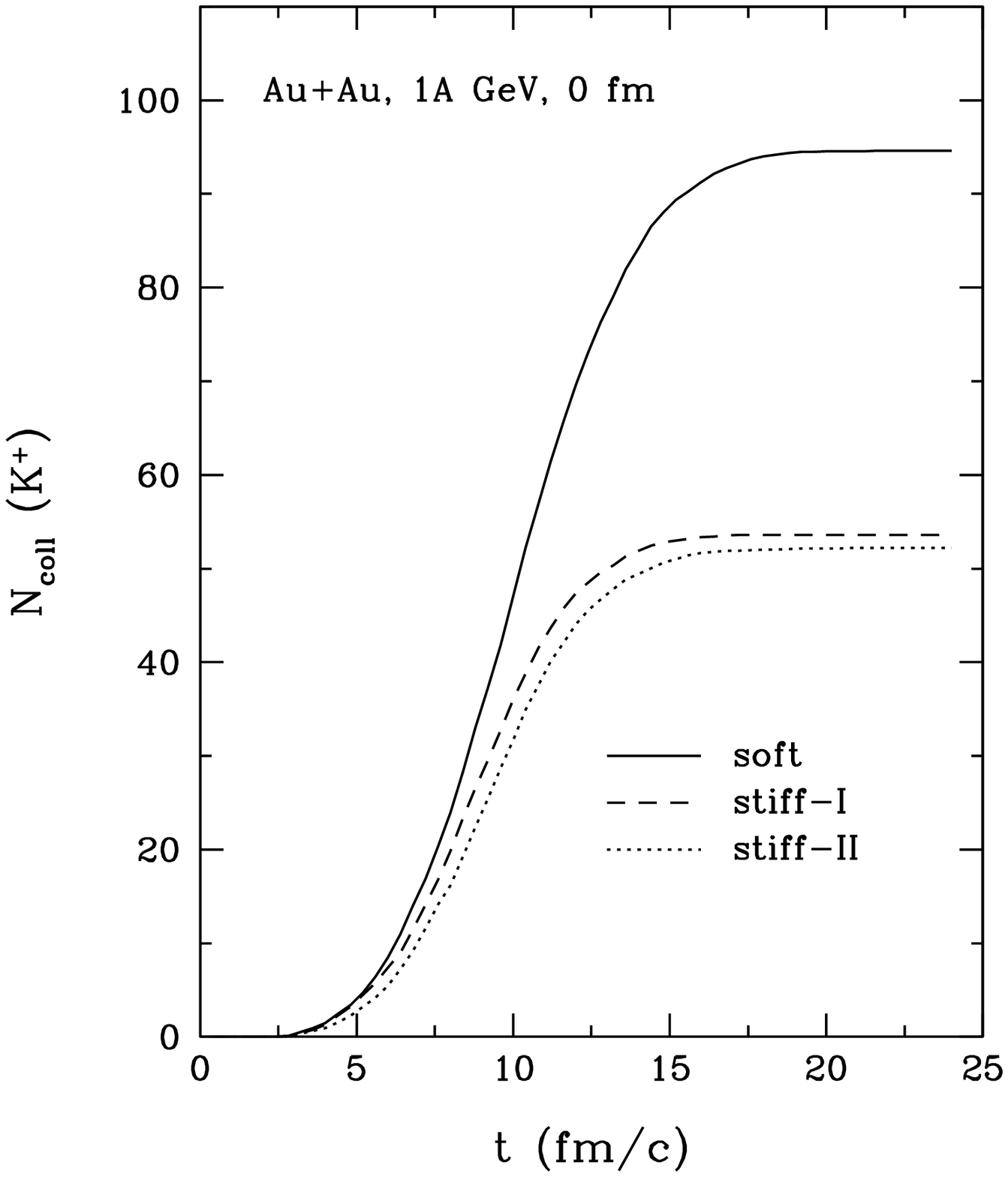,width=9.cm,height=9cm}}
}
%\vspace{-0.4cm}
\caption{Left: The nuclear equation of state in the non-linear
$\sigma-\omega$ model and in the Skyrme parameterization 
(taken from \protect\cite{li_ko}). 
Right: The total number of baryon-baryon collisions that have energies
above the kaon production threshold as a function of time for different
equations of state \protect\cite{li_ko} 
}
\label{eos_rbuu}
\end{figure}

Subthreshold K$^+$ production in relativistic nucleus-nucleus collisions
is one of the most promissing
probes to study the properties of nuclear matter
at densities far from the ground state density.
The sensitivity of kaon production on matter properties is based on
(i) the collective production processes via multiple interactions which are 
strongly enhanced in the dense phase of the collision and
(ii) the long mean free path of K$^+$ mesons.
Early transport calculations (BUU) predicted for central heavy ion collisions
at 0.7 AGeV that the K$^+$ yield obtained with a soft equation-of-state (EOS)
is about 2-3 times higher than for a stiff EOS \cite{aich_ko}.
This sensitivity to the EOS 
is expected to diminish for smaller collision systems or
beam energies above the kaon production threshold. 
These features have been reproduced recently by
relativistic transport models (RBUU) \cite{li_ko}.

\vspace{.4cm}
Fig.~\ref{eos_rbuu} (left) displays the compressional energy per nucleon  
for different parameterizations as used in the RBUU code. 
In the case of a soft EOS (solid line), much less kinetic energy is 
converted into compressional energy as compared to a stiff EOS (dashed line).  
Consequently, more thermal energy is available for particle production
for a soft than for a stiff EOS. 
This is demonstrated  in Fig.~\ref{eos_rbuu} (right) 
for central Au+Au collisions at 1 AGeV. 
RBUU calculations find 95  baryon-baryon collisions
with energies above the kaon production threshold when
assuming a soft EOS but only 54 of those collisions for a stiff EOS.
Forthermore, the average density at which kaons
are produced is lower for a stiff EOS (about 2.1 $\rho_o$)
than for a soft EOS (about 2.5 $\rho_o$) \cite{li_ko}.

\vspace{.4cm}
All transport models find consistently that the dominant channels contributing to
K$^+$ production are $\pi$N$\rightarrow$KY and $\Delta$N$\rightarrow$KYN
with Y=$\Lambda,\Sigma$ \cite{fuchs,cassing,brat}. 
The rate of these secondary collisions
depends not only on  the nuclear density and on
the ''thermal'' energy of the baryons, but also on  the abundance of
resonances and pions and on their in-medium cross sections.
An important input to these 
simulations are  parameterizations of
the elementary kaon production cross sections for the 
NN and $\Delta$N  channels. 
So far the  models used parameterizations for the cross section 
NN$\rightarrow$K$^+$YN 
which differ dramatically near threshold
\cite{ran_ko,schuer,sibirtsev,mueller}. 
Recent data on inclusive K$^+$ production
cross sections from pp collisions 
measured at COSY resolved the ambiguities \cite{cosy11}. 
New calculations which take into account a proper 
lifetime of the $\Delta$-resonance find that the $\pi$N channel 
is even more important than the $\Delta$N  channel \cite{fuchs,brat}.  

\vspace{.4cm}                                                                 
The transport model calculations reproduce 
reasonably well the gross properties of the 
experimental K$^+$ data \cite{fang_ko,cassing,hartnack}.
They are able to describe
double-differential K$^+$ production cross sections
as a function of bombarding energy and mass of the colliding nuclei
over several orders of magnitude
(within a factor of about 2).
Unfortunately, the  theoretical uncertainties on the absolute K$^+$
yields (resulting from the in-medium production cross sections) 
are still of the same size as the effect induced
by the different equations of state.
Nevertheless, some of the uncertainties can be  reduced when comparing
the results obtained for light and heavy collision systems or
for different bombarding energies \cite{maruyama,fang_ko,hartnack}.

\begin{figure}
\vspace{-0.5cm}
\hspace{ 0.cm}\mbox{\epsfig{file=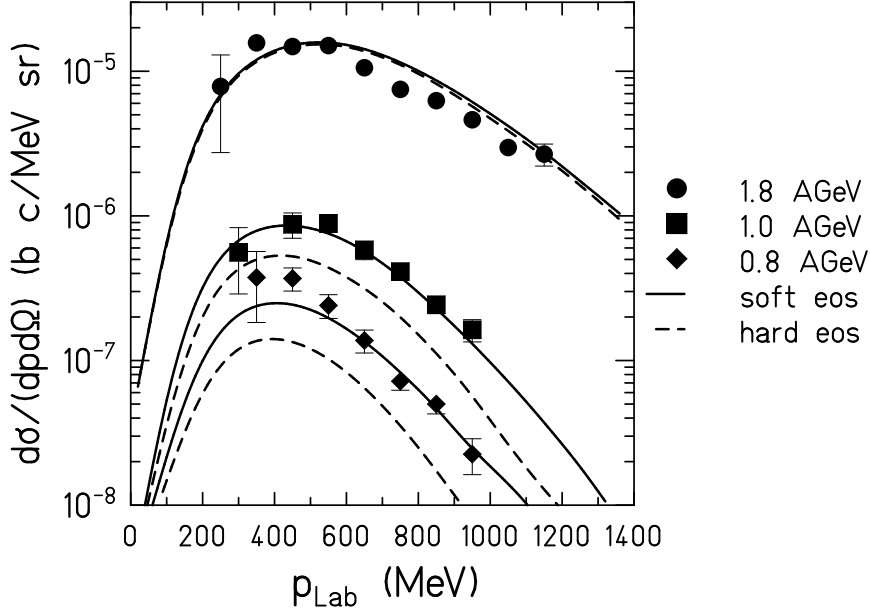,width=14.cm}}
\vspace{0.cm}
\caption{Double-differential K$^+$ production cross section measured in Ni+Ni 
collisions (0.8, 1.0, 1.8 AGeV) at $\Theta_{lab}$=44$^0$ as a
function of laboratory momentum \protect\cite{barth}.
The lines represent results of QMD calculations for a soft (solid line) and 
a hard (dashed line) equation of state \protect\cite{hartnack}.  }
\label{kp_ni_kaos}
\end{figure}

\begin{figure}
%\vspace{-1.2cm}
\begin{minipage}[t]{10.5cm}
\centerline{\mbox{\epsfig{file=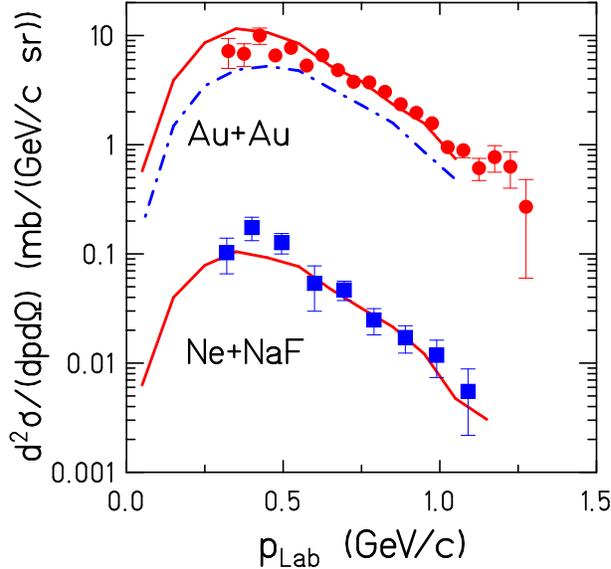,width=11.cm}}}
\end{minipage}
%\hspace*{0.2cm}
\begin{minipage}[t]{5.3cm}
\vspace{-7.cm}
\caption{Double-differential K$^+$ production cross section 
measured at 1.0 AGeV and  $\Theta_{lab}$=44$^0$ as a
function of laboratory momentum (circles: Au+Au \protect\cite{mang}, 
squares: Ne+NaF \protect\cite{ahner}).
The lines represent results of RBUU calculations for a soft (solid line) and
a hard (dashed-dotted line) equation of state \protect\cite{fang_ko}.}
\label{kp_ne_au}
\end{minipage}
\end{figure}

\vspace{.4cm}
Such comparisons are shown in  Fig.~\ref{kp_ni_kaos} and Fig.~\ref{kp_ne_au}
which present calculated and measured double-differential cross sections for
K$^+$ production measured at $\Theta_{lab}$=44$^o$. 
Fig.~\ref{kp_ni_kaos}
shows results for Ni+Ni collisions at 0.8, 1.0 and 1.8 AGeV
\cite{barth}. The data are compared with
results from a QMD calculation \cite{hartnack} for a soft (solid lines)
and a stiff equation-of-state (dashed lines). At a bombarding energy of
1.8 AGeV no effect of the EOS can be seen whereas at subthreshold
beam energies the calculations differ by a factor of two depending
on the EOS. The calculations use an old parameterization of the
elementary K$^+$ production cross section according to Ref.\cite{schuer}
and neglect momentum-dependent interactions. Taking those into account
would reduce the K$^+$ yield by a factor of about two. 
Furthermore, this calculation omits the reaction channel 
$\pi$N$\rightarrow$KY as well as
kaon rescattering and neglects the  kaon-nucleon potential.  
These processes, however, will not depend  strongly on the beam energy.
The absolute agreement of 
model predictions with the  data as shown in Fig.~\ref{kp_ni_kaos}  
is probably accidental
and should not be considered as a support for the underlying model 
assumptions. Nevertheless the calculations demonstrate that
when  "normalizing" the theoretical K$^+$ spectrum to the data taken 
at 1.8 AGeV,  the K$^+$ production cross sections measured  
at subthreshold beam energies support the assumption of a soft
equation-of-state.

%\vspace{.4cm}
Fig.~\ref{kp_ne_au} shows experimental results for 
two different system sizes: Ne+NaF  and  Au+Au  collisions
at 1.0 AGeV  \cite{ahner,mang}. The data are compared to relativistic
transport calculations (RBUU) with a soft (solid lines) and stiff EOS
(dashed line) \cite{fang_ko,li_ko}. 
For the light system Ne+NaF, the results for a stiff and a soft EOS 
coincide.
The model takes into account the momentum dependence of the 
NN interaction, kaon final state interactions and an in-medium
kaon-nucleon potential. However,  
the calculations use  a parameterization of the 
NN$\rightarrow$KYN process \cite{ran_ko} which was found to give a too large 
cross section
near threshold. Moreover, the process  $\pi$N$\rightarrow$KY is neglected.
Therefore, the absolute agreement of model results and data as shown in
Fig.~\ref{kp_ne_au} should not be overinterpreted.  
On the other hand, 
the theoretical uncertainties affect both the light and the heavy collision 
system in a similar way. Therefore, the relative agreement of the   
model calculation with both the Ne+NaF and the Au+Au data 
again favors  a soft EOS.

\vspace{.4cm}
As a summary of our discussion of this important aspect 
of nuclear physics we stress that 
according to transport models the  
K$^+$ abundances measured in central Au+Au collisions at energies
below the p+p$\rightarrow$K$^+$+X threshold constrains the compressibility of 
nuclear matter at high density to values below 300 MeV. 
This result is in agreement with 
calculations which explain simultaneously data on kaon production
and  on the directed flow of baryons in the same 
reaction \cite{cassing}.
However, estimates of the nuclear compressibility based on the suppressed 
pion production ask for a stiff eos \cite{stock}. We interprete this 
discrepancy as an incomplete understanding of pion production, propagation
ans absorption in a dense nuclear medium (see also section 3.2).

\section{Kaons and antikaons in dense nuclear matter}

The formation of a nuclear fireball in central nucleus-nucleus
collisions provides a possibility to study  
the properties of hadrons under extreme conditions. Inside the 
hot and dense nuclear medium, 
chiral symmetry (which is spontaneously broken in the vacuum) 
is expected to be partially restored
and the properties of hadrons may change considerably
\cite{weise}. A possible consequence of this effect is the 
modification of vector meson masses as studied in heavy ion collisions 
by dilepton experiments  \cite{ceres,dls}.

\begin{figure}
%\vspace{-0.6cm}
\begin{minipage}[t]{10.5cm}
\mbox{\epsfig{file=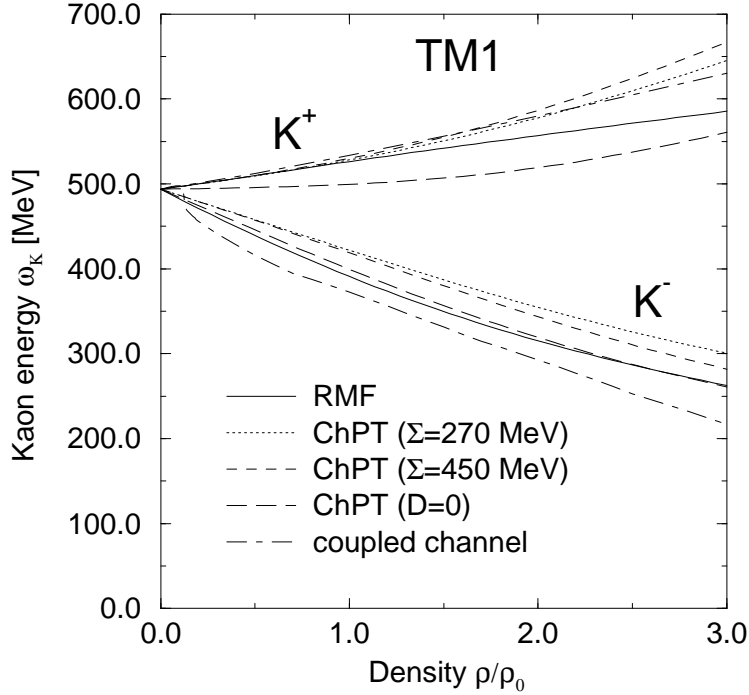,width=10.cm}}
\end{minipage}
%\hspace{0.6cm}
\begin{minipage}[t]{5cm}
\vspace{-4.cm}
\caption{The energy of kaons and antikaons in nuclear matter as a function
of density for a soft equation of state (taken from \protect\cite{schaffner2}.}
\label{ka_med_th}
\end{minipage}
\end{figure}

\vspace{.5cm} 
The K-meson is another promising candidate for the experimental 
study of in-medium modifications.
The properties of kaons and antikaons in dense nuclear matter
have been investigated using chiral perturbation theory
\cite{kaplan,brown}, chiral dynamics \cite{waas},  
a relativistic mean field model \cite{schaffner1} and a multi-channel 
approach \cite{lutz}. 
The calculations find a KN
potential which can be expressed by
an attractive scalar part and a repulsive vector part. 
The attractive kaon-nucleon s-wave interaction is related to explicit
chiral symmetry breaking by the large strange quark mass.
The vector potential changes its sign for antikaons and 
is attractive for these particles.
Therefore, the sum of the scalar and vector potential
- which is slightly repulsive for kaons - becomes strongly 
attractive for antikaons. These results are 
consistent with an analysis of kaon-nucleon scattering data
and kaonic atoms \cite{friedman1,friedman2}.
Fig.~\ref{ka_med_th} 
shows the in-medium kaon and antikaon effective mass as calculated by
different theoretical models as a function of nuclear density
(taken from \cite{schaffner2}).
A common feature of the different model 
calculations is that the kaon mass weakly increases
whereas the antikaon mass considerably decreases 
with increasing nuclear density. The latter effect may lead to
K$^-$ condensation in neutron stars \cite{brown1,knorren,fujii}
and possibly even to the formation of low mass black holes in the galaxy
\cite{brobet,li_lee_br,li_lee_br2}.

\vspace{.5cm}
The existence of a KN potential affects both the production and the propagation
of kaons and antikaons in nuclear matter. 
A weakly repulsive K$^+$N interaction should repel the K$^+$ mesons
from the nucleons. This effect is expected to modify the K$^+$ azimuthal 
emission pattern in such a way that it differs from 
the proton azimuthal distribution \cite{li_flow}.
First experimental results on the K$^+$ and $\Lambda$ directed flow were 
reported by the FOPI Collaboration \cite{ritman}. 
Fig.~\ref{kp_flow_ni_fopi2} 
shows the average transverse velocity projected onto
the reaction plane  $<p_x>$/m as a function of the normalized c.m. rapidity for
$\Lambda$ (top), K$^o_s$ (middle) and K$^+$ (bottom) in comparison
to protons (histogram). The $\Lambda$'s exhibit the same flow behaviour as
the protons whereas the kaons do not show any significant flow signal.
The K$^+$ data are compared to results of RBUU calculations
assuming different  in-medium kaon potentials:
no potential (dotted line), vector potential only (dashed line) and
scalar + vector potential (solid line) \cite{li_flow}.
The data seem to favor the assumption of a weakly repulsive
in-medium kaon potential. 

\begin{figure}
\vspace{0.cm}
\begin{minipage}[t]{10cm}
\vspace{-1.6cm}
\mbox{\epsfig{file=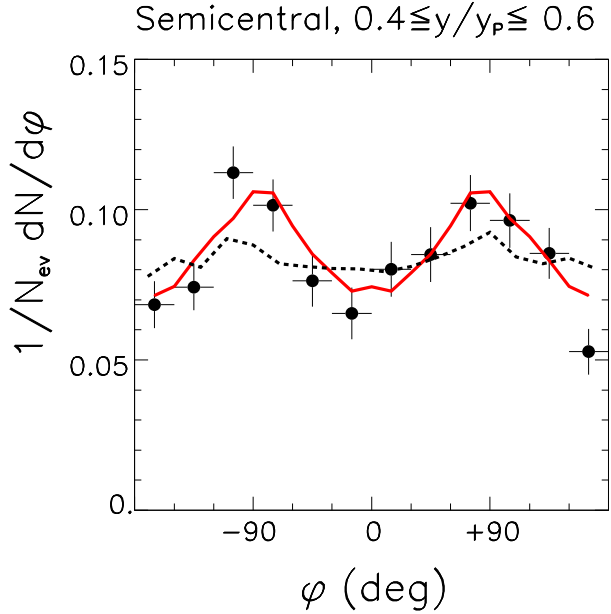,width=12.cm}}
\end{minipage}
\begin{minipage}[t]{5.5cm}
\vspace{1.cm}
\caption{K$^+$ azimuthal distribution for semi-central Au+Au collisions
at 1 AGeV. The kaons have transverse momenta
of 0.2 GeV/c$<p_T<$0.8 GeV/c and are measured in the range of
normalized rapidities of $0.4< y/y_{proj}< 0.6$ \protect\cite{shin}. 
The lines represent results of RBUUcalculations with K$^+$ rescattering 
(dotted line) and with K$^+$ rescattering and 
an in-medium KN potential (solid line) \protect\cite{li2}.}
\label{kp_azimut_semi}
\end{minipage}
\end{figure}

\vspace{.4cm}
According to transport calculations, the existence of a strong in-medium
kaon potential affects also the K$^+$ azimuthal angular distribution
at midrapidity. The experimental results are 
shown in Fig.~\ref{kp_azimut_semi} for semi-central Au+Au collisions
at 1 AGeV \cite{shin}. 
The data have been taken in a small interval of normalized rapidity 
0.4$< y/y_{proj}<$0.6. The solid line corresponds to
a RBUU calculation which
includes a strong in-medium KN potential whereas the dashed line is the
result without an in-medium KN potential \cite{li2}.
The strong KN potential is clearly needed
to reproduce the data.       

Due to the strongly attractive mean-field potential of antikaons, 
the azimuthal emission pattern of K$^-$ mesons is expected to be  
modified  dramatically and therefore should provide an unique signature of  
the in-medium effect
\cite{li_ko2}. The strong absorption of K$^-$ mesons - which would lead 
to a pronounced K$^-$ antiflow pattern near target rapidity - is predicted 
to be cancelled
by the strongly attractive in-medium antikaon-nucleon interaction. The resulting
K$^-$ azimuthal distribution at backward rapidities is expected to be
rather flat in contrast to the one of protons - if the KN potential exists.         
Up to now no data on the K$^-$ azimuthal emission pattern are available. 
Experiments on this issue are scheduled at SIS.  

\vspace{.4cm}
The modifications of the masses of kaons and antikaons in dense nuclear
matter will affect their yields differently. Therefore, the cross section 
ratio of kaons and antikaons is a sensitive probe of 
in-medium KN potentials \cite{li_ko_fang,cassing,brat}. 
As shown in Fig.~\ref{ka_med_th}, the antikaon effective
mass is predicted to drop significantly in dense nuclear matter while 
the K$^+$ effective mass increases weakly.  
This in-medium mass modification lowers the 
threshold for the antikaon production process
NN$\rightarrow$K$^+$K$^-$NN but increases it for kaon production via
NN$\rightarrow$K$^+$YN.
Consequently the K$^+$ production is expected to be slightly suppressed
whereas the K$^-$ production
is strongly enhanced as compared to free NN collisions. 
The latter effect should be very pronounced at subthreshold bombarding energies,
where the K$^-$ excitation function is very steep and
thus acts as an amplifier:
even a small mass reduction might result in a strong K$^-$ enhancement.
The in-medium reduction of the threshold for the production of a 
K$^-+$K$^+$ pair is due to the 
attractive KN potential only, as the vector parts of K$^+$ and K$^-$ 
cancel.

\begin{figure}
\vspace{0.cm}
\begin{minipage}[t]{9cm}
\vspace{-2cm}
\hspace{ 0.cm}\mbox{\epsfig{file=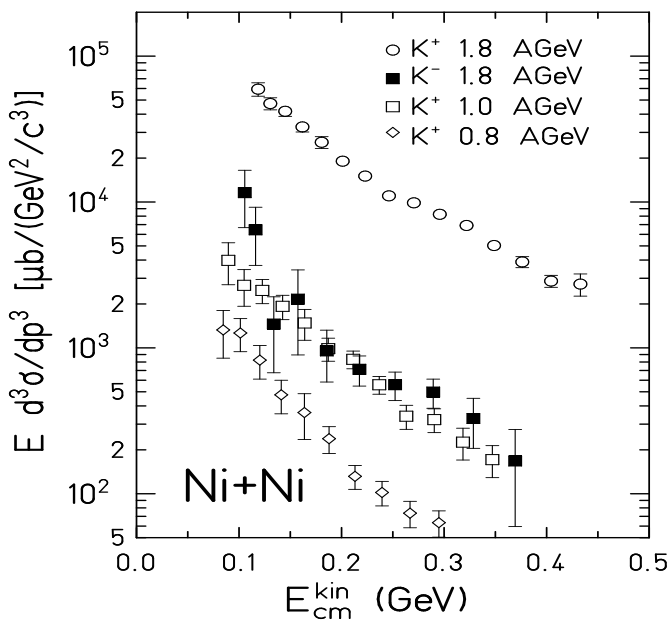,width=12.cm,height=10cm}}
\end{minipage}
\hspace{1cm}
\begin{minipage}[t]{5.3cm}
\hspace{-1.cm}
\caption{Inclusive invariant K-meson production cross section measured in 
Ni+Ni collisions at  $\Theta_{lab}$=44$^0$ \protect\cite{barth}. 
Open symbols: K$^+$ (0.8, 1.0, 1.8 AGeV), full symbols: K$^-$ (1.8 AGeV).
 }
\label{ka_ni_inv}
\end{minipage}
\end{figure}

\vspace{.5cm}                                                           
The available
data on subthreshold K$^-$ production in nucleus-nucleus
collisions have been presented in section 4.1.
The large K$^-$ yield measured in Ni+Ni
collisions at 1.85 AGeV by the FRS group \cite{li_ko_fang,kienle}
was reproduced by transport calculations (RBUU)  if  
an in-medium mass reduction of the K$^-$ meson is assumed \cite{li_ko_fang}.
The KaoS Collaboration found evidence for an enhanced K$^-$ yield 
in Ni+Ni collisions from a comparison to the K$^+$ yield at equivalent
beam energies  
\cite{barth}. 
Fig.~\ref{ka_ni_inv} 
shows the invariant K$^+$ production cross section (open symbols) as a function
of the K$^+$ kinetic c.m. energy measured in Ni+Ni 
collisions at 0.8, 1.0 and 1.8 AGeV
around midrapidity. The K$^-$ data (full symbols)  measured at
1.8 AGeV agree with the K$^-$ invariant cross section as measured in
Ni+Ni collisions at 1.85 AGeV by the FRS at
$\Theta_{lab}$=0$^o$ \cite{schroeter,gillitzer,kienle}.

The question is, whether there is experimental 
evidence  for an enhancement of the K$^-$ yield measured at 1.8 AGeV
and how to quantify it. 
The KaoS Collaboration used 
the K$^+$ cross section mesured at 1.0 AGeV as a reference for 
K$^-$ production at 1.8 AGeV. 
These two beam energies are ''equivalent'' 
in the sense that they allow to study
K$^+$ and K$^-$ production at the same Q-value \cite{barth}:

\vspace{.2cm}
Q(NN$\rightarrow$K$^+\Lambda$N) =$\sqrt s$ -$\sqrt s_{thres}$ =2.32
 GeV - 2.55 GeV = -0.23 GeV 

\vspace{.2cm}
Q(NN$\rightarrow$K$^+$K$^-$NN) =$\sqrt s$ -$\sqrt s_{thres}$ =2.63
 GeV - 2.86 GeV = -0.23 GeV 

\begin{figure}
\begin{minipage}[t]{10cm}
\vspace{-1.cm}
\mbox{\epsfig{file=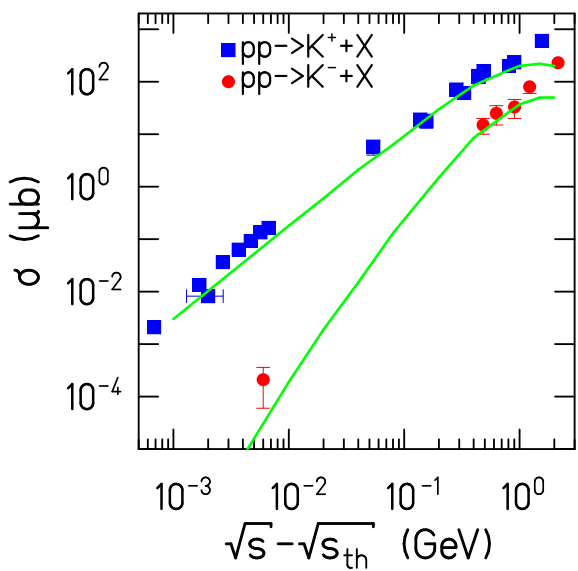,width=11.cm}}
\end{minipage}
\begin{minipage}[t]{5.5cm}
\vspace{1cm}
\caption{Kaon and antikaon production cross section measured in p+p collisions 
as a function of the energy above threshold (Q-value). The data are taken 
from \protect\cite{cosy11,hera,hogan}. The lines represent parameterizations
\protect\cite{sibirtsev,si_ca_ko}.   }
\label{ka_pp}
\end{minipage}
\end{figure}

\noindent                                                                        
The choice of the equivalent energies for the
comparison of K$^+$ and K$^-$  production  is meant to be a crude 
correction for the differences in accessible phase space. 
According to Fig.~\ref{ka_ni_inv} the
K$^+$ yield at 1 GeV/nucleon agrees roughly with the
K$^-$ yield at 1.8 GeV/nucleon.
Not only the cross section but also the dependence on A$_{part}$
is the same for K$^+$ and K$^-$ observed at equivalent energies 
(see Fig.~\ref{ka_mult_ni}).
This result for nucleus-nucleus collisions is quite
different from  the K$^-$/K$^+$
ratio for proton-proton collisions at equivalent beam energies. 
Fig.~\ref{ka_pp} 
shows the available data (including the most recent results from COSY)
on inclusive cross sections for K$^+$ and antikaon production
in proton-proton collisions  as a function of the energy
above threshold \cite{cosy11,hera,hogan}. The K$^+$ data exceed the antikaon
data by 1-2 orders of magnitude.   
The lines represent a recent parameterization of the elementary
kaon production cross sections \cite{sibirtsev,si_ca_ko}.

Before concluding on nontrivial in-medium effects one 
should check whether or not pion induced kaon production ($\pi$N$\rightarrow$K$^+$Y, $\pi$N$\rightarrow$K$^+$K$^-$N) causes an
enhancement of the K$^-$ yield with respect to the K$^+$ yield at
equivalent beam energies. Typical values for the cross sections
at the same energy above threshold
are $\sigma$($\pi^+$p$\rightarrow$K$^+$Y)$\approx$ 300 $\mu$b and 
$\sigma$($\pi^-$p$\rightarrow$K$^-$K$^+$n)$\approx$ 200 $\mu$b 
\cite{cugnon,si_ca_ko,efremov2,efremov,tsushima}. 
The pion momentum thresholds  are 900 MeV/c 
and 1500 MeV/c for the production of K$^+$Y  and K$^-$K$^+$ pairs,
respectively, on stationary protons. Assuming a mean momentum of 
400 MeV/c for the nucleons (in the c.m. system) the relevant pion momenta 
are 600 MeV/c and 1000 MeV/c.

\begin{figure}
%\vspace{-2.cm}
\begin{minipage}[t]{7.4cm}
\vspace{-2.cm}
\hspace{ 0.cm}\mbox{\epsfig{file=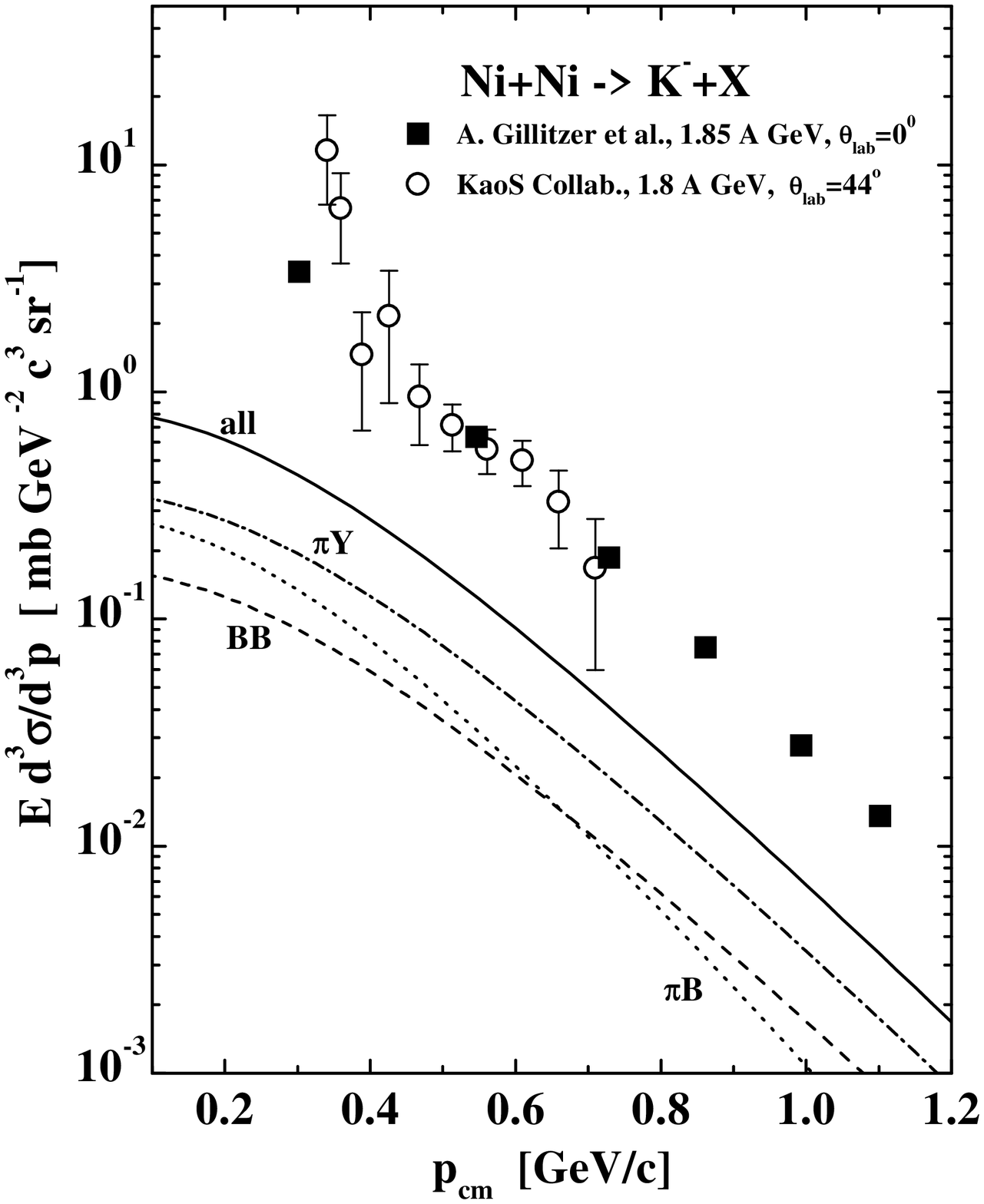,width=8.cm}}
%\vspace{-2.cm}
\caption{Inclusive invariant K$^-$ production cross section for Ni+Ni 
collisions as a function of the c.m. momentum measured at  
1.85 AGeV and $\Theta_{lab}$=0$^0$ (full squares \protect\cite{kienle})
and at   1.8 AGeV and $\Theta_{lab}$=44$^0$ 
(open circles \protect\cite{barth}). 
The lines represent 
results of a RBUU calculation \protect\cite{cassing} without including 
in-medium effects (bare K$^-$ mass).
The partial  K$^-$ production channels are
baryon-baryon (BB, dashed line),  
pion-baryon ($\pi$B, dotted line) and  pion-hyperon 
($\pi$Y, dashed-dotted line) collisions. 
The solid line is the  sum of all channels.       }
\label{km_ni_nom}
\end{minipage}
\hspace{0.7cm}
\begin{minipage}[t]{7.4cm}
\vspace{-2.cm}
\mbox{\epsfig{file=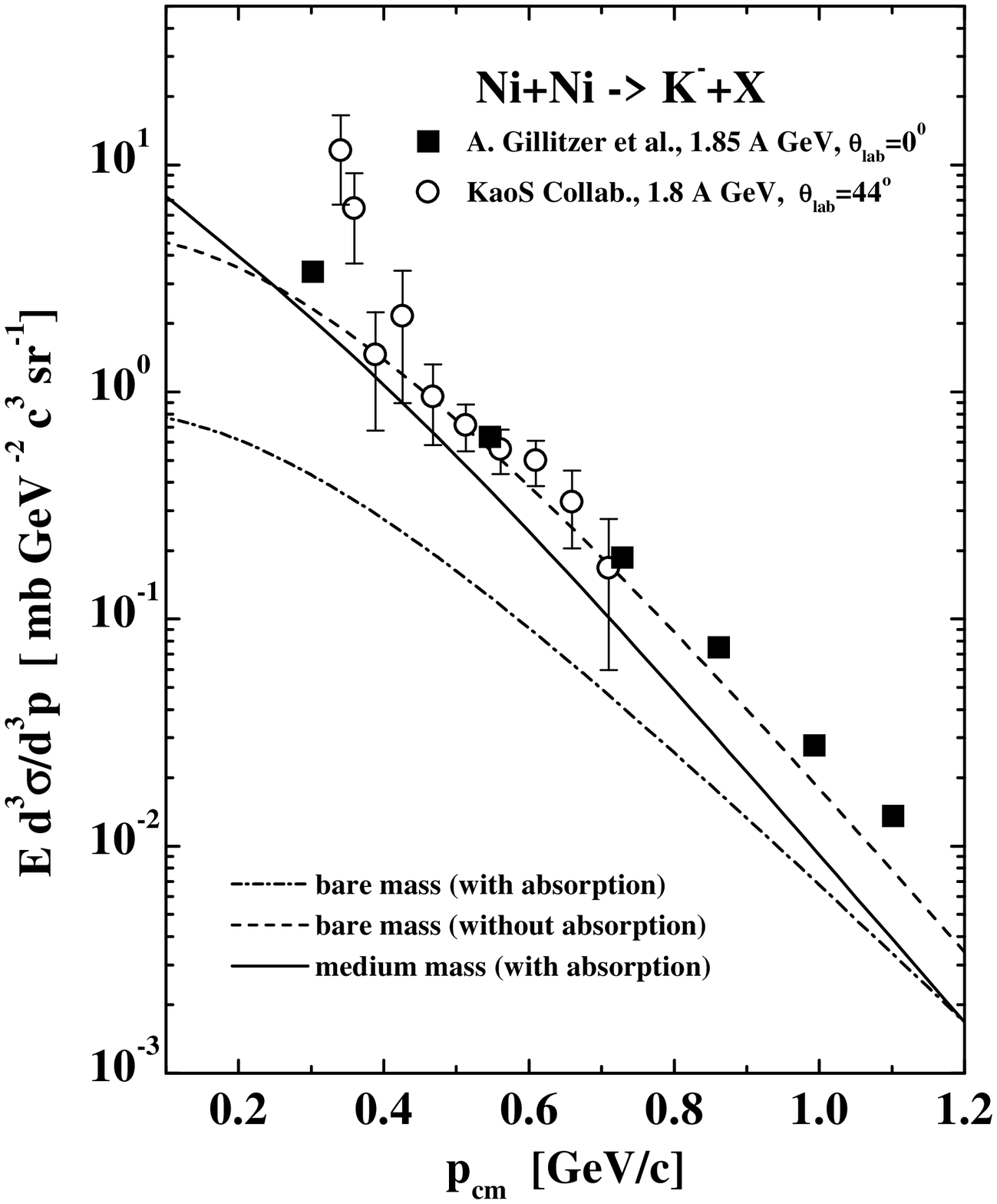,width=8.1cm}}
%\vspace{-2.cm}
\caption{Inclusive invariant K$^-$ production cross section for Ni+Ni 
collisions as a function of the c.m. momentum measured at  
1.85 AGeV and $\Theta_{lab}$=0$^0$ (full squares \protect\cite{kienle})
and at   1.8 AGeV and $\Theta_{lab}$=44$^0$ 
(open circles \protect\cite{barth}). 
The lines represent 
results of a RBUU calculation \protect\cite{cassing}:
bare K$^-$ mass with absorption (dashed-dotted), 
bare K$^-$ mass without absorption (dashed) and
in-medium K$^-$ mass with absorption (solid, $\alpha$ = 0.2).
}
\label{km_ni_med}
\end{minipage}
\vspace{0.4cm}
\end{figure}

From the pion spectra measured in  Ni+Ni collisions one can estimate
that for a beam energy of 1 AGeV the pion yield in the c.m. system 
above 600 MeV/c
is about 10 times higher than the pion yield above 1000 MeV/c 
obtained for a beam energy of 1.8 AGeV \cite{wagner1}. 
According to these numbers,
the pion induced K$^-$ yield from Ni+Ni collisions at 1.8 AGeV
is expected to be a factor of about 15 smaller than the pion induced 
K$^+$ yield from Ni+Ni collisions at 1.0 AGeV.  
Therefore, it is very unlikely that pion induced kaon production
accounts for the observed K$^-$/K$^+$ ratio in Ni+Ni collisions
at equivalent energies. The above estimation is roughly confirmed 
by RBUU calculations which predict that for Ni+Ni at 1.8 AGeV 
pion-baryon
collisions contribute only 5-10\% to the K$^-$ yield (including
K$^-$ absorption but without in-medium effects, see Fig.~\ref{km_ni_nom}) 
whereas more than 90\% of the K$^+$ yield in
Ni+Ni at 1.0 AGeV is due to $\pi$N and $\Delta$N collisions                 
\cite{cassing,brat}.

\begin{figure}
%\vspace{-1.cm}
\begin{minipage}[t]{10cm}
\hspace{ 0.cm}\mbox{\epsfig{file=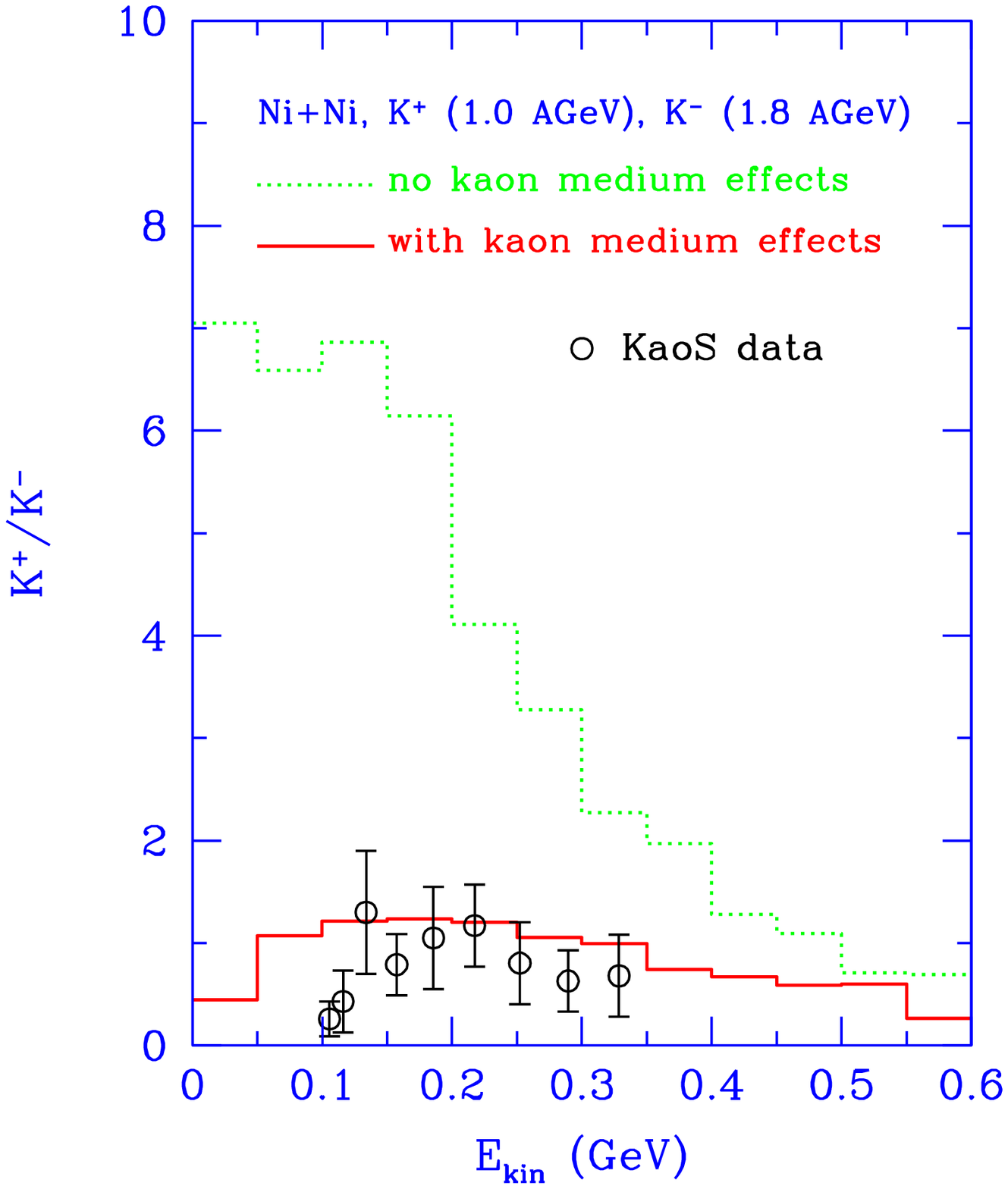,width=11.5cm,height=10cm}}
\end{minipage}
\begin{minipage}[t]{5.5cm}
\vspace{-5.cm}
\caption{K$^+$/K$^-$ ratio as a function of c.m. kinetic energy
for Ni+Ni collisions at equivalent beam energies (K$^+$ measured at 1.0 AGeV
and K$^-$ at 1.8 AGeV, $\Theta_{lab}$=44$^0$) \protect\cite{barth}. 
The lines show results of a RBUU calculation without (dotted) and with
medium effects (solid) \protect\cite{li_lee_br,li_lee_br2}.     
}
\label{ka_rat_rbuu}
\end{minipage}
\end{figure}

From the above arguments we conclude, that 
the K$^-$/K$^+$ ratio found in Ni+Ni collisions 
at equivalent energies cannot be understood in terms of a superposition
of free baryon-baryon or pion-baryon collisions producing the kaons. 
The large K$^-$ yield is even more surprising as K$^-$ mesons
are strongly absorbed in the nuclear medium
by strangeness exchange reactions like
K$^-$N$\rightarrow$Y$\pi$ with Y=$\Lambda,\Sigma$  \cite{dover}, whereas
the K$^+$ meson can hardly be absorbed due to its anti-strange quark
content and the charge
exchange reaction K$^+$n$\leftrightarrow$K$^0$p
does not lead to K$^+$ losses in isospin symmetric nuclear matter.
On the other hand, 
the strangeness exchange reaction $\pi$Y$\rightarrow$K$^-$N,
which is the ''inverse'' of K$^-$ absorption,
might be an additional source of in-medium K$^-$ production \cite{ko1}.   
Indeed, RBUU calculations claim that (for bare kaon masses, see 
Fig.~\ref{km_ni_nom}) 
the process  $\pi$Y$\rightarrow$K$^-$N  is the most important 
K$^-$ production channel in Ni+Ni collisions at 1.8 AGeV \cite{cassing}. 
This calculations is based on the parameterization of 
the elementary kaon production cross sections as shown in Fig.~\ref{ka_pp}.
Nevertheless, the sum of all K$^-$ production channels
considered by the transport calculations is about a factor of 4-5 below
the measured data points, if in-medium effects on the K$^-$ mass 
are omitted (see Fig.~\ref{km_ni_nom}).  In contrast, 
Fig.~\ref{km_ni_med} demonstrates 
that reasonable agreement with the experimental
data is achieved if an in-medium reduction of the K$^-$ mass is assumed
according to  m$^*_K(\rho)$ = m$^o_K$ (1 - $\alpha\rho/\rho_o$)
with $\alpha$=0.24 \cite{cass2}. As demonstrated in 
Fig.~\ref{ka_med_th}, various calculations
of the kaon selfenergy in nuclear matter find  similar reductions 
of the in-medium K$^-$ mass with increasing nuclear density.

\vspace{.4cm}
The kaon in-medium modification and its  effect on the kaon yields
from Ni+Ni collisions was also studied by the Stony Brook 
group using a RBUU transport code
\cite{li_lee_br,li_lee_br2}. The result of the calculation is presented in 
fig.~\ref{ka_rat_rbuu}. 
It shows the K$^+$/K$^-$ ratio for equivalent beam energies 
as a function of kaon energy without (dotted line) and with kaon medium
effects (solid line).
In order to get agreement with the KaoS
data  (symbols) the authors varied the density dependence of the kaon 
potentials. 
Based on the kaon in-medium properties as constrained by the heavy ion data 
the authors found that the critical density for 
K$^-$ condensation is about  3 $\rho_o$. This effect is predicted to 
limit the maximum  possible mass of neutron stars to about 1.5 solar masses
\cite{li_lee_br,li_lee_br2}.

\section{Strange meson production at ultrarelativistic energies}

Even at ultrarelativistic beam energies, 
the information about strange particle production is to a large extend 
contained in the $K$- and $\overline{K}$-mesons.
Only the strange quarks carried by the hyperon-antihyperon pairs are 
not accounted for.  
The number of strange quarks in those hyperons which originate 
from associated production can 
be inferred from the difference in number between $K$ and $\overline{K}$
(e.g. at high energy
$N_{hyp}=2.0\cdot(N_{K^+}-N_{K^-})$ for isospin symmetric systems
with the factor 2.0 accounting for ($N_{K^0}-N_{\overline{K^0}}$)).

\begin{figure}[hpt]
\vspace{0.cm}
%\hspace{1.5cm}
%\begin{turn}{-90}
\hspace{ 0.cm}\mbox{\epsfig{file=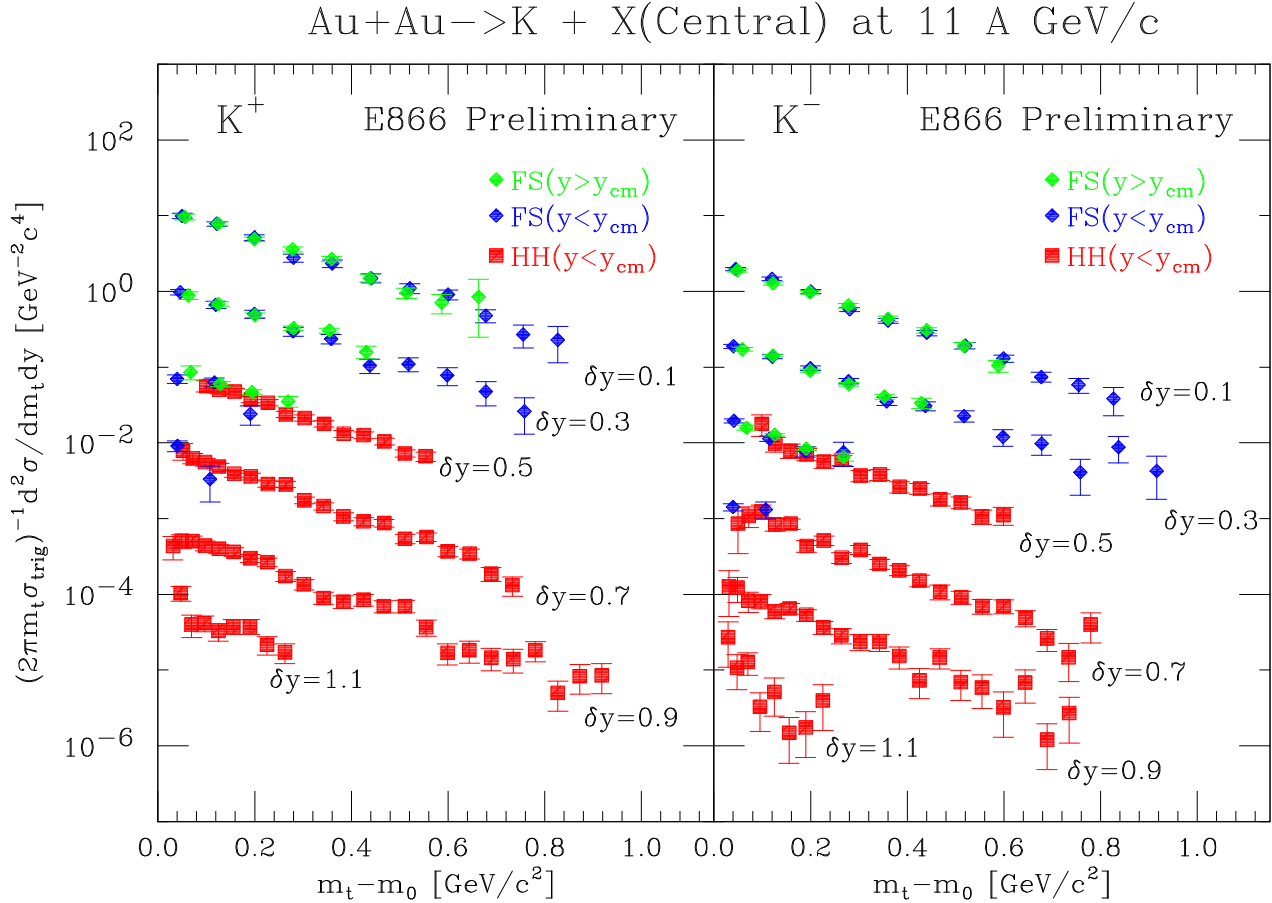,width=12.cm,angle=90}}
%\end{turn}
\vspace{0.cm}
\caption{The m$_T$ distributions of K$^+$ and K$^-$ mesons from
central Au+Au collisions at 10 AGeV in different rapidity bins
($\delta$y=$|$y-y$_{cm}|$). Starting with $\delta$y=0.1 at the top, each
successive spectrum is divided by 10 \protect\cite{e802}.
}
\label{ka_mt_e802}
\end{figure}
The rapidity distribution of 
K-mesons have already been dealt with in chapter 2. It was shown that for 
A+A collisions at AGS and SPS energies the kaon rapidity distributions
are significantly wider than expected for a thermal source.

\begin{figure}
%\vspace{-2.cm}
\hspace{ 1.5cm}\mbox{\epsfig{file=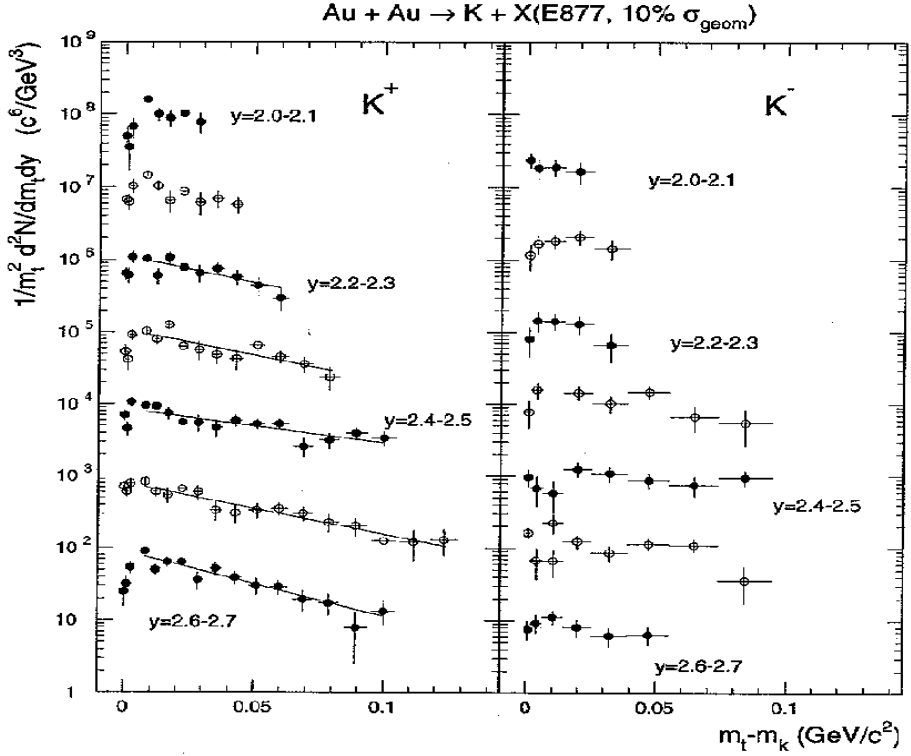,width=12.cm,height=10.cm}}
%\end{turn}
\vspace{0.cm}
\caption{Transverse mass spectra for K$^+$ (left) and K$^-$ (right)
for central Au+Au  collisions at 10 AGeV beam energy.
The data are presented in rapidity bins of 0.1 units widths successively 
}
\label{ka_mt_e877}
\end{figure}

\vspace{.4cm}
Charged kaon spectra have been measured at the AGS for the (nearly) symmetric
systems Si+Al and Au+Au in both peripheral and central collisions.
Fig.~\ref{ka_mt_e802} 
presents m$_T$ distributions of K$^{\pm}$ mesons from central 
Au+Au collisions at 10 AGeV in different rapidity bins
\cite{abbot,e802}. The spectra are 
well described by a single exponential and the inverse slope parameters
are 170 - 200 MeV both for K$^+$ and K$^-$. These values are similar to 
the ones obtained in Si+Au collisions. 
Fig.~\ref{ka_mt_e877} 
shows K$^+$ and K$^-$ transverse mass spectra for very small values
of m$_T$ for the same collision system \cite{lacasse}. 
In the range of m$_T$-m$_K<$0.1, the inverse slope parameter 
has values of 60 to 90 MeV. The  two data sets  overlap in rapidity at
y=2.4.    
Both experimental results are still preliminary. It will be interesting to see 
whether the significant differences in slope parameters at small and large 
m$_T$ persist. If so, a kaon low-p$_T$ enhancement would emerge.

\begin{figure}
\vspace{0.cm}
\centerline{
\hspace{ -1.cm}\mbox{\epsfig{file=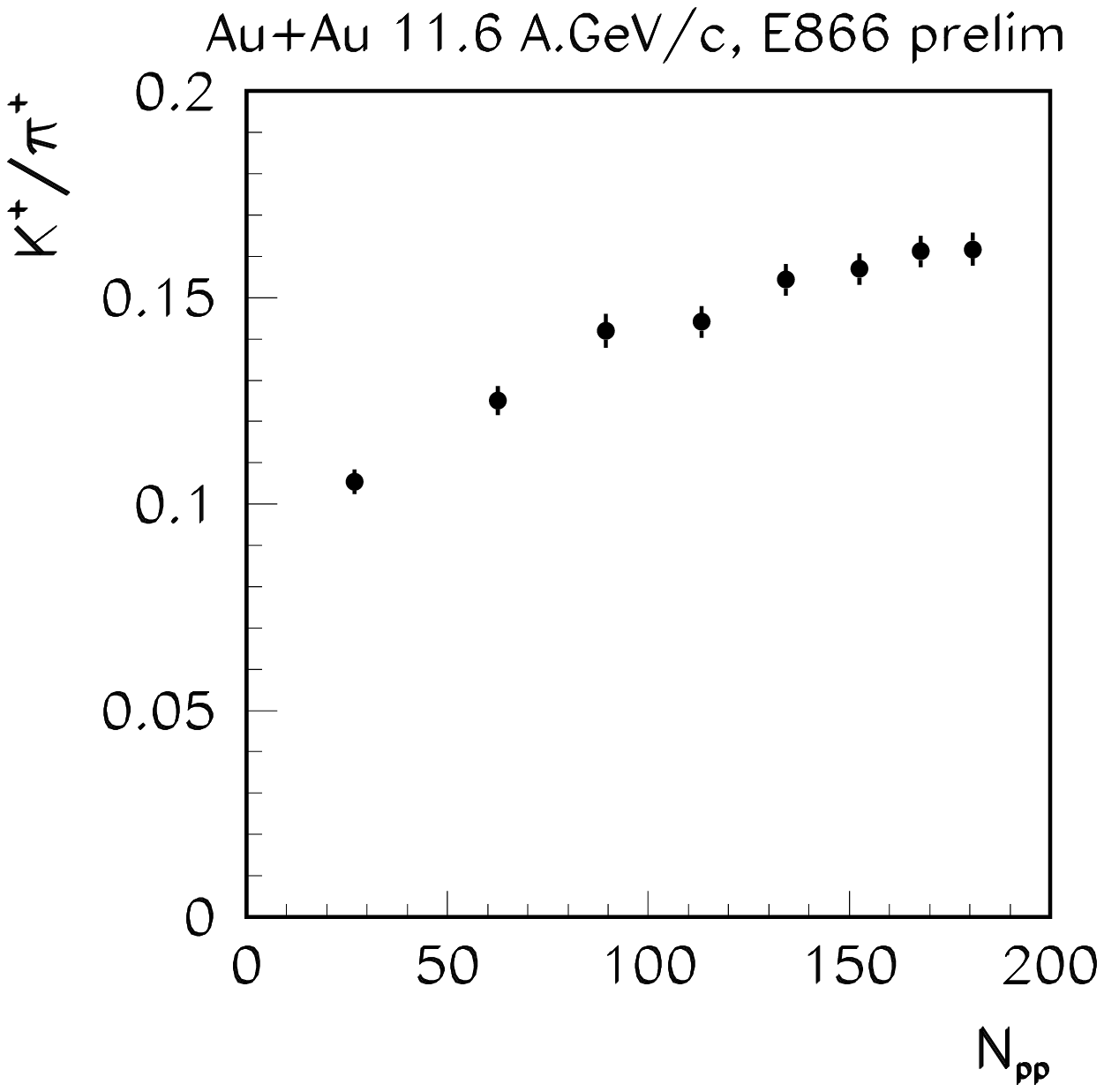,width=7.cm}}
\hspace{ 1.cm}\mbox{\epsfig{file=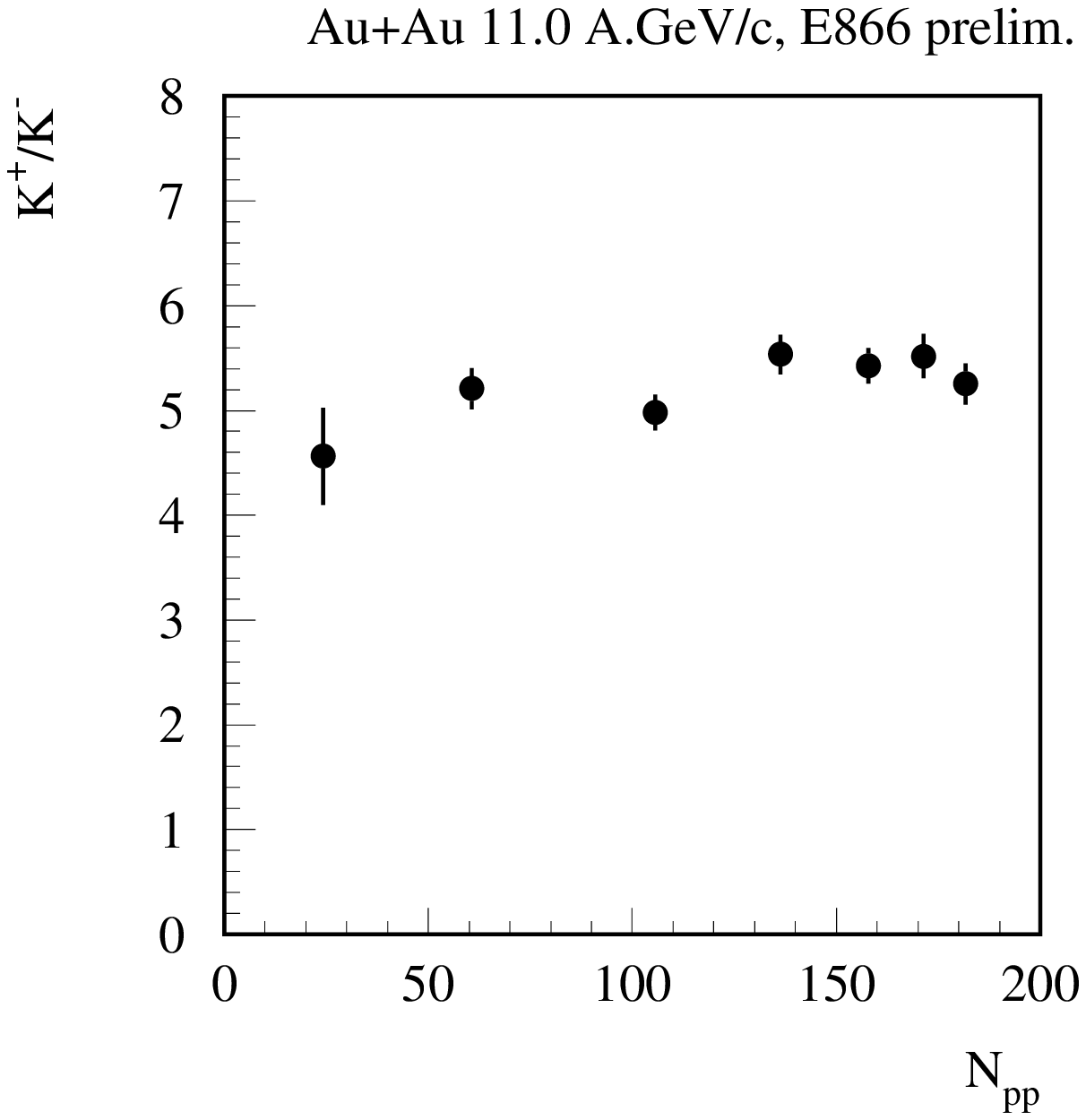,width=7.cm}}
}
\vspace{0.cm}
\caption{Ratio of K$^+/\pi^+$ (left) and K$^+$/K$^-$ (right)
fiducial yields as a function of the number of projectile participants 
for Au+Au collisions at 10.7 AGeV beam energy 
(preliminary, taken from \protect\cite{og97}).
}
\label{kp_pi_ags}
\end{figure}

\vspace{.4cm}
The most remarkable result  is a ''strangeness enhancement'' seen in
the  K$^+$/$\pi^+$ ratio \cite{e802,og97}. 
Fig.~\ref{kp_pi_ags} 
shows particle ratios measured at mid-rapidity as a function
of centrality in Au+Au collisions at 10.7 AGeV \cite{og97}.  
The K$^+$/$\pi^+$  ratio (left part)
increases from about 0.105 in peripheral  Au+Au 
collisions to about 0.16 in central collisions. It is interesting to 
note that the K$^+$/K$^-$ stays approximately constant as a function
of centrality (right part of Fig.~\ref{kp_pi_ags}). One would expect that due 
to absorption the relative K$^-$ yield decreases with increasing 
centrality as it was found for the antiprotons \cite{e802}.

The corresponding K$^+$/$\pi^+$ 
ratio in nucleon-nucleon collisions is 0.05 for 4$\pi$
multiplicities (which is only slightly (10\%) 
lower than the midrapidity value).
Microscopic models interpret 
the three times higher K$^+$/$\pi^+$ ratio in central A+A collisions
as the result of  interactions
of the produced pions with the fast incoming nucleons\cite{RQMD,MHof}.
''Strangeness enhancement'' then means that    
strange mesons can populate the larger  
phase space which is opened by multiple hadron-hadron 
collisions in heavy systems.    
Remember that the different centrality dependence of the  
K$^+$ and the $\pi^+$ multiplicity measured in Au+Au reactions 
at SIS energies (see Fig~\ref{kp_mult_au}) was also
explained by secondary meson-baryon collisions.    

\begin{figure}
%\vspace{-1.cm}
\hspace{ 1.5cm}\mbox{\epsfig{file=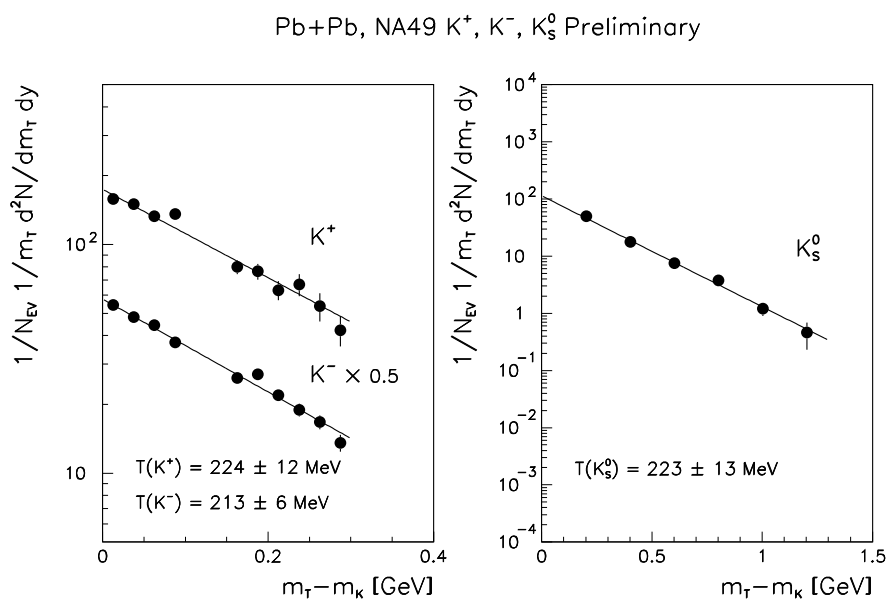,width=13.cm}}
\vspace{-0.5cm}
\caption{The transverse mass spectra of K$^+$ and K$^-$ mesons
(2.5 $<$ y $<$ 3.3) and of K$^0_s$ (2.0 $<$ y $<$ 2.7)
per unit of rapidity for Pb+Pb collisions at 158 AGeV
\protect\cite{bormann}. }
\label{ka_mt_sps}
\end{figure}

\begin{figure}[hpt]
\vspace{-1.cm}
%\hspace{1.5cm}
\begin{minipage}[t]{9cm}
\mbox{\epsfig{file=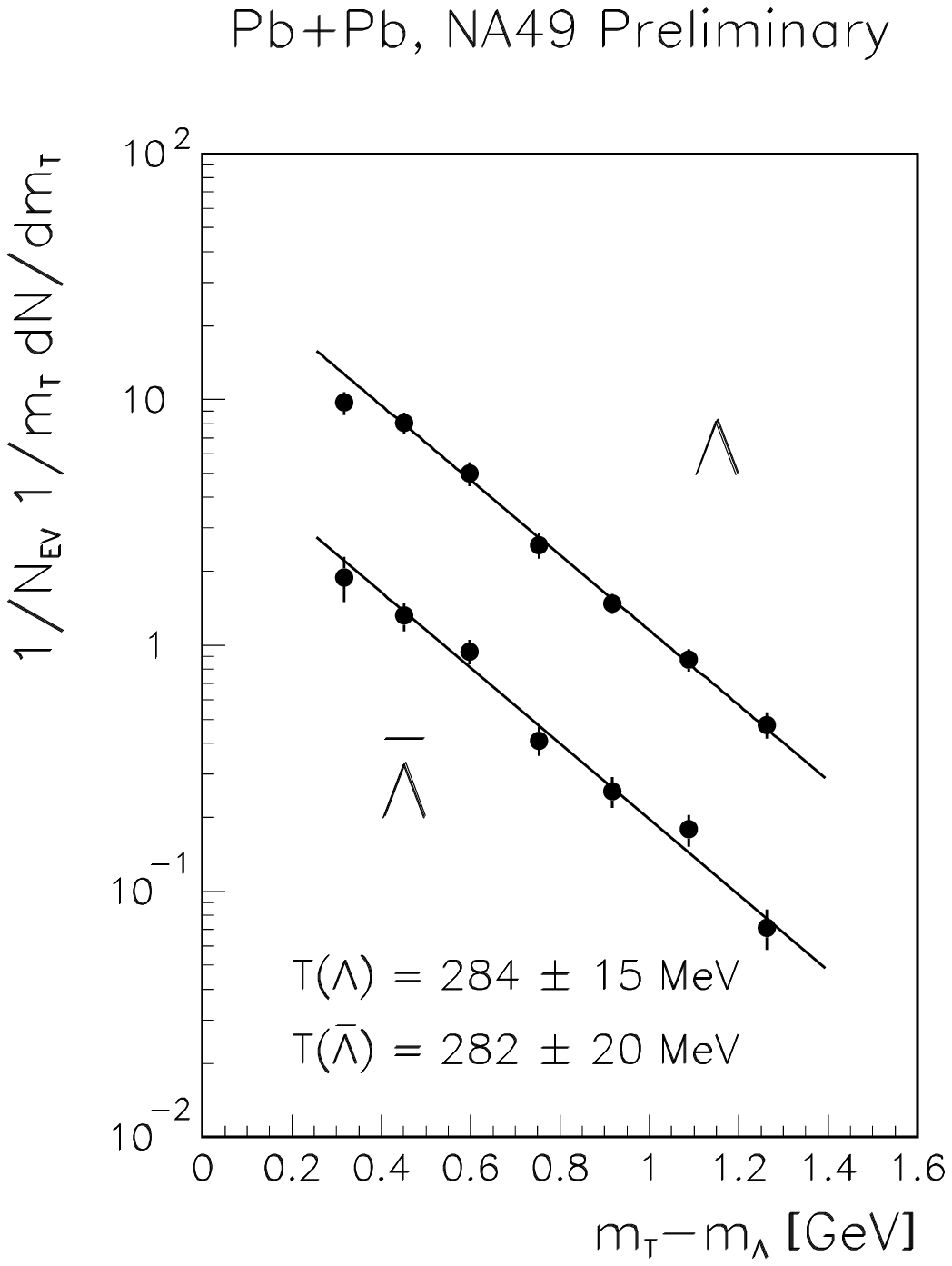,width=10.cm,height=11cm}}
\end{minipage}
\begin{minipage}[t]{5cm}
\vspace{-5.cm}
\caption{The transverse mass spectra of $\Lambda$ and $\overline\Lambda$
(2.6 $<$ y $<$ 3.8) for Pb+Pb collisions at 158 AGeV. 
The spectra are divided by the width of the 
rapidity interval \protect\cite{bormann}.
}
\label{lambda_mt_sps}
\end{minipage}
\end{figure}

At the highest energies presently available for heavy ion beams (200 AGeV)
the data on symmetric
systems comprise S+S \cite{Bae93,Alb94}
and Pb+Pb collisions \cite{Jones}.
As an example
Fig.~\ref{ka_mt_sps} and Fig.~\ref{lambda_mt_sps}  
show preliminary transverse mass spectra of kaons and
Lambdas from Pb+Pb collisions at 158 AGeV measured by the NA49 Collaboration. 
All spectra can be described by a single exponential
function in m$_T$. The slope factors are listed in Table \ref{slopmt}.
The large values of about 200 MeV and larger (see $\Lambda$ and
$\overline \Lambda$ for Pb+Pb) suggest that the spectral slopes are affected 
by a transverse or radial flow.

\begin{table}[htb]
\caption{Inverse slope parameters T fitted to the the transverse mass
distributions according to d$\sigma$/dm$_T$ $\propto$ m$_T$ exp(-m$_T$/T)
The data are taken from \protect\cite{Bae93,Alb94,Jones}.}

\vspace{0.5cm}
\begin{center}
\begin{tabular}{|c|c|c|c|c|c|}
\hline
reaction & $\Lambda$& $\overline \Lambda$& K$^0_S$& K$^+$ & K$^-$ \\
\hline
p + $^{32}$S& 182$\pm$17 & 132$\pm$18 &  205$\pm$16 & 227$\pm$52 & 286$\pm$106\\
\hline
$^{32}$S+S & 204$\pm$17 & 180$\pm$24 &  210$\pm$16 & 227$\pm$15 & 270$\pm$37\\
\hline
$^{32}$S+Ag & 234$\pm$17 & 221$\pm$24 &  231$\pm$17 & 254$\pm$31 & 181$\pm$28\\
\hline
$^{32}$S+Au & 240$\pm$18 & 223$\pm$22 &  227$\pm$18 & -  & - \\
\hline
$^{208}$Pb+Pb & 293$\pm$10 & 288$\pm$15 &  223$\pm$13 &224$\pm$12 & 213$\pm$6 \\
\hline
\end{tabular}
\end{center}
\label{slopmt}
\end{table}

\begin{figure}
%\vspace{-1.cm}
\begin{minipage}[t]{10cm}
\mbox{\epsfig{file=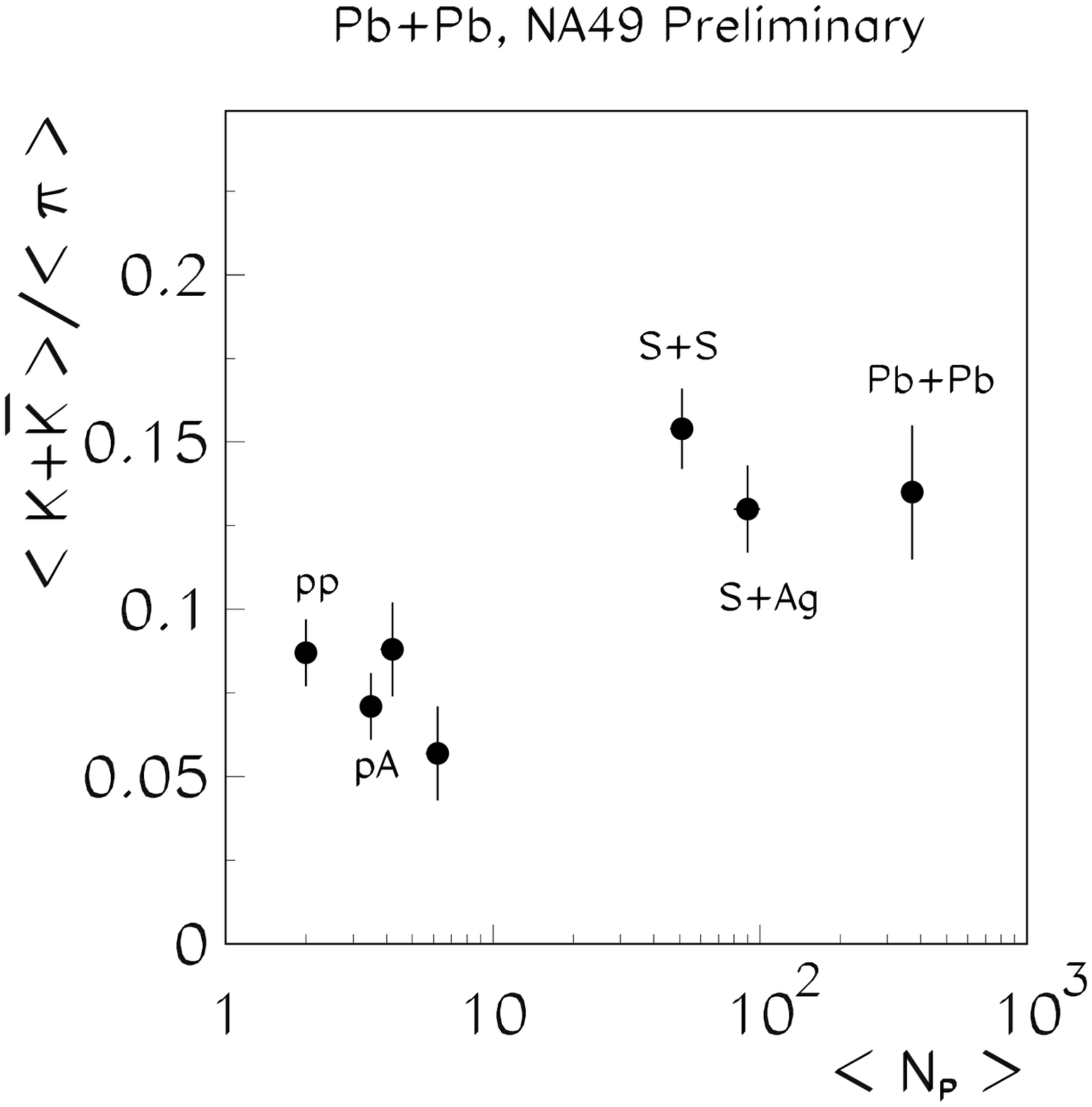,width=10.cm}}
\end{minipage}
\begin{minipage}[t]{5cm}
\vspace{-6.cm}
\caption{The K/$\pi$ ratio, defined as the sum of the 
multiplicities of all kaons (K$^+$, K$^-$, 2K$^0_s$) divided by all pions 
($\pi^+, \pi^-, \pi^0$), as a function of the number of participant nucleons 
for Pb+Pb collisions at 158 AGeV \protect\cite{bormann}.
}
\label{ka_pi_np_sps}
\end{minipage}
\end{figure}

Both at midrapidity and in 
4$\pi$, strange particle yields at SPS energies 
are enhanced by a factor of about two
when normalized to the pions and 
compared to nucleon-nucleon collisions \cite{GR2}.
In terms of K-mesons this finding is summarized in 
Fig.~\ref{ka_pi_np_sps}
in which the K/$\pi$ ratio is given as a function of the number of 
participating nucleons for p+p, p+A, S+A and Pb+Pb collisions \cite{bormann}.   
However, as already mentioned above, this kind of strangeness enhancement
has been observed also at SIS and AGS energies where it was explained
by multiple interaction of participants. The kaon to pion ratio is 
an ambiguous measure of ''strangeness'' enhancement as one compares 
particles with different masses which populate  phase space differently. 
The phase space available in A+A collisions is enhanced as compared to 
the one available in p+p collisions 
due to multiple nucleon-nucleon interactions. This ''new'' phase space
is preferentially populated by particles with high energies or large masses.
For example, at SIS energies 
the yield of high-energy pions increases (like the K$^+$ yield)
much faster than the low-energy pion yield when increasing the mass of the 
collision system \cite{muentz1}. 
Therefore, one should consider  for example 
the kaon to eta ratio or the kaon to high-energy pion  ratio 
as an  experimental indicator for strangeness enhancement. In these ratios,
simple phase space effects largely cancel out. A possible candidate in
this respect is also the $\overline \Lambda$/$\overline p$ ratio which will 
be discussed in the next chapter.       
On the other hand, model calculations  take kinematical 
effects into account but are not able to reproduce the strangeness data 
for the SPS energies \cite{sorge95,werner93}.  

The $\phi$ meson is a special strange particle as it consists 
of a strange and antistrange quark. First results from S+W collisions
confirmed the strangeness enhancement also in this observable
\cite{abreu}. Recently preliminary data on $\phi$s produced in central
Pb+Pb collisions have been presented \cite{friese}; both the shape of 
the transverse momentum distribution, which indicates a rather 
large slope factor of 
$\approx$ 350 MeV  and the large 4$\pi$ yield fit into the systematics
of a correlation between slope factor and particle mass and the
relative strange particle enhancement.

\begin{figure}
%\vspace{-1.5cm}
\hspace{ 1.cm}\mbox{\epsfig{file=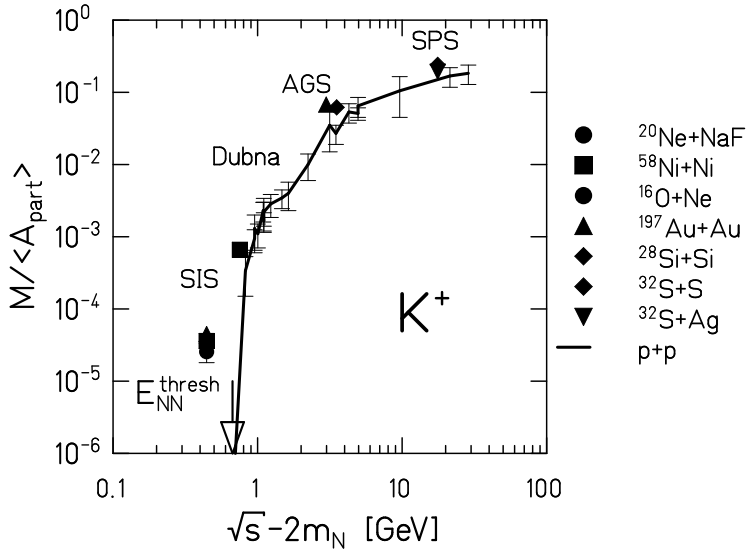,width=12.cm}}
\vspace{-0.5cm}
\caption{K$^+$ multiplicity per average number of participants
for A+A (full symbols) and p+p collisions (solid line with error bars)
as a function of the energy available in the nucleon-nucleon system. 
The data are taken 
from \protect\cite{GR2,mang,barth}. 
}
%\vspace{-1.5cm}
\label{kp_apart_exci}
\end{figure}

\begin{figure}
\hspace{ 1.cm}\mbox{\epsfig{file=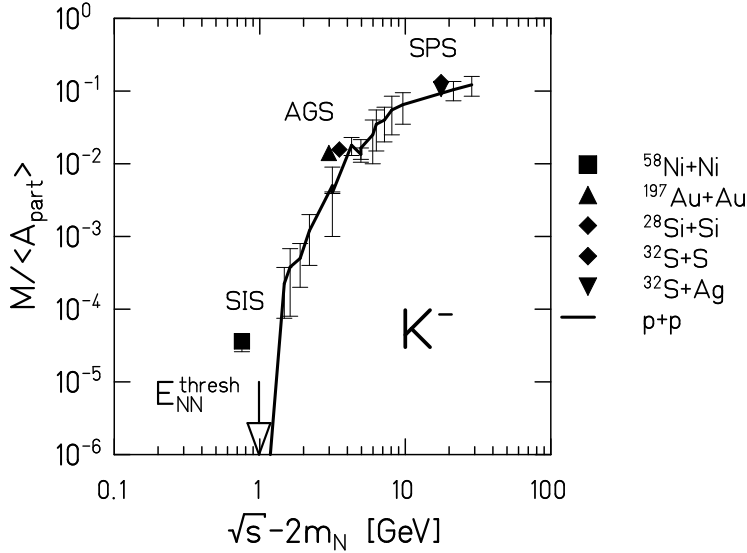,width=12.cm}}
\vspace{-0.5cm}
\caption{K$^-$ multiplicity per average number of participants
for A+A (full symbols) and p+p collisions (solid line with error bars)
as a function of
the  energy available in the nucleon-nucleon system. The data are taken
from \protect\cite{GR2,barth}.
}
\label{km_apart_exci}
\end{figure}

In the following we discuss the excitation function of kaon 
and antikaon production in nuclear collisions from SIS to SPS energies.   
Fig.~\ref{kp_apart_exci} and Fig.~\ref{km_apart_exci}  
show  the measured K$^+$ and K$^-$
multiplicities per participating nucleons as a function of the
beam energy which is expressed as the sum of the kinetic energies 
of two nucleons
in their c.m. system. This is the energy  available for the production
of new particles. The data are taken from \cite{GR2,mang,barth}.
The kaon yields per participating nucleon are larger in   
A+A collisions (full symbols) 
than in  p+p  collisons (solid line with error bars)
for all beam energies.
The difference between A+A and p+p data can be interpreted as
an increase of the effective beam energy in A+A collisions. 
This energy shift scales approximately with the
c.m. kinetic energy of the colliding nucleons
E$^{NN}_{kin}$ (= $\sqrt s$-2m$_N$):
$$\frac{M^{pp}_K}{A_{part}}(E^{NN}_{kin}) =
\frac{M^{AA}_K}{A_{part}}(\alpha\times E^{NN}_{kin})$$
with $\alpha$ = 1.3 - 2. This scaling behaviour suggests that in A+A collisions
the average available energy in  a nucleon pair (which
produces a K meson) is 1.3 - 2 times larger than in a free p+p
collision independent of the energy regime.

In section 2.3.1 we have shown that in  p+p collisions at SPS energies the 
protons lose on  average nearly half of their kinetic energy by particle 
production. In A+A collisions the situation is more difficult 
to interpret.
The rapidity distributions (normalized to beam rapidity) 
of the final state protons from 
central A+A collisions look similar  
at all beam energies suggesting  a similar stopping power. 
In contrast, the inelasticity as 
defined in section 2.3.1 (mainly by pion production)
is increasing by a factor of about 10 from  SIS to  SPS energies 
in central collisions of heavy nuclei. We attribute this discrepancy to the
reabsorption of pionic (i.e. produced) energy  
by the nucleons. This effect is most pronounced at low beam energies. 
K$^+$ mesons cannot be reabsorbed and thus represent an ideal probe
for the energy available for particle production in p+p and in A+A collisions.
Our finding that $\alpha$ is between  1.3 and  2 (independent of beam energy) 
suggests that the nuclear stopping power in Au+Au collisions is up to 
a factor of 2 higher than in p+p collisions at all energies.

It is surprising that 
$\alpha$ is the same for K$^+$ and K$^-$ mesons. This  means that 
K$^-$ absorption is not visible in spite of the
large cross section  for strangeness exchange reactions 
K$^-$N$\rightarrow\pi\Lambda$ (or $\Sigma$)  of 40-70 mb
depending on relative momentum.  Unfortunately,  
the large uncertainties on the K$^+$ and K$^-$ yields in the p+p data 
preclude any far reaching conclusion. This deficiency of the p+p data
asks for new experiments with the aim to provide a solid basis for
quantitative comparisons with A+A data. 

\clearpage

\chapter{Production of etas, antibaryons and multistrange hyperons}
\section{$\eta$ mesons}

The meson production experiments at the BEVALAC concentrated on
charged mesons. At SIS, the TAPS collaboration started a program to
measure neutral mesons with a photon detector \cite{taps}. 
The main goal was to study $\pi^0$ and $\eta$-meson 
production at beam energies near the threshold which
is E$^{th}_{NN}$= 1254 MeV for free nucleon-nucleon collisions.
Eta mesons originate almost exclusively from the decay of the
N$^*$(1535) resonance and thus are sensitive to the abundance of this
resonance.

\begin{figure}
\begin{minipage}[t]{7.4cm}
\mbox{\epsfig{file=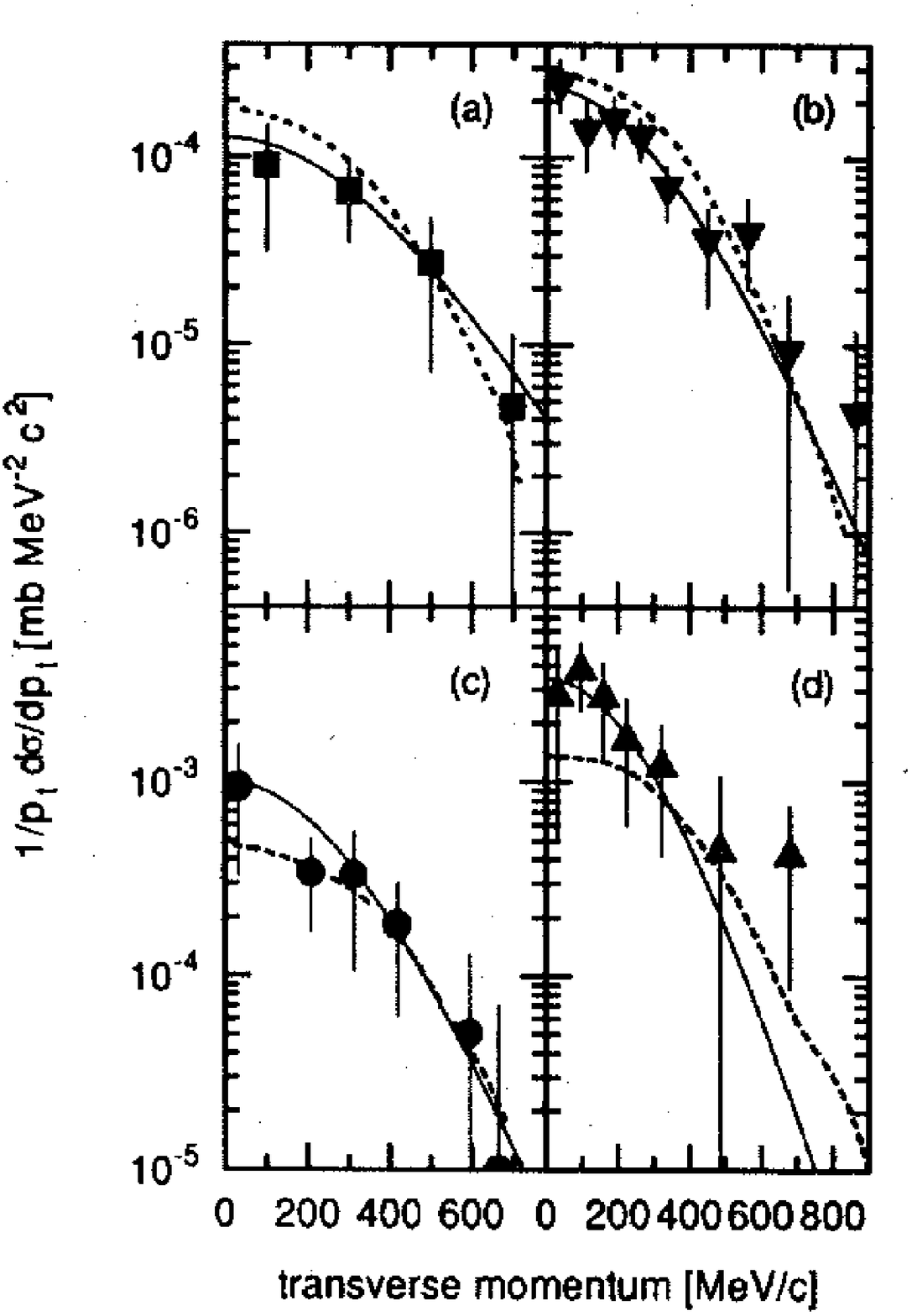,width=8cm,height=11.cm}}
\caption{ Transverse momentum distributions at midrapidity for the 
systems Ar+Ca at 1.0 (a) and 1.5 AGeV (b), Kr+Zr (c) and Au+Au (d)
at 1.0 AGeV \protect\cite{berg}. The solid lines represent fits with 
a thermal, isotropic source at midrapidity while the dashed curves 
show results of BUU calculations \protect\cite{ehehalt}. }
\label{eta_spec_berg}
\end{minipage}
\hspace{0.5cm}
\begin{minipage}[t]{7.4cm}
\mbox{\epsfig{file=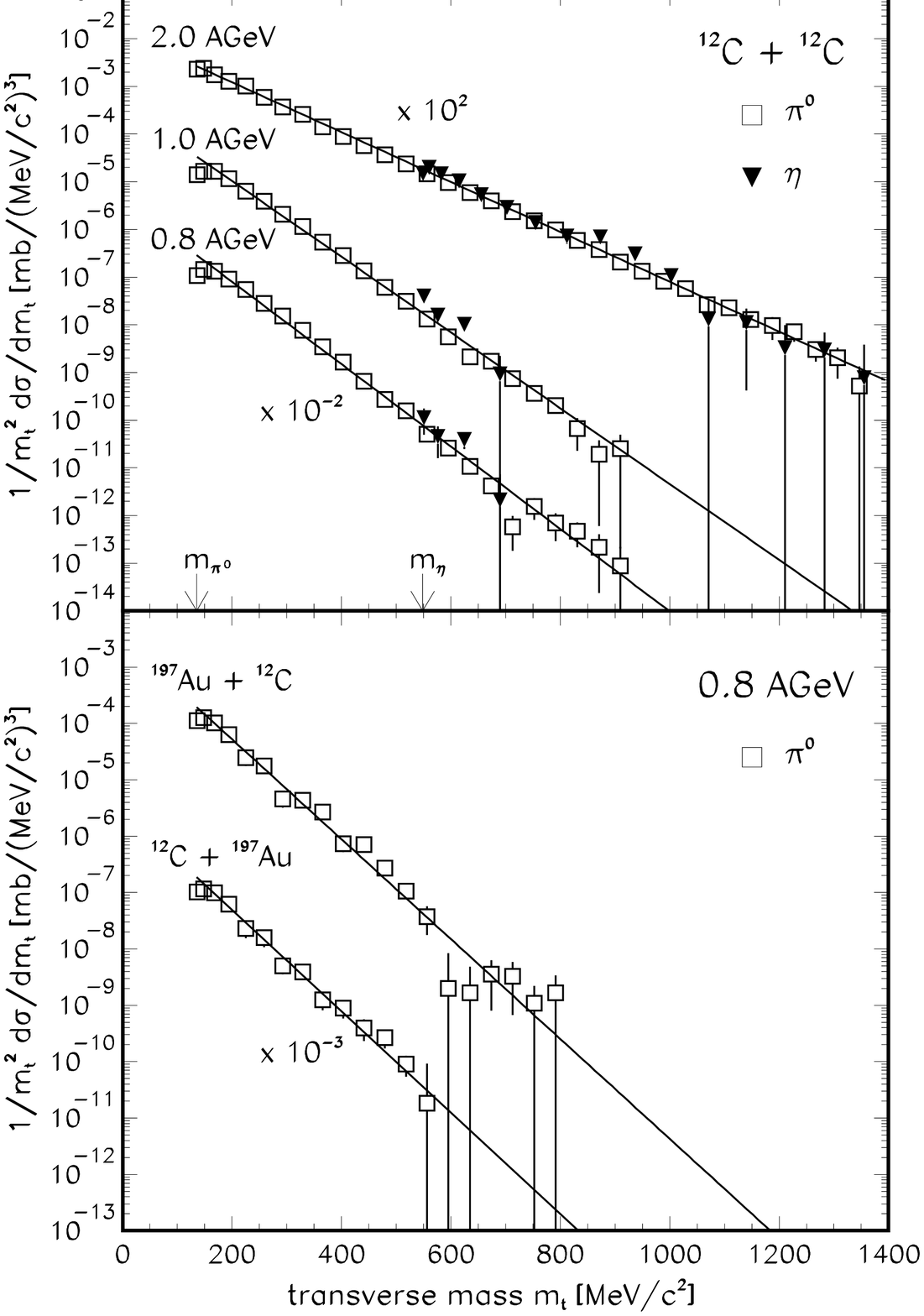,width=7.cm}}
\vspace{0.1cm}
\caption{Transverse mass spectra of $\pi^0$ and $\eta$ mesons measured in the
system C+C  at beam energies of 0.8, 1.0 and  2.0 AGeV within intervals of 
rapidity of 0.42-0.74, 0.42-0.74 and 0.8-1.08, respectively (upper frame).
Transverse mass spectra of $\pi^0$ are measured in C+Au and Au+C at 0.8 AGeV
at rapidities of 0.42-0.74 (lower frame). The lines represent fits 
to the pion data according to $d\sigma/dm_T\propto m^2_T exp(-m_T/T)$
which corresponds to the assumption of a thermal and isotropic meson
source \protect\cite{averbeck}.  
}
\label{eta_pi_cc}
\end{minipage}
\end{figure}

\vspace{.4cm}
Fig.~\ref{eta_spec_berg} shows the first results on
the $\eta$ transverse momentum distribution  measured at
midrapidity for Ar+Ca at 1.0 and 1.5 AGeV, Kr+Zr and  Au+Au at 1.0 AGeV
\cite{berg}.
The simultaneous measurement of $\eta$ and $\pi^0$ mesons allow to determine
the $\eta/\pi^0$ ratio with reduced systematic errors.
After extrapolation to full phase space, the  $\eta$/$\pi^0$ ratios
are found to be 1.3$\pm$0.8 \% at  1.0 AGeV  and
2.2$\pm$0.4 \% at 1.5 AGeV.
From the $\eta/\pi^0$ ratio the relative abundance of $\Delta$(1232) and
N$^*$(1535) at their respective freeze-out has been estimated.
Assuming $\Delta\rightarrow\pi$N to be the dominant pion production channel
and taking into account the appropriate isospin factors together
with a branching ratio of 40\% for the N$^*$(1535) decay into the $\eta$
channel, the TAPS collaboration found a N$^*$(1535)/$\Delta$(1232) ratio
of (1.1$\pm$0.6)\% and (1.8$\pm$0.3)\% at 1.0 and 1.5 AGeV, respectively
\cite{berg}.

\begin{figure}
\vspace{0.cm}
\begin{minipage}[t]{10cm}
\hspace{ 0.cm}\mbox{\epsfig{file=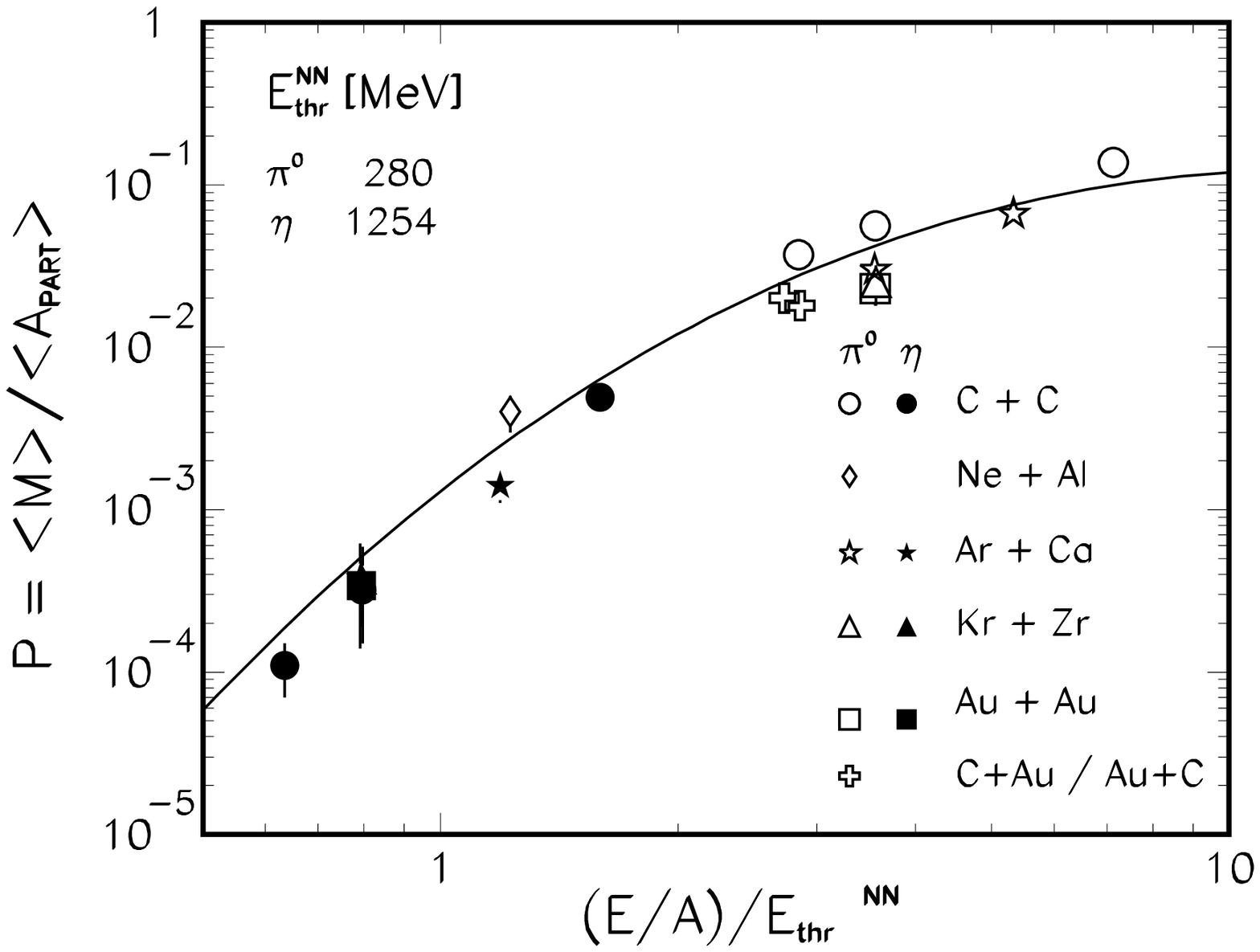,width=10.cm}}
\end{minipage}
\hspace{0.5cm}
\begin{minipage}[t]{5cm}
\vspace{-5.cm}
\caption{Average $\pi^0$ and $\eta$ multiplicities per average number
of paticipants in A+A collisions as a function of the bombarding energy,
corrected for energy loss in the target and normalized to the production
thresholds. The picture is taken from \protect\cite{averbeck}. }
\label{eta_pi_exci}
\end{minipage}
\end{figure}

\vspace{.4cm}
Recently, $\eta$ mesons were measured in the light system  C+C
at different bombarding energies. Fig.~\ref{eta_pi_cc} shows the
production cross sections for $\pi^0$ and $\eta$ mesons
in a Boltzmann representation as a function of the transverse mass
\cite{averbeck}.
The $\eta$ and $\pi^0$ spectra taken at 2.0 AGeV
fall on top of each other if plotted 
as a function of the transverse mass  m$_T$.
Such a m$_T$ scaling is expected (but no proof) for a thermally
and hadrochemically equilibrated source.
In any case it is worthwhile to note that the pions perfectly follow a
Boltzmann distribution although they are emitted from a source
which on average consists of 6 (participant) nucleons only.

Fig.~\ref{eta_pi_exci} presents the neutral meson production probability per
participant nucleon a as a function of the bombarding energy
for different collision systems \cite{averbeck}.
The beam energy is normalized to the respective threshold energy for
free NN collisions.
The eta and pion data follow a common trend which means that their
production probability just depends on the available energy.
Charged pions also fit into this picture.

\begin{figure}[hpt]
\vspace{0.cm}
\begin{minipage}[t]{10cm}
\hspace{ 0.cm}\mbox{\epsfig{file=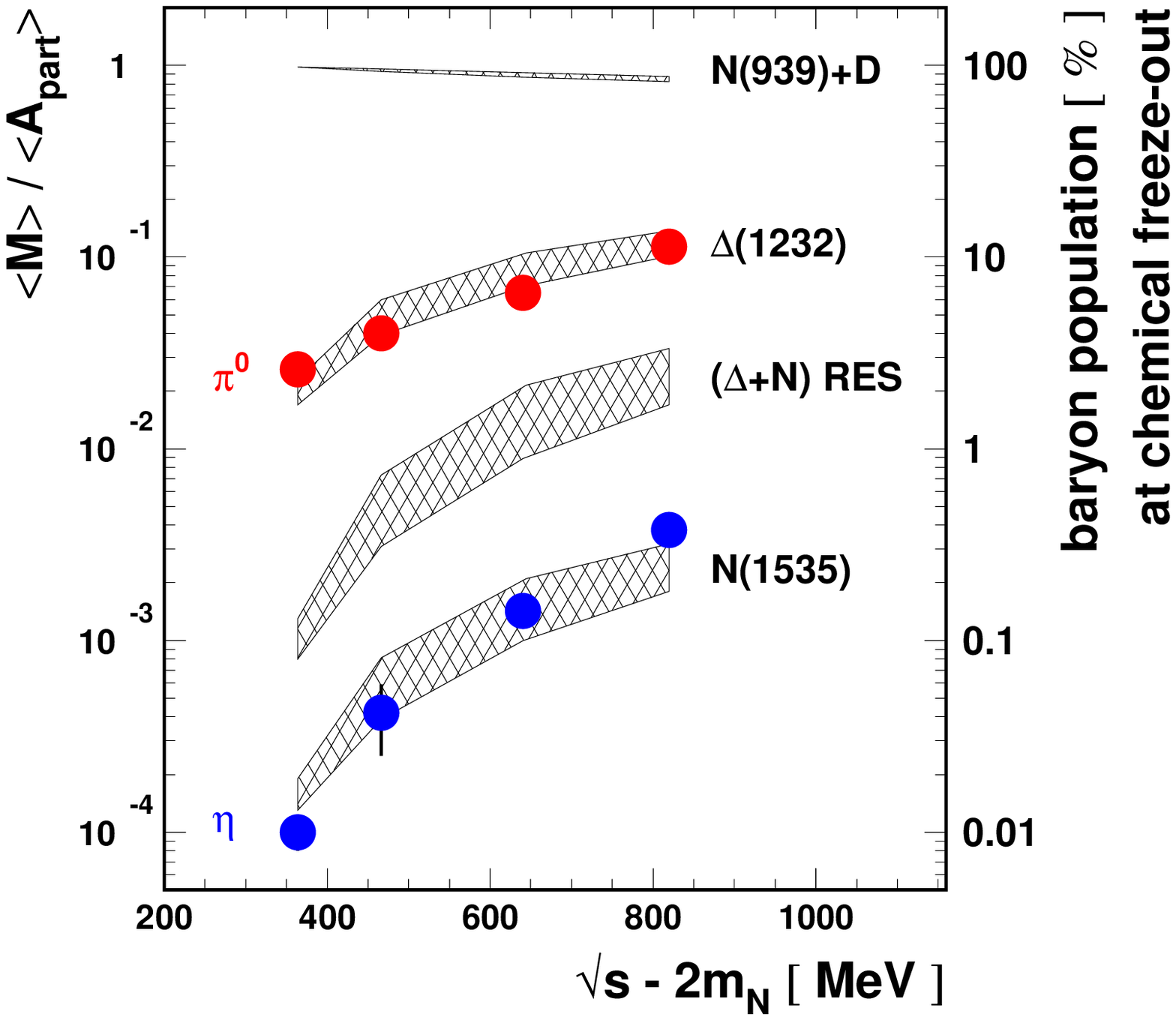,width=9.5cm,height=8cm}}
\end{minipage}
\hspace{0.2cm}
\begin{minipage}[t]{5.5cm}
\vspace{-8.cm}
\caption{Inclusive $\pi^0$ and $\eta$ multiplicities per average number 
of participants (left scale) as a function of the energy available 
in the nucleon-nucleon system. The hatched areas show the relative 
populations of nucleons and deuterons, of the $\Delta$(1332) resonance,
of the N(1535) resonance and of the sum of all remaining $\Delta$ and N
resonances (right scale), as obtained from a hadron-gas model 
\protect\cite{averbeck2}.  
}
\label{reson_abund}
\end{minipage}
\end{figure}

The abundances of resonances like $\Delta$(1232) and N$^*$(1535) as 
derived from the pion and eta yields are in agreement with the assumption of
thermal and hadrochemical equilibrium. This is demonstrated in 
Fig.~\ref{reson_abund}
which compares the meson yields (full symbols)  to the 
resonance population at chemical freeze-out (shaded areas) for different
available energies. The resonance population was
calculated for a given temperature T and baryochemical potential $\mu_B$
\cite{averbeck2}.
The parameter set (T, $\mu_B$) was determined from particle production 
ratios $\pi$/A$_{part}$, $\eta/\pi^0$ and from production rates of 
$\pi$, $\Delta$, p, d  for different collisions systems and 
bombarding energies \cite{averbeck2} (see chapter 2). 
Up to 15\% of the nucleons are found to be excited to resonance states at 
freeze-out for the  largest value of 
available energy (which corresponds to a beam energy of 2 AGeV).      

\begin{figure}[hpt]
\begin{minipage}[t]{10cm}
\hspace{ 0.cm}\mbox{\epsfig{file=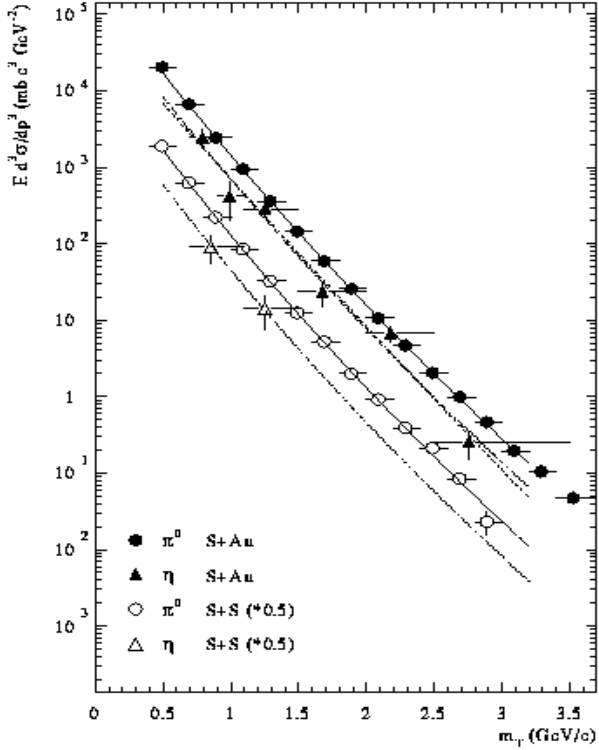,width=10.cm,height=11cm}}
\end{minipage}
\begin{minipage}[t]{5.5cm}
\vspace{-5.cm}
\caption{ Invariant cross sections of $\pi^0$ and $\eta$ mesons averaged
over the rapidity interval of $ 2.1 \leq y \leq$ 2.9 as a function of 
transverse mass for 200 AGeV S+Au and S+S minimum bias data. The S+S data
have been scaled by a factor of 0.5 for better presentation 
\protect\cite{wa80}.
}
\label{eta_pi_wa80}
\end{minipage}
\end{figure}

\vspace{.4cm}
Eta production was also measured at SPS energies by the 
WA80 Collaboration. 
In  S+Au and S+S collisions at 200 AGeV/c the 
$\eta/\pi^0$ cross section ratio was found to be 0.12$\pm$0.04  \cite{wa80}.
Fig.~\ref{eta_pi_wa80} 
presents the invariant cross sections of $\pi^0$ and $\eta$ mesons 
averaged over the rapidity interval of 2.1 $\leq$ y $\leq$ 2.9 as a function
of transverse mass.  
Etas and pions roughly follow  m$_T$ scaling with an 
inverse slope parameter of T=226$\pm$9 MeV with a slightly smaller
value for the pions. This deviation suggests a radial expansion velocity 
at freeze-out of approximatelly $\beta$=0.3-0.5.

\section{Antibaryons}

The production  of a proton-antiproton pair in 
a proton-proton collision requires
a beam energy of 5.6 GeV. Nonetheless, antiprotons were
observed at the BEVALAC/LBL in 1.65 and 2 AGeV Si+Si and 2 AGeV Ne+NaF
collisions \cite{shor1}, at the SIS/GSI with Ne and Ni beams
of 1.66 - 1.93 AGeV on various targets \cite{schroeter} and
in C+C collisions at 3.65 AGeV  at the JINR in Dubna \cite{baldin}.
In order to suppress sufficiently the pions and antikaons - which are
typically 7 and 3 orders of magnitude more abundant than antiprotons -  
the measurements at LBL and GSI  have been performed at 0 degree with a
small-acceptance beam-line spectrometer.

\vspace{.4cm}
Although antiproton production is extremely subthreshold at BEVALAC/SIS 
energies, it seems to follow an universal trend.  
It was pointed out that the antiprotons fit
into a scaling pattern together with the pions and kaons
when plotting the invariant cross sections over the energy needed to 
produce the particles \cite{shor1,shor2}.  This energy is the sum of the
particle c.m. kinetic energy and the c.m. threshold energy: 
E$^*$ = E$^{c.m.}_{kin}$ + E$_{thresh}$. The threshold energy
E$_{thresh}$ is the minimum energy required to produce the particle 
in a NN collision and is equal to m$_{\pi}$ for pion production, 
m$_K$+m$_{\Lambda}$-m$_N$ for K$^+$ production, 2m$_K$ for K$^-$ 
production and 2m$_p$ for antiproton production.  
The invariant particle production cross sections fall on an exponential 
E$d^3\sigma/dp^3\propto$exp(-E$^*$/T) with T the inverse slope parameter
depending on beam energy and on the size of the colliding system.
This scaling was  found for Si+Si(Al) collisions both
at BEVALAC  and at AGS energies \cite{shor2}. 
Surprisingly enough, the universal scaling holds also for antiprotons which 
are expected to be strongly absorbed.
The FRS/GSI data confirmed the scaling law for Ne+NaF collisions but
found deviations for the heavier Ni+Ni system \cite{kienle}.

\begin{figure}
\vspace{0.cm}
\hspace{ 0.5cm}\mbox{\epsfig{file=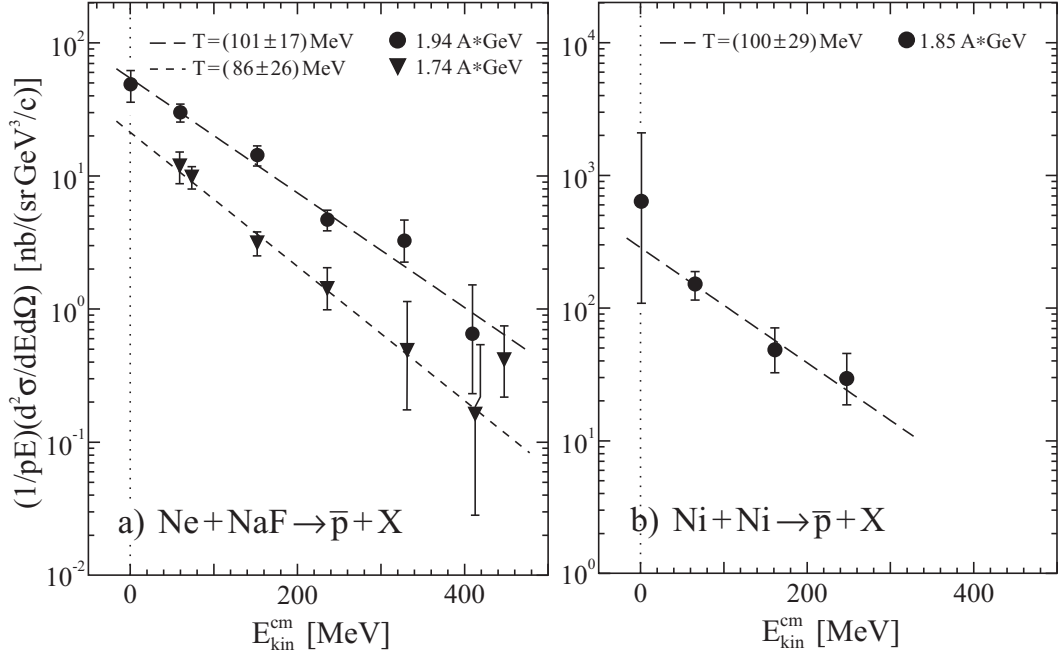,width=14.cm}}
\vspace{0.cm}
\caption{ Invariant $\overline p$ production cross sections as a function of
the c.m. kinetic energy for the reactions Ne+NaF at 
1.74 and 1.94 AGeV (left) and Ni+Ni at 1.85 AGeV (right) 
\protect\cite{kienle}. The dashed lines are exponential fits to the data
with inverse slope parameters T as indicated.  
}
\label{pbar_spec_frs}
\end{figure}

\vspace{.4cm}
The phenomenon of  deep subthreshold particle production opens 
the possibility to study   
(i) the mechanisms of energy accumulation via  
multiple hadron-hadron collisions or short-range correlations etc. and (ii) 
in medium modifications of cross sections, effective masses, thresholds etc. 
The Boltzmann distributions and the scaling pattern 
of the produced particles suggest a thermal 
origin. Therefore, thermal fireball models have been proposed to explain the 
measured particle yields. 
As an example we mention the ansatz of Ref. \cite{koch_dov}.
The authors assume that 
only nucleons and $\Delta$ resonances are in thermal and chemical equilibrium.
Kaons, antikaons and antiprotons are not equilibrated  but 
created (and possibly absorbed) in thermal baryon-baryon collisions.
The particle abundances
are calculated by rate equations. The model uses free
production cross sections and allows for a cooling of the fireball.
The pions are not considered to be in equilibrium but are created via decays 
of $\Delta$ resonances after freeze-out. Therefore, $\pi$N collisions are 
omitted in this approach.     
The thermal equilibrium of baryons guarantees the population of the very high
energy tails of the Boltzmann distribution. These energies
are required for deep subthreshold processes.  
The K$^-/p$ and $\bar p$/p ratios measured in Si+Si collisions at 2 AGeV 
at the BEVALAC \cite{shor1} could be reproduced by this model.     
Antiproton production in C+C collisions could not be explained; this was 
considered as an indication for a non-equilibrated system.

\vspace{.4cm}
Fig.~\ref{pbar_spec_frs} shows the invariant $\bar p$ production cross sections
as a function of the $\bar p$ c.m. kinetic energy  measured
in Ne+NaF and Ni+Ni collisions by the FRS group at SIS \cite{kienle}.
The spectral slopes follow exponentials as indicated by the dashed lines.  
The inverse slope parameters  are 101$\pm$17 MeV and 86$\pm$26 MeV 
for Ne+NaF collisions (at 1.94  and 1.74 AGeV)
and 100$\pm$29 MeV for Ni+Ni collisions at 1.85 AGeV.  
These values agree with the ones for the corresponding K$^-$ mesons
and for the high-energy tail of the $\pi^-$ mesons \cite{kienle}.
The data shown in Fig.~\ref{pbar_spec_frs} 
allow to extract information on the mass 
dependence of antiproton production and on the excitation function. 

The ratio of the invariant cross sections for Ni+Ni$\rightarrow\bar p$+X and 
Ne+NaF$\rightarrow\bar p$+X was found to be
$\sigma_{Ni}^{\bar p}$/$\sigma_{Ne}^{\bar p}$ = 9$\pm$3 
\cite{kienle}. When correcting for the geometrical reaction cross section
one obtains  the $\bar p$ multiplicities 
M$^{\bar p}$=$\sigma^{\bar p}/\sigma_R$ with $\sigma_R$ = 2.7 b for Ni+Ni 
and 1.3 b for Ne+NaF. 
Hence the ratio of $\bar p$ multiplicities is   
M$_{Ni}^{\bar p}$/M$_{Ne}^{\bar p}$ = 4.3$\pm$1.4. 
According to a geometrical model,
the average number of participants is $<$A$_{part}>$ = 10.2 for impact 
parameter integrated Ne+NaF collisions and $<$A$_{part}>$ = 29.2 for Ni+Ni
\cite{schroeter}. These numbers show that the $\bar p$ multiplicity
increases somewhat stronger than the number of participants.

\vspace{.4cm}
However, the measured $\bar p$ yields (and their ratios) 
reflect a subtle interplay of production and absorption  processes. 
The main problem in the interpretation of the antiproton data is the
large probability for $\bar p$ annihilation. 
The antiproton annihilation cross section in $\bar p$p collisions for  a 
$\bar p$ momentum of 1 GeV/c is about 70 mb \cite{land_boern} corresponding
to a mean free path of $\lambda$=0.9 fm in normal nuclear matter.
As the antiproton is detected under $\Theta_{lab}$=0$^o$
one can assume, that it has to propagate through a piece of nuclear
matter which has the dimension of the radius of the target nucleus
(R=1.2$\times$A$^{1/3}$ fm). 
According to exp(-R/$\lambda$) 
only about 3\%  of the primordial $\bar p$ yield reaches the
detector for Ne+NaF collisions and about 0.6\% for Ni=Ni collisions.  
These numbers agree reasonably well with the result of a RBUU calculation
which finds a $\bar p$ survival rate of about 1\% for Si+Si
collisions at a $\bar p$ momentum of 1 GeV/c \cite{li_ko_fa_ze}.  
When taking the estimated losses into account, the absorption corrected  
(primordial) multiplicity ratio will be 
M$_{Ni}^{\bar p}$/M$_{Ne}^{\bar p}$ $\approx$ 21$\pm$7.      
Now one can estimate the A$_{part}$-dependence
of $\bar p$ production corrected for absorption: 
M$^{\bar p}$$\propto$A$_{part}^{\alpha}$ with $\alpha$=2.9$^{+0.3}_{-0.4}$. 
This large value of $\alpha$ is evidence for  
antiproton production via multiple interaction of participants.

\vspace{.4cm}
The comparison of experimental $\bar p$ production cross 
sections to model calculations is very difficult  
due to the strong $\bar p$ absorption in the nuclear environment.
Nevertheless it is worthwhile to study this observable because 
antiproton production at beam energies far below the threshold
is extremely sensitive to medium effects. Transport models
studied antiproton production in nucleus-nucleus collision by
taking into account the effect of the antiproton scalar and 
vector selfenergy in the nuclear
medium. This effect leads - similarly as for antikaons - to a strong decrease
of the in-medium $\bar p$ effective  mass with increasing nuclear density
and hence to a significant enhancement of the 
$\bar p$ yield \cite{li_ko_fa_ze,batko,teis}.

\begin{figure}
\vspace{0.cm}
\begin{minipage}[t]{10cm}
\hspace{ 0.cm}\mbox{\epsfig{file=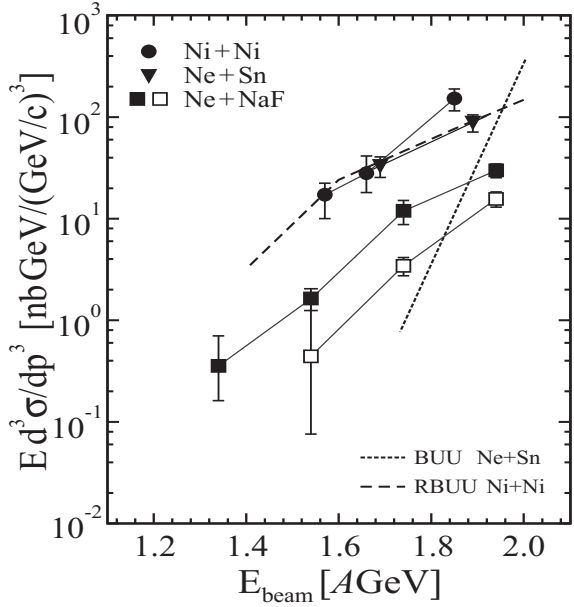,width=7.5cm,height=8cm}}
\end{minipage}
\begin{minipage}[t]{5cm}
\vspace{-8.cm}
\caption{Excitation functions for $\overline p$ production for Ni+Ni,
Ne+Sn and Ne+NaF collisions with respect to the beam energy 
\protect\cite{kienle}. The Ni+Ni data are taken
at p$_{cm}$=0.34 GeV/c, the Ne+NaF data at p$_{cm}$= 0.34 GeV/c (full
squares) and 0.56 GeV/c (open squares), the Ne+Sn data at 
p$_{cm}$= 1.5 GeV/c. The lines represent calculations
without (dotted line, BUU) and with in-medium mass reduction of the 
$\overline p$ (dashed line, RBUU)
\protect\cite{teis}.  }
\label{pbar_exci}
\end{minipage}
\end{figure}

The  comparison of experimental and calculated excitation functions 
(for a given reaction system) is a possibility
to bypass the problem of absorption as the energy dependence of 
$\bar p$ production    
should not be influenced by the $\bar p$ annihilation.
The fact, that antiprotons are produced very far below threshold 
by highly collective processes should result in a very steep excitation
function. This is demonstrated by the dotted line in 
Fig.~\ref{pbar_exci}
which is the prediction of a BUU calculation. The data, however, show a much
flatter slope. This trend is reproduced by a RBUU calculation which considers
an in-medium mass reduction of the antiproton (dashed line in 
Fig.~\ref{pbar_exci} \cite{teis}).

\begin{figure}
\vspace{0.cm}
\begin{minipage}[t]{10cm}
%\begin{turn}{180}
\mbox{\epsfig{file=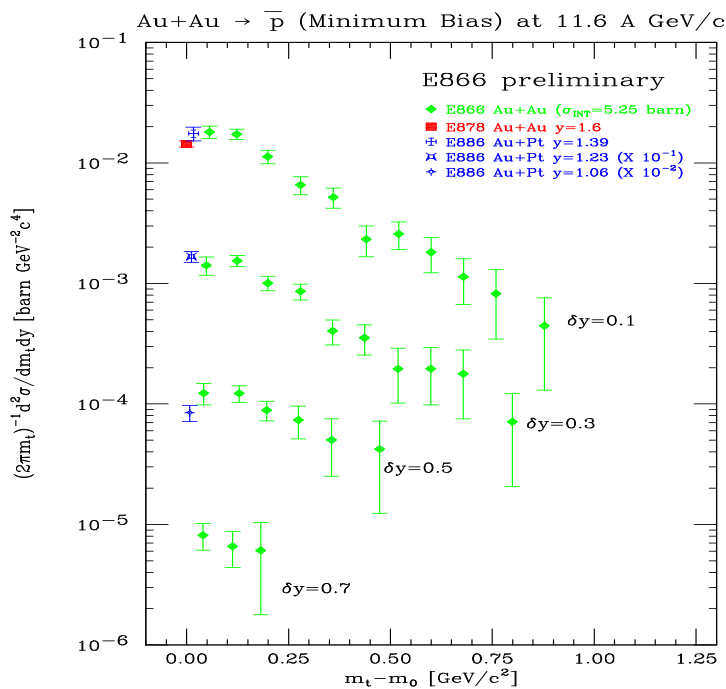,width=9.5cm,height=9.cm,angle=180}}
%\end{turn}
\end{minipage}
\hspace{0.5cm}
\begin{minipage}[t]{4.5cm}
%\vspace{-5cm}
\caption{Invariant cross sections of $\overline p$ from Au+Au collisions
at 10.7 AGeV in different rapidity bins ($\delta$y=$|$y-y$_{cm}|$). The
topmost spectrum is absolutely normalized, while each successive spectrum
is divided by 10. Taken from \protect\cite{e802}.  
}
\label{pbar_e866}
\end{minipage}
\end{figure}

\begin{figure}
\vspace{0.cm}
\begin{minipage}[t]{11cm}
\mbox{\epsfig{file=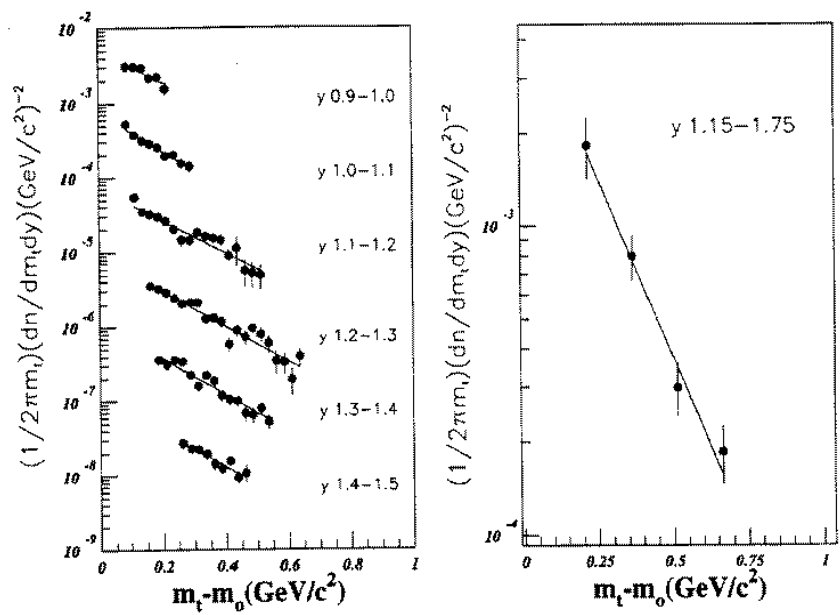,width=11.cm,height=8cm}}
\end{minipage}
\hspace{1cm}
\begin{minipage}[t]{4.cm}
\vspace{-5.cm}
\caption{Transverse mass spectra of $\bar p$ (left) and
$\bar \Lambda$ (right) for Si+Au central collisions at 13.7 AGeV
\protect\cite{wu859}. 
The solid lines represent exponentials fitted to the data. 
}
\label{p_bar_lam_bar_ags}
\end{minipage}
\end{figure}

\vspace{.4cm}
As mentioned above, the scaling of pion, 
kaon and antiproton production with the 
available energy E$^*$ holds also at AGS energies, namely 
for Si+Al at 13.7 AGeV \cite{shor2}. 
This finding indicates 
that the particles are produced according to the available phase space
which just depends on beam energy (and on the mass of the collision system).
The scaling is also roughly fulfilled for Au+Au collisions at
10.7 AGeV \cite{e802}. 
However, in the heavy system the spectral slopes 
increase with particle mass,  
Fig.~\ref{pbar_e866} shows the antiproton production cross sections
measured for this system at different rapidity bins. The spectra follow
an exponential with an inverse slope parameter of about 250 MeV. This
value is larger than the ones for kaons and antikaons 
(170 - 200 MeV, see Fig.~\ref{ka_mt_e802})    
but may be slightly smaller than the one for protons (250 - 300 MeV)
\cite{e802}. These large  inverse slope parameters
indicate that the particles (including antiprotons) participate in 
a collective transverse expansion of the fireball.

Antiproton production has been also studied as a function of centrality
in Au+Au collisions at 10.7 AGeV \cite{e802}. 
In contrast to the K$^+$/$\pi^+$ and the 
K$^-$/$\pi^+$ ratios which increase with increasing number of partricipants,
the $\bar p$/$\pi^+$ ratio decreases. This effect has been attributed 
to an enhanced  absorption of $\bar p$'s in central collisions 
\cite{sako866}.

\vspace{.4cm}
In ultrarelativistic nucleus-nucleus collisions, the study of 
antibaryon production such as $\bar p$ and $\bar \Lambda$ may shed light
on the mechanisms of antiquark production.  In particular, the 
ratio $\bar \Lambda$/$\bar p$ is a measure of the the $\bar s$/$\bar u$
ratio  which  is expected to approach unity in the deconfined phase. 
Therefore, a $\bar \Lambda$/$\bar p$ ratio which is 
enhanced in A+A collisions over the pp value  
($\approx$ 0.25) has been considered as a signature for the 
transient existence of a deconfined phase. 
The search for such an enhancement 
motivates the measurement of $\bar p$ and $\bar \Lambda$ both at the 
AGS and the SPS.  
Fig.~\ref{p_bar_lam_bar_ags} shows the transverse mass spectra of 
antiprotons (left) and 
anti-Lambdas (right) for Si+Au central collisions at 13.7 AGeV 
\cite{wu859}. The inverse slope parameters of the exponential functions 
fitted to the data (see solid lines in Fig.~\ref{p_bar_lam_bar_ags}) 
are about 180 MeV for the 
antiprotons and about 190 MeV for the anti-Lambda. 
The number of $\bar \Lambda$ is obtained by fitting the invariant mass of  
$\bar p$-$\pi^+$ pairs and correcting it for the branching ratio of 64.2\%
for this decay  channel.  The experimentally 
measured dn/dy ratio of $\bar \Lambda$/$\bar p$ is found to be
1.02$\pm$0.21$\pm$0.10.  
When correcting the $\bar p$ yield for the contamination due to the 
$\bar \Lambda$ decay, the $\bar \Lambda$/$\bar p$ ratio 
is 2.9$\pm$0.9$\pm$0.5 which is higher than predicted for a system
in chemical equilibrium. It rather hints at absorption effects which are 
more pronounced for $\overline p$ than for $\overline \Lambda$ baryons.

\begin{figure}
%\vspace{-2.cm}
\begin{minipage}[t]{10cm}
\mbox{\epsfig{file=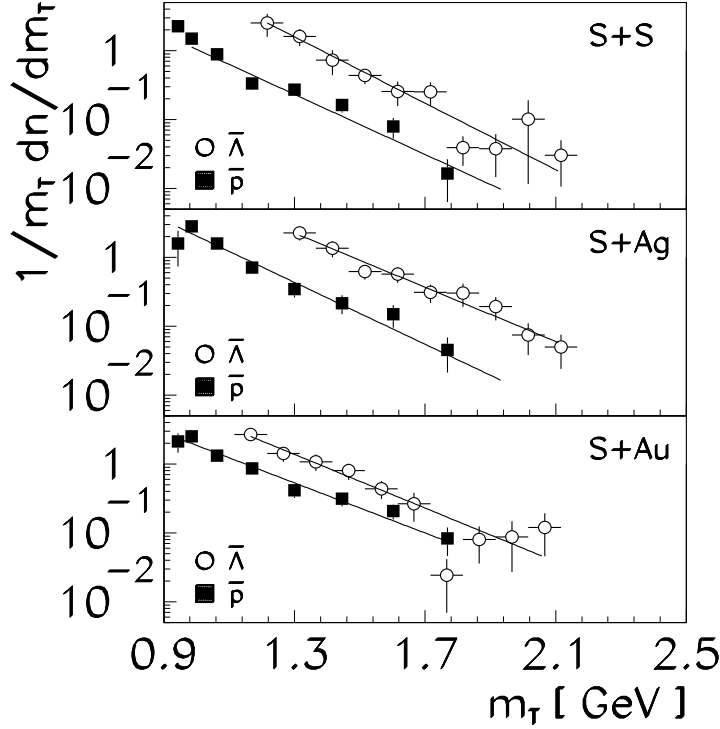,width=10.cm,height=11cm}}
\end{minipage}
\begin{minipage}[t]{5.5cm}
\vspace{-6.cm}
\caption{Transverse mass spectra of $\bar p$ and $\bar \Lambda$
for S+S, S+Ag and S+Au central collisions at 200 AGeV \protect\cite{na35}.
The solid lines correspond to exponentials fitted to the data.
}
\label{p_bar_lam_bar_sps}
\end{minipage}
%\vspace{-0.5cm}
\end{figure}

\vspace{.4cm}
A transverse mass spectrum of antiprotons from Pb+Pb at 158 AGeV  
has already been shown in section 2 (Fig.~\ref{mt_na44}). 
The large inverse slope
parameter of 278$\pm$9 which agrees with the one for protons
(289$\pm$7) was interpreted as a signature for transverse flow \cite{na44}.
Fig.~\ref{p_bar_lam_bar_sps} 
presents transverse mass spectra of $\bar p$ and $\bar \Lambda$
for S+S, S+Ag and S+Au central collisions at 200 AGeV \cite{na35}.  
The solid lines correspond to exponentials fitted to the data with
inverse slope parameters around 200 MeV. Similar values  
have been reported for antiprotons near midrapidity (175$\pm$6 for central 
S+S and 215$\pm$ for S+Pb collisions at 200 AGeV \cite{na44_95}).  
The $\bar \Lambda$/$\bar p$ ratio was found to be 1.9$^{+0.7}_{-0.6}$ for
central S+S and 1.1$^{+0.4}_{-0.3}$ for central S+Au collisions. 
These values are significantly larger than the ones determined for 
N+N (0.25$\pm$0.1), p+S (0.5$\pm$0.1) and p+Au (0.3$\pm$0.1) collisions.

Any interpretation of the ratios of antibaryon multiplicities should take
their multiplett structure as well as feeding from weak decays into account.
The antinucleons are an isospin doublet. Antiproton yields, therefore,
should be doubled to obtain antinucleon yields (in isospin symmetric
systems). The simple hyperons consist of $\Lambda,\Sigma^0,\Sigma^+,
\Sigma^-$. Normally only $\Lambda$s are detected, which contain all
$\Sigma^0$s because of the fast electromagnetic decay of the latter.
In NN interactions at SPS energies the relative abundance of
($\Lambda+\Sigma^0)/(\Lambda+\Sigma^0+\Sigma^++\Sigma^-$) was determined to be
0.6 \cite{wrob}. Thus the measured
$\Lambda$'s have to be scaled by a factor of 1.6 to obtain all hyperons.
The weak decays of 
multistrange hyperons lead
almost exculsively to $\Lambda$-hyperons. Since it is difficult to
identify the $\Lambda$'s stemming from $\Xi$-hyperon decays, the measured
$\Lambda$ yield is normally an overestimate of the true yield.
This effect is strongly amplified for antibaryons since the ratios
$\overline{p}/\overline{\Lambda}/\overline{\Xi}/\overline{\Omega}$
are much smaller than $p/\Lambda/\Xi/\Omega$ (see next section).
                                                 
\vspace{.4cm}
The significance of antibaryon yields and ratios for an assessment of
the differences between p+p and A+A collisions is weakened by their
large absorption  and strangeness exchange reactions in hadronic matter.
In-medium modifications of these cross sections may complicate the problem
even more. A realistic treatment  of these effects is a challenge for 
theory in general and for transport models in particular. There are two
further drawbacks of  the antibaryon observables: the feeding 
from heavier antihyperons and the incomplete coverage of phase space
in experiments. A large fraction of the data on antibaryons was obtained
from light projectiles (Si, S) on heavy targets in limited ranges of 
rapidities. Since baryon densities have a strong forward-backward 
asymmetry in these collision systems, antibaryon absorption will exhibit
a similar qualitative dependence. Thus it is dangerous to compare 
target rapidity data (Si+Pb) with results from S+Au forward from midrapidity.
Furthermore, baryon densities in central collisions are higher at AGS than at
SPS energies. Therefore it is again questionable to directly compare 
data on antibaryon yields and ratios for these two energy regimes. 
Our remark made earlier on the $\overline \Lambda$/$\overline p$ ratio
being indicative of the early abundance of $\overline s$ quarks relative
to ($\overline u$+$\overline d$) quarks holds only if the disappearance    
rate of antihyperons is similar to the absorption rate of antinucleons,
or if chemical equilibrium governs the $\overline \Lambda$/$\overline p$ ratio.

\vspace{.4cm}
Up to now there is no final answer to the question whether
the strangeness enhancement observed at SPS energies 
(i.e. K/$\pi$ or $\overline \Lambda$/$\overline p$ ratio)
can be understood in a purely
hadronic scenario as it is the case at SIS and
AGS energies  or whether new processes have to be invoked.
Since nuclear collisions are true multibody interactions involving the strong
force at low momentum transfer, a rigorous theoretical treatment of the
dynamical details of heavy ion collisions is not possible. Therefore
phenomenological models are used to predict the outcome of a nuclear collision.
In most of these models the elementary cross sections between hadrons are
used to calculate what happens to the incoming nucleons and produced
particles. Such models are necessarily based on many, to some extend
arbitrary, assumptions. It is beyond the scope of this survey to enter into a
discussion on the ingredients of the different models. Here we only
point out that they can describe all the features of meson production from SIS
to AGS energies albeit with sometimes different and contradicting assumptions
(for the special K$^-$ enhancement below threshold see chapter 4).
At SPS energies, however, 
the experimental data can be reproduced
by model calculations only if new collective
phenomena are invoked which imply either the fusion of strings 
(RQMD \cite{sorge95}, VENUS \cite{werner93})
or a modification of the strange quark content in the sea quarks 
(DPM \cite{capella}).

\section{Multistrange hyperons}

In a baryon-rich environment hyperons are formed more easily
than antihyperons. Hyperons may contain more than one strange quark.
The $\Xi^{0,-}$ and $\Omega^-$ are made up of two strange (+ a light)
and three strange quarks, respectively. Their production requires
the creation of two and three $s\overline{s}$ pairs and is therefore
strongly  suppressed. To first order an $\overline{\Omega}$ is produced
with the same probablilty as the $\Omega$, because both are made up
of newly created quarks only. The $\overline{\Xi}$ is suppressed relativ
to the $\Xi$, because an additional light antiquark has to be created
in the formation of the former. In A+A collisions one therefore expects
that the production probability behaves like $\overline{\Omega}/\Omega 
> \overline{\Xi}/\Xi > \overline{\Lambda}/\Lambda$. 
It has been speculated that in the relatively high $s\overline{s}$
density environment reached in a Quark-Gluon-Plasma, the multistrange
hyperons will be more enhanced than the simple $\Lambda$ and $\Sigma$
particles \cite{raf,koch} since the
hyperon abundance should be proportional to the density of the s-quarks
raised to the power n$_s$ with n$_s$
being the number of strange quarks in                                                                 
the respective hyperon. 
A similar behavior is found by microscopic model calculations 
which introduce in addition to the purely hadronic processes chromoelectric
flux tubes (colour ropes) \cite{sorge95}
and by the transchemistry models \cite{alcor} which describe 
the hadronization of quark matter.

\begin{table}
\caption{Particle multiplicities and ratios 
as measured in central Pb+Pb collisions at 158 AGeV
by NA49 in  comparison to the results of ALCOR model calculations 
\protect\cite{alcor}.
}
\vspace{0.5cm}
\begin{center}
\begin{tabular}{|c|c|c|c|}
\hline
particle type or ratio& NA49& reference & ALCOR\\
\hline
h$^-$& 680 & \cite{Jones}&  730\\
\hline
K$^0_s$ & 66 & \cite{bormann} & 63\\
\hline
participant baryons & 384 & \cite{App98}& 390\\
\hline
$\overline{\Lambda}/\Lambda$&0.2& \cite{bormann}& 0.3\\
\hline
K$^+$/K$^-$& 1.8& \cite{bormann}& 2.0\\
\hline
(K+$\overline{K}$)/$<\pi>$&0.13& \cite{bormann}& 0.13\\
\hline
$\Xi^-/\Lambda$&0.13$\pm$0.04&\cite{odyn97}&0.13\\
\hline
$\overline{\Omega}/\Omega$&0.42$\pm$0.12&\cite{holm97}&1.5\\
\hline
$\Omega^-/\Xi^-$& 0.19$\pm$0.04 &\cite{holm97}&0.18\\
\hline
$\overline{\Omega}/\overline{\Xi}_+$&0.30$\pm$0.09&\cite{holm97}&0.37\\
\hline
$(\Omega^-+\overline{\Omega})/(\Xi^-+\overline{\Xi}_+)$&0.21$\pm$0.03&
\cite{holm97}&0.25\\
\hline
\end{tabular}
\end{center}
\label{pmult}
\end{table}

\begin{figure}
%\vspace{-2.cm}
\begin{minipage}[t]{9cm}
\mbox{\epsfig{file=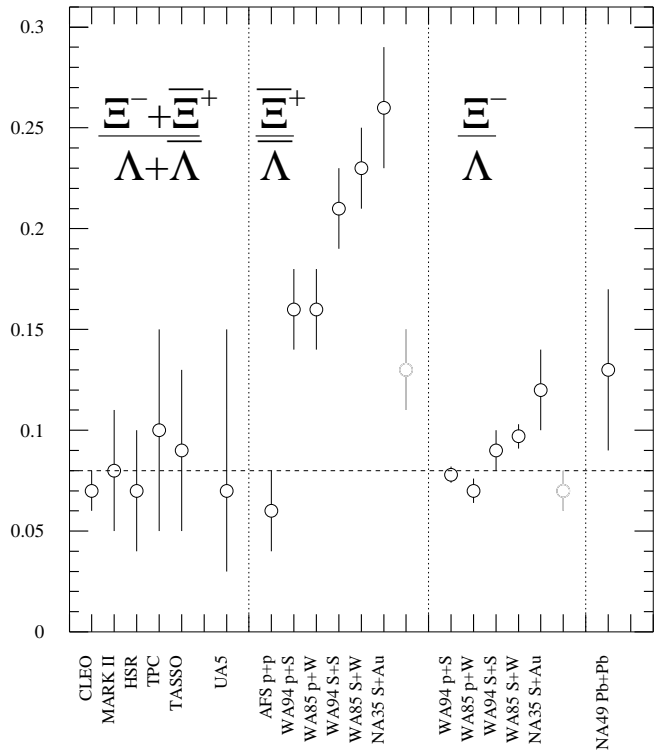,width=8.5cm}}
\end{minipage}
\hspace{0.5cm}
\begin{minipage}[t]{5.5cm}
\vspace{-4.5cm}
\caption{Comparison of $\Xi^-/\Lambda$ and 
$\overline\Xi^+$/$\overline\Lambda$ ratios obtained in different p+p,
p+$\overline p$, e$^+$+e$^-$ and A+A collisions at CERN SPS energies  
(taken from \protect\cite{odyn97}).}
\label{xi_rat_sps}
\end{minipage}
\end{figure}

Multistrange hyperons are difficult to detect because of their
complicated decay topology: for example, a $\Xi^-$ with $c\tau$=4.9 cm
decays into a $\Lambda\pi^-$ leading to a kink in the trajectory of a
negatively charged particle. The $\Lambda$ with $c\tau$=7.9 cm leaves
no visible track but creates the characteristic V$^0$ toplogy when it decays
with the V$^0$ pointing to the kink. The $\Omega^-$ has the same decay
topology with the $\pi^-$ replaced by a K$^-$ meson. In high energy
nuclear collisions these topologies are burried in the high multiplicities of
the charged particle tracks. Experimental data on multistrange hyperons
were published on p+$^{32}$S and  
$^{32}$S+$^{32}$S \cite{seyboth97}
for transverse momenta above 1 GeV/c. Preliminary data on $\Xi$ production
in $^{32}$S+$^{32}$S collisions  from NA35
at lower p$_T$ have recently been shown \cite{retyk97}.
New preliminary results from Pb+Pb collisions obtained by WA97 and NA49
also become available \cite{holm97,odyn97}
Some  relevant particle ratios including multistrange hyperons observed 
in Pb+Pb collisions are
summarized in Fig.~\ref{xi_rat_sps} \cite{seyboth97}. 
The ratios of multistrange to normal hyperons increase
with   system size and are larger for antihyperons than for hyperons.
In Table \ref{pmult},  
the experimental numbers for Pb+Pb collisions are compared 
to the predictions of the ALCOR model \cite{alcor}.

Particle ratios including multistrange antihyperons are affected by similar 
problems  as those including ''normal'' antihyperons and antinucleons.
The large strangeness exchange cross sections will influence strongly
the abundance of antihyperons  and it will be difficult or even impossible 
to calculate the different contributions to the particle yields in 
a nonequilibrium situation.  Only if the particle abundances come close to 
chemical equilibrium values one can hope to predict them quantitatively.
Microscopic model studies may help to verify whether such conditions  
are reached in central A+A collisions.

\chapter{Conclusions}
Hadronic particle production is an important and at high energies the dominant
process in nucleus-nucleus collisions. Therefore its study and understanding
is essential for the physics of highly compressed and heated 
hadronic matter. 
The fundamental theory for the description
of hadronic particle production is Quantum Chromodynamics. 
However, within the framework of this
theory soft processes cannot be calculated, neither for
nucleon-nucleon nor nucleus-nucleus collisions. Therefore,
phenomenological models are invoked to describe and interprete the
experimental results. These models  are especially helpful to identify
and understand differences in the particle production
characteristics between N+N and A+A collisions. The combination of
phenomenology and experiment in the analysis of nuclear collision data
offer the possibility to investigate
fundamental questions in strong interaction physics like
the nuclear equation of state (EOS),
in-medium modifications of hadron properties and
the deconfinement phase transition.
Other interesting topics concern thermal
and chemical equilibrium as well as energy and baryon densities reached
in nuclear collisions.

\vspace{.4cm}
The measured particle yields are in accordance with the assumption of
a fireball being close to chemical equilibrium, both
for A+A and p+p collisions with the exception that in the
p+p system strange particles are suppressed. The widths of rapidity 
distributions agree
with those from a thermal source at low beam (SIS) energies but are much larger
at high (CERN SPS) energies, again both for A+A and p+p reactions.
These observations at high beam energies  can be explained by
the formation of strings and their fragmentation into
hadrons which are equally distributed in phase space. This process  seems
to be a common feature to both p+p and A+A collisions.
In  central collisions of heavy nuclei and at all energies, the
transverse momentum distributions of the produced particles
show evidence for a transverse expansion of the fireball.

The pion phase-space distributions
suggest that the  bulk of the pions freeze out in a late and dilute  stage
of the  thermal fireball. At low (SIS) energies, however,
an azimuthally anisotropic emission pattern (varying with pion energy)
was found which is explained by rescattering of pions
from the spectators.  This observation indicates that
high-energy pions stem from an early, dense and hot phase of the reaction.

\vspace{.4cm}
The enhanced production of strange particles
has been observed at all beam energies.
At low (SIS) and intermediate (AGS) beam energies, the kaon to pion ratio
increases with the number of participating nucleons.
This experimental finding is a signature for the important role
of collective effects:
the energy needed to create a kaon is accumulated in sequential processes
involving more than two nucleons.
On the other hand, at high (SPS) energies the kaon to pion ratios stay
remarkably constant in the range from 50 to 300 participating nucleons.

\vspace{.4cm}
At beam energies below 1.6 AGeV,
collisions between high-energy pions
(or $\Delta$ resonances) and nucleons
are considered to be  the dominant process 
of ''subthreshold'' K$^+$  production.
This process depends sensitively on the baryon density
and on the thermal energy of the fireball and hence  on
the compressibility of nuclear matter.
The determination of the EOS far from normal nuclear density
is an old  challenge in nuclear physics
and is of great importance also for astrophysics.
Heavy ion collisions at beam energies of 1-2 AGeV create a
fireball with baryon densities of 2-3 $\rho_0$ at moderate temperatures
and thus  are well suited to study the properties of nuclear matter far
from its ground state.
The analysis of available kaon production data by transport calculations 
favors a soft EOS. 
Systematic experimental studies of K$^+$  production
for light and heavy collision systems and at different beam energies
are needed to  constrain considerably the range
of possible values for the nuclear compressibility.

\vspace{.5cm}
According to calculations based on effective chiral Lagrangians 
the  in-medium mass of the antikaon is predicted
to drop to about half its vacuum value at baryon densities of 2 $\rho_0$.
Experimental evidence for an enhanced K$^-$ production yield in
nucleus-nucleus collisions confirms the underlying concept.
Furthermore, the in-medium potentials of kaons and antikaons are expected to
influence considerably their azimuthal emission pattern.
Absence of directed in-plane flow of K$^+$ mesons  and an enhanced
K$^+$ out-of-plane emission have been explained by a repulsive KN in-medium
potential. A crucial test of the concept of in-medium KN potentials
will be the azimuthal distribution of antikaons which will be measured
in future experiments with KaoS and FOPI at SIS/GSI.

\vspace{.5cm}
From SIS to SPS energies a dramatic  change of the hadrochemical
composition of the  created fireball takes place:  a transition from the
baryonic to the pionic regime.  The pion/baryon ratio increases from about
0.1 - 0.2 at beam energies of 1-2 AGeV 
to about 1 at 10 AGeV and up to 5 at 100-200 AGeV.
The strangeness/baryon ratio increases  from
0.001 at 2 AGeV to 0.07 at 10 AGeV and up to 0.4 at 200 AGeV.
The experiments performed so far at AGS and SPS have
shown that the nuclei are still nearly stopped
creating energy densities as high as ten times the nuclear ground state
density.
Such an energy density is one of the necessary conditions for the creation
of the Quark-Gluon-Plasma. Some of the hadronic signals, which were
predicted to indicate its transient existence,  have also been seen.
Measurements of the J/$\Psi$ suppression as seen in $\mu^+\mu^-$ pairs also
ask for new nonhadronic phenomena. The question, whether the
QGP is already seen at SPS or even AGS energies has so far no commonly
accepted answer.
The study of the  phase transition from a state of deconfined quarks
and gluons
back to the hadronic world will be  the major aim of   new experiments at
new accelerators at BNL (STAR, PHENIX etc. at RHIC) and CERN
(ALICE at LHC).

\end{document}